\documentclass[11pt,a4paper]{report}
\usepackage{amsmath}
\usepackage{setspace}
\usepackage{xypic}
\usepackage[all]{xy}
\usepackage{latexsym}
\usepackage{theorem}
\newtheorem{teorema}{Theorem}[section]
\newtheorem{definicion}[teorema]{Definition}
\include{amslatex}
\newtheorem{proposicion}[teorema]{Proposition}

\newtheorem{corolario}[teorema]{Corollary}
\textheight
22 cm \textwidth 14 cm \oddsidemargin 7.5 mm \topmargin 0 mm
{\theorembodyfont{\rmfamily}
}
{\theorembodyfont{\rmfamily}
}

\numberwithin{equation}{section}

\include{amsmath}

\begin{document}
\paragraph{}
\paragraph{}
\paragraph{}
\begin{center}
{\LARGE {\bf AVERAGED DYNAMICS OF 
ULTRA-RELATIVISTIC CHARGED
PARTICLE BEAMS}}

\paragraph{}
\paragraph{}
\paragraph{}

{\large by
\paragraph{}
Ricardo Gallego Torrom\'{e}}

\paragraph{}
\paragraph{}

\paragraph{}
\paragraph{}
This thesis submitted for the degree of Doctor of Philosophy
\end{center}

\bigskip
\thispagestyle{empty}
\newpage
\begin{center}

{\bf Abstract}
\end{center}
\paragraph{}
In this thesis, we consider the suitability of using the
charged cold fluid model in the description of ultra-relativistic beams.
The method that we have used is the following. Firstly, the necessary notions of
kinetic theory and differential geometry of second order differential
 equations are explained. Then an
  averaging procedure is applied to a connection associated with the
  Lorentz force equation. The result of this averaging is an affine connection on
   the space-time manifold. The corresponding geodesic equation defines the
   averaged Lorentz force equation. We prove that for ultra-relativistic
   beams described by narrow distribution functions, the solutions of both
    equations are similar. This fact justifies the replacement of the Lorentz force
    equation by the simpler {\it averaged Lorentz force equation}. After this, for
     each of these models we associate the corresponding kinetic model, which
      are based on the Vlasov equation and {\it averaged Vlasov equation} respectively.
       The averaged Vlasov equation is simpler than the original Vlasov equation. This fact allows us to prove
        that the differential operation defining the averaged charged cold
         fluid equation is controlled by the {\it diameter of the distribution function},
          by powers of the {\it energy of the beam} and by the time of evolution $t$. We show that
          the Vlasov equation and the averaged Vlasov equation have similar
           solutions, when the initial conditions are the same. Finally, as an application
            of the {\it averaged Lorentz force equation} we re-derive the beam dynamics
             formalism used in accelerator physics from the Jacobi equation of the averaged Lorentz force equation.

\thispagestyle{empty}
\newpage
{\bf Acknowledgements.}

I would like to express my gratitude to Dr.
Volker Perlick for the many discussions on the material of this thesis, his advice and
 patience in the supervision of the thesis. I would also like to acknowledge to
 the members of the mathematical physics group of the physics department at Lancaster University
 for many discussions and advice in some matters concerning the material presented in this memoir
 and to the Cockcroft Institute
for support.
 My friends at Lancaster have also contributed to make this thesis possible and I express to them my gratitude.
\thispagestyle{empty}
\newpage
\paragraph{}
\paragraph{}
\paragraph{}
This thesis is dedicated to my parents, Luis and Elvira and to my brothers Luis Miguel, Julio and Elvira.
\thispagestyle{empty}
\tableofcontents{}
\chapter{Introduction}
\section{Motivation of the thesis}

Current models of classical electrodynamics of charged point particles contain logical inconsistencies that arise when
back-reaction effects are considered. For example, the standard theory of back-reaction
is based on the Lorentz-Dirac equation [1-5]. However, it is well known that the Lorentz-Dirac equation is
problematic from a physical point of view: some of its solutions
contain pre-acceleration effects; others are run-away solutions. This is the case for
 a large class of initial conditions. This peculiarity
of the Lorentz-Dirac equation is due to the fact that it is a third order
 differential equation.

A possible
solution to the problems of the Lorentz-Dirac equation is the theory
 proposed by Landau and Lifshitz in [2] (recently reviewed for instance in reference [5]).
From the analysis of this question performed in reference [5], one extracts the following conclusion:

{\it The charged point particle description is valid iff the changes in the
acceleration of the particle occur over time scales longer
 than the characteristic time parameter
$t_0=\frac{2}{3}\frac{q^2}{m}$}.

The parameter $q$ is the charge and $m$ is the mass of the point particle
 and one has assumed units such that the permeability of the vacuum $\epsilon_0$ is $1$
 and the speed of light $c$ is also set equal to $1$. If
the point particle approximation condition holds, the Landau-Lifshitz reduction
of order procedure can be applied to the
Lorentz-Dirac equation to obtain a second order differential
equation as an approximation, free from pathological solutions. In
 this regime, the Landau-Lifshitz equation can be considered an appropriate
approximation of the Lorentz-Dirac equation.

Despite solving the problem of the Lorentz-Dirac equation, there are
several reasons why the solution proposed by Landau and
Lifshitz is not completely satisfactory:
\begin{enumerate}
\item The {\it order reduction procedure} is an ad-hoc procedure (although
consistent with the point particle picture).

\item The Landau-Lifshitz equation is the leading order term approximation of the Lorentz-Dirac equation.
 Therefore, it is not a fundamental equation.

\item The characteristic time $t_0$ is
proportional to $\frac{q}{m}q$. Therefore, let us consider a
physical system with a large number of identical charged particles
performing a {\it collective motion}. The prototype example is the motion of a
bunch of particles in an accelerator machine.
It can happen that the behavior of the system is coherent and that one has to
read the factor $q$ as the total charge of the bunch and $m$ as the total mass of the whole bunch.
 Under these conditions, it is natural to consider that the factor $\frac{q}{m}$ remains the same
as for an individual charged point particle, but $q$ increases proportionally with the number of particles.
 Then for intense beams of particles, the point charge approximation and the reduction of
 order procedure will break down.
\end{enumerate}

Even if the first two points can be {\it covered} under the interpretation of
classical electrodynamics as the limit of the fundamental quantum electrodynamics, the third
 point has relevance for us.
The energies and luminosity achieved in modern particle accelerators can push to
 the validity of present models of electrodynamics its limits. This is basically
because one is dealing with
bunches containing a large number of charged particles, which can reach $10^{9}-10^{11}$
particles per bunch, moving together in a small
 phase-space domain (all the particles are concentrated around a center of
  mass, in position and velocity).

  Since the possible
effect discussed in point $3$ is additive, for modeling systems
like those bunches of particles, one needs an alternative description to Lorentz-Dirac and Landau-Lifshitz models.

In this context, fluid models have been used to study the dynamics of ultra-relativistic beams of charged
particles. One of these models is the proposal contained in [6]. In that work, it was shown how
 to do an asymptotic analysis of the charged cold fluid model.
 The main claim in [6] was that the
model proposed provides a self-consistent description of the fully coupled dynamics
 of a bunch of particles with the electromagnetic field. The reason for this is the
 smoothness properties of the fields, compared with the discrete and singular
   character of the point particle description behind the Lorentz-Dirac equation.

However, the use of the charged cold fluid model was not
justified in [6]. This justification is necessary, because of the discrete nature
 of a bunch of particles. Therefore, prior to the use of this model, one has to
address the following question:

{\it When is it a good approximation, in the regime of
ultra-relativistic dynamics,
 to describe the interaction of a large number of charged point particles
with the total (external and associated) electromagnetic field by a charged cold fluid model? }

A simplified version, is the following question:

{\it When is it a good approximation, in the regime of
ultra-relativistic dynamics,
 to describe the interaction of a large number  of charged point particles
with an external electromagnetic field by a charged cold fluid model? }

In the present thesis we address this second question. In
particular, we present an {\it averaged description} of the
collection of charged point particles, defining a {\it mean velocity vector
field}. The averaging operation is interpreted from a kinetic theory
point of view, introducing the $one$-particle distribution function as
a solution of the Vlasov equation [7,8] and the associated
averaged Vlasov equation, that we will introduce later. Our final result in this direction is contained in
{\it theorem 5.3.7}, which can be stated in words in the following way:

{\it For {\it narrow} distribution functions and in the {\it ultra-relativistic regime},
one can usehe t charged cold fluid model as a good approximation to the Vlasov model,
in the dynamical description of ultra-relativistic bunches of charged particles.
 The error of the approximation
  is of the same order as the area of the support of the distribution function in velocity space.}

What this result means is that, in this regime, if the Vlasov equation holds, the differential equation
 defining the charged cold fluid equation holds approximately. Also, we note that a precise statement is involved,
 requiring some technical assumptions which we will discuss in the appropriate place.

In order to achieve the above results, a connection associated with the Lorentz force
 connection will be introduced (the Lorentz connection).
We introduce an averaged version of this connection (the averaged Lorentz connection).
 The main advantage of this technique is that the averaged connection is simpler than
 the original one. This allows us to perform calculations whose results are not easily
 obtainable using to do in any other way.

\subsection{Other results of the thesis}
Another application of the theory of the averaged Lorentz model is the following. After introducing the
 Jacobi equation of an affine connection, we discuss the Jacobi equation associated with the averaged
 Lorentz connection.
  Then we prove that the linear dynamics used in accelerator physics [9-11] is an approximation to the
    Jacobi equation of the averaged Lorentz connection. Based
 on this interpretation, we define a notion of reference trajectory in
  beam dynamics in terms of observable quantities. Also, using the averaged Lorentz equation,
 we provide observable consequences of the collective nature of the bunch of particles.

  We have also considered the question of how the gauge invariance principle
 affects the interpretation of the Lorentz force equation as an Euler-Lagrange equation
 of a functional. This led us to a precise definition of semi-Randers space.

 Finally, we should mention that during the analysis of the main problem
 considered in this thesis and its mathematical formalization we found
  a generalization of the notion of {\it connection} in differential
 geometry. We call this object {\it almost (projective) connection}. This is described in the {\it appendix}.

\section{Structure of the thesis}

In this {\it chapter} we introduce
 some notation and conventions that we will follow through this thesis.

In {\it chapter 2}, an introduction to relativistic kinetic theory is provided, following reference [7].
 Then we define the charged cold fluid model and consider the asymptotic method developed in [6].
 We will state the main problem considered in this thesis and give a short out-line of the strategy to solve it.

In {\it chapter 3}, we introduce the theory of non-linear
connections defined by a second order differential system [12,13].
 We also introduce the formalism and the notion
of averaged connection, following the method already contained in ref. [22]. In particular we
define the average of linear connections acting on sections of some
relevant bundles (the pull-back bundles $\pi^*{\bf T}^{(p,q)}{\bf M}$). The data that we
 need to determine these connections is a system of second order differential equations called semi-spray.
 Those connections are obtained basically from the structure of the corresponding differential equations.
\paragraph{}
{\bf The original content of the Thesis constitutes { chapters} 4, 5, 6, 7 and { appendix 4}}.
\paragraph{}
In {\it chapter 4}, the notion of semi-Randers space is introduced and discussed
as a geometric description of the interaction of charged point particles with an external
electromagnetic field [14, 15]. Then the Lorentz connection is obtained.
Following the theory described in {\it chapter 3}, the corresponding averaged Lorentz connection  is determined.
It turns out that,
if the dynamics happen in the ultra-relativistic limit and the support of the probability
distribution function $f$ is narrow (in a sense to be specified), the solutions of the Lorentz force
 equation can be approximated by the solutions of the averaged Lorentz force equation.
We give an estimate of the approximation
 as a function of the time of evolution, the energy of the system and the diameter of the distribution.

In {\it chapter 5} it is proved that under the same assumptions as in {\it theorem 4.6.6},
 the relativistic charged cold fluid
model can be obtained as an approximation from a kinetic
model. The method that we follow to obtain this conclusion is the following.
 First, we introduce the averaged Vlasov equation and compare it with the original
  Vlasov equation. In particular we prove that, for the same initial conditions,
 both models have similar solutions
   in the ultra-relativistic regime when the distributions functions are
    narrow. After this,
    we use the averaged Vlasov model to give a bound on the acceleration of
 the main velocity vector field of the averaged Vlasov model.
     This bound is given in terms of the diameter of the distribution function,
 the energy and the time evolution.
      Then we prove that the mean velocity field associated with the solution
      of the Vlasov equation is similar to the mean velocity field obtained from the solution of the
       averaged Vlasov equation. This fact finishes the proof of {\it theorem 5.3.7},
        which is the answer to the main problem considered in this thesis.

In {\it chapter 6} we use the Jacobi equation of the averaged
connection to provide a geometric formulation of the transverse and
longitudinal linear beam dynamics. Particular examples illustrate the general
 formalism. We obtain from the Jacobi equation of the averaged Lorentz connection
 the equations of motion of the transverse dynamics in magnetic dipole and
 quadrupole fields. In a similar way, the longitudinal dynamics in a constant
 and alternating electric field are obtained from the Jacobi equation.
 Although these are known examples, they illustrate the usefulness of the Jacobi equation
 of the averaged connection in beam dynamics.
Corrections to the ordinary dynamics coming from {\it collective effects} are
considered. As an application of the formalism
 we can provide a definition of {\it reference trajectory} that
  by construction is given in terms of observable quantities.

In {\it chapter 7} we discuss some of the results presented in this thesis as well
as perspectives for further developments.

The present thesis work has produced the following articles and pre-prints [14], [15] and [16]:
\begin{enumerate}
\item R. Gallego Torrom\'{e}, {\it On the Notion of Semi-Randers
Spaces}, arXiv:0906.1940.

\item R. Gallego Torrom\'{e}, {\it Geometric Formulation of the
Classical Dynamics of Charged Particles in a External
Electromagnetic Field}, arXiv:0905.2060, submitted.

\item R. Gallego Torrom\'{e}, {\it Fluid Models from Kinetic Theory
using Geometric Averaging}, arXiv:0912.2767, submitted.

\item R. Gallego Torrom\'{e}, {\it Averaged Lorentz Dynamics
 and an Application in Plasma Dynamics}, arXiv:0912.0183, accepted to publish
 in the Proceedings of the XVIII Fall Meeting in Geometry and Physics, American Physics Society.

\end{enumerate}

\section{General conventions used in the thesis}

For the main physical applications in this thesis, the space-time structure will be a flat
four dimensional manifold
endowed with a Lorentzian metric $\eta$ with signature $(+,-,-,-)$.
However, some results and techniques are valid for arbitrary dimension, signature and curvature.
 In these cases,
it is explicitly stated. Sometimes we will require, to simplify the calculations,
that the metric $\eta$ is flat. When this is the case, it will be indicated.
In all cases we assume that the space-time manifold {\bf M} is time-oriented.

Einstein summation convention is considered for any identical and repeated
covariant and contravariant indices, if the contrary is not stated.
 All Latin indices run from $0$ to $n-1$, where $n$ is the dimension of the space-time manifold.
 Vector notation is used for the spatial components (with respect to a given frame) of a vector.
 Indices are lowered using the metric $\eta_{ij}$ and raised using the inverse
 metric $\eta^{ij}$, unless anything else is stated.
The exterior product and the exterior derivative are
normalized as in reference [17].

We have adopted the following convention
 for the physical parameters and constants appearing in the models,
\begin{displaymath}
\, q=1,\, m=1,\, \epsilon_0=1, \mu_0=1,\, c=\frac{1}{\sqrt{\epsilon_0\mu_0}}=1,
\end{displaymath}
where $\epsilon_0$ ${\mu_0}$ are the dielectric and magnetic
permeability constants of the vacuum; $m$ is the mass and $q$ the charge of the species of
 particles that we are considering.

We also use the following convention [3, pg 618]:
\begin{equation}
\vec{D}=\vec{E}+\vec{P},\quad \vec{H}=\vec{B}-\vec{M}.
\end{equation}
$\vec{P}$ is the polarization vector and $\vec{M}$ is the
magnetization of the medium. Since we are considering that the bunch of particle propagates in the vacuum,
 we have that $\vec{P}=0,\,
\vec{M}=0$. Hence,  one has $\vec{E}=\vec{D},\quad \vec{H}=\vec{B}.$

The Maxwell equations are
\begin{equation}
\vec{\nabla}\cdot\vec{E}=\rho,\quad
\vec{\nabla}\times \vec{B}=\big(\vec{J}+\frac{\partial
\vec{E}}{\partial t }\big),
\end{equation}
\begin{equation}
\vec{\nabla}\cdot\vec{B}=0,\quad
\vec{\nabla}\times \vec{E}=-\frac{\partial \vec{B}}{\partial t
}.
\end{equation}
The Maxwell equations can also be written in a covariant form in the following way:
\begin{equation}
\partial_i {\bf F}_{jk}\,+\partial_k{\bf F}_{ji}\,+\partial_j\,{\bf F}_{ik}=0,
\quad \quad ^{\eta}\nabla_i {\bf F}^i\,_j=\eta_{kj}J^k,
\end{equation}
where $^{\eta}\nabla_i$ is the covariant derivative associated to the Levi-Civita
connection along the direction $e_i$.

 The electromagnetic tensor is described by a $2$-form, that in
a local frame determines the following matrix:
\begin{displaymath}
{\bf F}_{ij}(x)=\left(
\begin{array}{cccc}
  0 & E_1 (x) & E_2(x) &  E_3(x) \\
  -E_1(x) & 0 & -B_3(x) & B_2(x) \\
  -E_2(x) & B_3(x) & 0 & -B_1(x) \\
  -E_3(x) & -B_2(x) & B_1(x) & 0 \\
\end{array}
\right).
\end{displaymath}
The Lorentz force is written as
\begin{equation}
\vec{F}:=q\big(\vec{E}+{\vec{v}}\times
\vec{B}\big),\quad
\vec{v}=\frac{d\vec{\sigma}}{dt},\quad \vec{F}=m\frac{d(\gamma \vec{v}}{dt}.
\end{equation}
The parameter $\tau$ is the proper-time along $\sigma$ associated with the metric $\eta$.
In covariant formalism, the Electromagnetic field
is described by a $2$-form ${\bf F}=\,{\bf F}_{ij}\, dx^i\wedge dx^j$.
The Lorentz force equation for a mass $m=1$ and a charge $q=-1$ is
\begin{equation}
\frac{d^2 \sigma^i}{d\tau^2}=\,-{\bf F}^i\,_j \,\frac{d \sigma^j}{d
\tau}.
\end{equation}
There are several categories of metric structures where we will work. The
most general is semi-Riemannian category [47]. The results stated in
this category refer to structures $\{\eta(x)\}$ which are non-degenerate,
 symmetric bilinear forms for each fixed $x\in {\bf M}$, and are smoothly
 defined on the $n$-dimensional manifold {\bf M}.

The second most general category refers to Lorentzian manifolds. In
this case the metrics $\{\eta(x)\}$ are billinear, symmetric forms with signature
$(+,-,-,-)$ for fixed $x\in {\bf M}$ and are smoothly
 defined on the manifold is four dimensional space-time because physical
 reasons. Those results be defined may also be defined for $n$-dimensional
 space-times with signature $(+,-,...,-)$.

The third category of geometry is Minkowski geometry. In this
case the metric is the Minkowski metric but the manifold considered
is a domain inside the space ${\bf R}^4$. In this category it makes sense
to speak of global inertial frames: a global inertial frame $(e_0, e_1, e_2,
e_3)$ on {\bf M} such that the metric $\eta$ has the following metric
components:
\begin{displaymath}
(\eta_{ij})=\left(
\begin{array}{cccc}
  1 & 0 & 0 &  0 \\
  0 &-1 & 0 & 0 \\
  0 & 0 &-1 & 0 \\
  0 & 0 & 0 &-1 \\
\end{array}
\right).
\end{displaymath}
The contraction operation is defined on the following way: given a
tangent vector $W\in {\bf T}_x{\bf M}$, there is an homomorphism on
the space of covariant tensors over $x$ denoted by:
\begin{align*}
\iota_W:{\bf T}^{(0,p)}_x{\bf M} &\longrightarrow {\bf
T}^{(0,p-1)}_x{\bf M }\\
T &\mapsto \iota_W T
\end{align*}
given by
\begin{displaymath}
\iota_W
T(X_1,...,X_{p-1}):=T(W,X_1,...,X_p),\quad \forall X_i \in {\bf
T}_x{\bf M},\quad W\in {\bf T}_x{\bf M}.
\end{displaymath}
A similar definition applies pointwise to sections of $\mathcal{F}(M)$-multilinear maps of vector fields
contracted with a given vector field
given by
\begin{displaymath}
\iota_W
T(X_1,...,X_{p-1})(x):=T(x)(W,X_1,...,X_p),\quad \forall X_i \in \Gamma({\bf
T}{\bf M}),\quad W\in \Gamma({\bf T}{\bf M}).
\end{displaymath}

When we write down the results, we try to formalize in the largest
category possible. Generally speaking, results from {\it chapter 3}
fall into the category of semi-Riemannian metrics (indeed, some
of them are even more generic that for metric structures). Results in
{\it chapter 4} fall into this category of semi-Riemannian
category, except for the main comparison results, where we explicitly use the flatness property
of the metric $\eta$ in some of the calculations.

In {\it chapter 5}, the results depend on the results of
{\it chapter 4}. Therefore, although some of them are formulated for
semi-Riemannian manifolds, the main results are formulated for compact
domains of the Minkowski space.

In {\it chapter 6}, the main results are formulated for Minkowski
space, since we have in mind to apply the geometric formalism to describe the
behavior of beams of particles in accelerators, where gravitational effects are usually neglected.

In {\it chapter 7}, we point out the general conclusions of this thesis as well as open problems proposed.

\chapter{Fluid and kinetic models for ultra-relativistic beams}

In this chapter we consider some basic notions that we will use later.
 In the same way, we introduce additional notation. There is also a short introduction
  to fluid models, kinetic models and to the asymptotic model described in [6].

\section{Basic relativistic kinetic theory}

In this {\it section} we review some elementary notions of the covariant kinetic theory
which are relevant for our work. We mainly follow the notation of reference [7].
We will consider collision-less processes and detailed balance processes.

\subsection{Intrinsic covariant formalism for relativistic \\kinetic theory}

In this thesis, the kinetic models are based on the following general assumptions:
\begin{enumerate}
\item The space-time manifold {\bf M} is $4$-dimensional and it is
endowed with a Lorentzian metric $\eta$. The signature of the metric
is $(1,-1,-1,-1)$, and the space-time is time orientable [7] and
oriented. In general,
 the metric $\eta$ is not flat. However we will use the flatness condition on
the metric $\eta$ in some of the calculations.

The Lorentzian metric $\eta$ has an associated Levi-Civita connection $^{\eta}\nabla$.
Also, $\eta$ determines the Hodge star operator
\begin{align*}
\star:{ \Gamma} {\bf \wedge}^{p} {\bf M} &\longrightarrow { \Gamma}
{\bf \wedge}^{4-p} {\bf M}\\
\omega_{i_1...i_p}\, e^{i_1}\wedge ...\wedge e^{i_p} &\mapsto
\omega_{i_1...i_p} \,e^{i_{p+1}}\wedge ...\wedge
e^{i_{4-p}}\,\epsilon^{i_1...i_p}\,_{i_{p+1}...i_{4-p}}.
\end{align*}
${ \Gamma} {\bf \wedge}^{p} {\bf M}:=\{\omega:{\bf M}\rightarrow
{\bf \wedge}^{p} {\bf M}\}$ is the set of smooth sections of the
vector bundle ${\bf \wedge}^{p} {\bf M}\rightarrow {\bf M}$, with
${\bf \wedge}^{p} {\bf M}$ the bundle of smooth $p$-forms over
${\bf M}$; $\epsilon^{i_1...i_p}\,_{i_p...i_{4-p}}$ is the total
skew-symmetric symbol, where the indices are raised using the
Lorentzian metric $\eta$ and with $\epsilon_{0123}=1$. $\omega_{i_1...i_p} e^{i_1}\wedge
...\wedge e^{i_p}$ is an arbitrary $p$-form expressed in a dual basis $\{e^0,...,e^3\}$
 of an orthonormal basis $\{e_0,...,e_3\}$.

The dual of a
vector field is the $1$-form defined pointwise by the relation
\begin{displaymath}
({V})^{\flat}(x)=\eta(V(x),\cdot).
\end{displaymath}
Similarly, the dual of a $1$-form is a vector field defined pointwise by the relation
\begin{displaymath}
\omega^{\sharp} (x)=\eta^{-1} (\omega, \cdot),
\end{displaymath}
where $\eta^{-1}$ is the bilinear form
\begin{displaymath}
\eta^{-1}:{\bf T}^*{\bf M}\times {\bf T}^*{\bf M}\longrightarrow {\bf R}
\end{displaymath}
\begin{displaymath}
(\omega, \phi)\mapsto \eta^{-1}(\omega, \phi):=\eta(\omega^{\sharp},
\phi^{\sharp}).
\end{displaymath}

\item The electromagnetic field is encoded in the $2$-form {\bf F}, while the excitation
field is encoded in the $2$-form ${\bf G}$ and the current
density is a $3$-form {\bf J}, all living on {\bf M}. They satisfy
Maxwell's equations, which can be written in terms of differential forms as
\begin{equation}
d{\bf F}=0,\quad\quad d \star {\bf G}={\bf J};
\end{equation}
 $d:\bigwedge^p{\bf M}\longrightarrow \bigwedge^{p+1}{\bf M}$ is the exterior
 derivative operator acting on forms. This is a
coordinate free form of the equations $(1.3.4)$.

\item The relation between ${\bf F}$ and ${\bf G}$ is given by the constitutive relations.
 We assume that these relations are linear and in particular we put ${\bf F}={\bf G}$, since
  the electromagnetic medium that we are considering is the vacuum.

\item The matter content of the models consists of a collection of identical charged point particles.
 The trajectory $\sigma(s)$ of each particle follows the Lorentz force equation
\begin{equation}
\frac{d y^i}{d \tau}={\bf F}^i\,_j (\sigma(\tau))\,y^j,\quad
y^j=\frac{d\sigma^j(\tau)}{d\tau},
\end{equation}
where $\tau$ is the proper-time associated with the trajectory.
The parameter $\tau$ is such that $\eta(\frac{d\sigma(\tau)}{d\tau},\frac{d\sigma(\tau)}{d\tau})=1$.

\item The support of the one-particle distribution function $f(x,y)$ is in the
$7$-dimensional unit hyperboloid bundle,
\begin{equation}
{\bf \Sigma}:=\{(x,y),\, x\in {\bf M},\,y\in {\bf T}_x{\bf
M},\,\eta(y,y)=1,\, y^0 >0\, \}.
\end{equation}
A global coordinate system on ${\bf \Sigma}$ is $(x^a,y^{i}),\,
a=0,1,2,3,\,i=1,2,3$, induced from any natural coordinate system on ${\bf TM}$.
In the unit hyperboloid, $y^0$ is given as a function of
$y^{1},\,y^2,\,y^3$ and $x^a$. The manifold ${\bf \Sigma}_x:=\{y\,\in {\bf T}_x{\bf
M},\,\eta(y,y)=1,\, y^0 >0\, \}$ is called the unit hyperboloid over $x$.

\item There is defined a volume form on the unit hyperboloid bundle ${\bf \Sigma}$.
 This volume form is obtained in terms of the metric $\eta$. On the
 tangent space ${\bf TM}$ there is a volume $8$-form
\begin{displaymath}
\sqrt{|det \eta|}\,dy^0\wedge\cdot\cdot\cdot dy^3\wedge
dx^0\wedge\cdot\cdot\cdot \wedge dx^3,\quad d^4(x)=dx^0\wedge\cdot \cdot\cdot dx^3,
\end{displaymath}
with $\sqrt{\eta}$ the determinant of the matrix associated to the
metric in a given coordinate system. The isometric embedding
$e: {\bf \Sigma} \hookrightarrow {\bf M}$ induces a volume form
on the manifold ${\bf \Sigma}$.
  We denote this volume form by $dvol(x,y)\wedge d^4 x$. Since the space-time manifold
 {\bf M} is $4$-dimensional, the volume form $dvol(x,y)\wedge d^4 x$ is
  a $7$-form.

   The volume form $dvol(x,y)$ on ${\bf \Sigma}_x$ is obtained contraction of $dvol(x,y)\wedge d^4x$
 on the orthogonal frame $\{e_0,...,e_3\}$: $d^4x(e_0,...,e_3)=1$.

\item The Liouville vector field $^L\chi$ is tangent to the hyperboloid
${\bf \Sigma}$.  Using the conventions of {\it section 1.3}, the
Liouville vector field $^L\chi$ can be written using local
coordinates as
\begin{equation}
^L\chi =y^i\partial_i
+(F^{i}\,_j\,y^j-\Gamma^{i}\,_{jk}\,y^jy^k)\frac{\partial}{\partial
y^{i}},\quad i,j,k=0,1,2,3.
\end{equation}
{\bf Remark} Note that we have adopted the {\it extrinsic} formalism, where $y^0$
is considered an independent coordinate. Later we will explain the relation
between the intrinsic and extrinsic formalism, and that they are equivalent for our purposes.

\item The one-particle distribution function $f(x,y)$ is defined
over ${\bf \Sigma}$ and satisfies the equation
\begin{equation}
^L\chi(f)=0.
\end{equation}
Equation $(2.1.5)$ corresponds to the Vlasov equation in plasma physics and kinetic theory.
 The one-particle distribution function $f(x,y)$ is introduced as the probability
density of finding a particle at the point $x\in{\bf M}$ with
velocity vector $y\in {\bf T}_x{\bf M}$ [7]. This interpretation was
supported in [7] using balance arguments and assuming that $f(x,y)$
is continuous. We will also assume additional smoothness and regularity conditions for $f(x,y)$.

\end{enumerate}

\subsection{Extrinsic formulation of the kinetic model}

There is an
alternative description
 of a kinetic model to the intrinsic one. In this alternative description the
 calculations are performed on the whole tensor
 bundle ${\bf TM}$ and then the results are restricted to the unit hyperboloid.
 One uses the constraint $\eta(y,y)=1$ when it is necessary.
Note that the action of $^L\chi$ is on the ring of smooth
functions of the hyperboloid ${\bf \Sigma}$, since $^L\chi$ is a tangent vector
 to the unit hyperboloid ${\bf \Sigma}$. This follows from the fact that
 $^L\chi|_{{\bf \Sigma}} \cdot(\eta(y,y))=0$ and that
the function $\eta(y,y)=\eta_{ij}(x) y^iy^j$ generates a foliation of {\bf TM}.
 A formal proof of this fact can be found for instance in [18].

We will also define later an averaged Liouville vector field
$<\,^L\chi>$. This vector field lives
 on {\bf M} rather than on the tangent bundle {\bf TM}. However,
 it defines a second order differential equation and therefore a Liouville
  equation. The flow of $<\,^L\chi>$ does not preserve the function $\eta(y,y)(x):=\eta_{ij}(x)y^i y^j$.
 Indeed, it is not guaranteed that the flow will preserve a {\it structure}.
Therefore for the study of these kind of flows, it is more convenient to adopt the external formalism.

Using the volume form $dvol(x,y)$
  one can obtain the velocity moments of the distribution $f(x,y)$.
Therefore one can define moments of the distribution function, which
are the expectation values of polynomials on $y$. With these
moments, one can define the mean velocity field, the
 covariant kinetic energy-momentum tensor and the covariant energy-momentum flux tensor:
\begin{equation}
 V^i(x)=\frac{1}{\int_{{\bf \Sigma}_x} f(x,y)\,
dvol(x,y)}\int_{{\bf \Sigma}_x} y^i f(x,y) \,dvol(x,y).
\end{equation}
\begin{equation}
T^{ij}(x)=\frac{1}{\int_{{\bf \Sigma}_x} f(x,y) \,dvol(x,y)}\int_{{\bf \Sigma}_x}
y^i y^j f(x,y) \,dvol(x,y),
\end{equation}
\begin{equation}
Q^{ijk}(x)=\frac{1}{\int_{{\bf \Sigma}_x} f(x,y) \,dvol(x,y)}\int_{{\bf \Sigma}_x}
y^i y^j y^k\,f(x,y) \,dvol(x,y),
\end{equation}
The balance equation for the number of particles implies the
relations [7]
\begin{equation}
^{\eta}\nabla_i V^i (x)=\frac{1}{\int_{{\bf \Sigma}_x} f(x,y)\,
dvol(x,y)}\int_{{\bf \Sigma}_x} \,^L\chi(f)\, dvol(x,y),
\end{equation}
\begin{equation}
^{\eta}\nabla_j T^{ij}(x)=\,{\bf F}^i\,_j\,V^j+\frac{1}{\int_{{\bf
\Sigma}_x} f(x,y)\, dvol(x,y)}{\int_{{\bf \Sigma}_x} y^i\, ^L\chi(f)\,
dvol(x,y)}.
\end{equation}
Since $f$ follows the Liouville equation $(2.1.5)$, one obtains
\begin{displaymath}
^{\eta}\nabla_i V^i (x)=0,\quad ^{\eta}\nabla_j T^{ij}(x)=\,{\bf
F}^i\,_j\,V^j .
\end{displaymath}

\section{Relativistic charged cold fluid model}

We introduce some geometric and physical objects that we need in the description of
 the asymptotic expansion of the relativistic cold fluid model proposed in
[6]. The electromagnetic field is encoded in the $2$-form ${\bf F}$,
which is a solution of the Maxwell equations $(2.1.1)$. The external
electromagnetic field ${\bf F}$ is created by the external current density
${\bf J}$ such that in the space time regions that we will consider, one has that
${\bf J}(x)=0$. The current density $\mathcal{ J}$ describes a
system of charged point particles which also contributes to the total
electromagnetic field. The whole dynamics is non-linear and one
needs additional
 information to completely determine the dynamics. There are two additional pieces of information:
\begin{enumerate}
\item One has to postulate the dynamic equation for the current density ${\mathcal{J}}$.
Examples for these joint dynamics are the Maxwell-Lorentz system,
Maxwell-Vlasov, Klimontovich-Maxwell's system [8, {\it section
2.5}]) and Maxwell-Lorentz-Dirac system [5].

\item
In order to completely determine the system, constitutive relations between
${\bf F}$ and ${\bf G}$ are needed.
We assume that the constitutive relations are ${\bf G}=\epsilon_0 {\bf F}$.
We have adopted units such that $\epsilon_0=1$.
\end{enumerate}

In flat regions, the metric $\eta$ admits a set of translational
Killing vectors $\{K_{i},\,i=0,1,2,3\}$, $\mathcal{L}_{K_{i}}\eta=0.$

Using differential forms, one can write conservation laws in a
geometric way. For any vector field $W$ on ${\bf M}$ there is an
associated {\it drive $3$-form} [17]:
\begin{displaymath}
\tau^{em}_W =\frac{1}{2}(\iota_W {\bf F}\wedge \star {\bf F}-\iota_W \star
{\bf F}\wedge {\bf F}).
\end{displaymath}
For the case of Killing vector fields, the exterior derivative of
$\tau^{em}_W$ is
\begin{displaymath}
d\tau^{em}_W =-\iota_W {\bf F}\wedge {\bf J},
\end{displaymath}
where $\iota_W {\bf F}$ is the contraction of the vector field $W$ with the $2$-form {\bf F}.
In the region outside of the sources ${\bf  J}=0$,
\begin{equation}
d\tau^{em}_W=0.
\end{equation}
In the presence of matter, equation $(2.2.1)$ has to be
generalized. For example, let us consider a model for matter
described by a time-like vector field $V$. Then for a {\it dust}, the stress-energy tensor is
\begin{equation}
T (x)= \mathcal{ N} {V}^{\flat}(x)\otimes {V}^{\flat}(x).
\end{equation}
$\mathcal{ N}$ is a regular scalar density field and the velocity
field is normalized, $\eta(V,V)=1$. The current density
$\mathcal{J}$ is proportional to the velocity field:
\begin{equation}
\mathcal{ J}=\mathcal{ N}\star ({V})^{\flat}.
\end{equation}
and then
$d\mathcal{J}=0$.
Combined with the assumption of the total momentum conservation
$d(\tau^{em}_W +\star \iota_W T)=0$
implies the field equation of motion for the fluid:
\begin{equation}
^{\eta}\nabla_V {V}(x)=(\iota_V{\bf F})^{\sharp}
\end{equation}
as a balance equation [17, pg 242-243], [19].

The
dynamics of the relativistic charged cold fluid model is described by the
following coupled system of differential equations,
\begin{equation}
d\mathcal{\bf F}=0,\quad d\star {\bf F}=-\rho \star
{V}^{\flat},\quad ^{\eta}\nabla_V {V}(x)=(\iota_V{ \bf F})^{\sharp},\quad
\eta(V,V)=1.
\end{equation}
Although this is a complete model, in this thesis we will work with external electromagnetic fields.
 In this context, the mathematical and physical analysis are highly simplified.

\section{Asymptotic expansion of the relativistic charged cold fluid model}

Let us consider the
following $1$-parameter family of differential forms and vector
fields:
\begin{equation}
V^{\epsilon} =\sum^{+\infty}_{n=-1} \epsilon^n V_n,\quad
\rho^{\epsilon} =\sum^{+\infty}_{n=1} \epsilon^n \rho_n,\quad {\bf
F}^{\epsilon} =\sum^{+\infty}_{n=-1} \epsilon^n {\bf F}_n ,
\end{equation}
where
\begin{equation}
V_n \in {\Gamma {\bf TM}},\quad \rho_n\in  \Gamma\wedge ^0{\bf
M},\quad {\bf F}_n \in \Gamma\wedge ^2{\bf M}
\end{equation}
and $\epsilon$ is a small parameter.
Substituting these expansions in equation $(2.2.5)$ and equating terms of equal in  $\epsilon$,
 one obtains enough conditions to
determine the fields $(2.3.2)$ inductively [6]. For instance, the leading
order terms are  the vector field $V_{-1}$ and the $2$-form
${\bf F}_{-1}$ such that:
\begin{displaymath}
^{\eta}\nabla_{V_{-1}} {V}_{-1}=\iota_{V_{-1}} {\bf F}_{-1},\quad d{\bf F}_{-1}=0,\quad d\star {\bf
F}=0,\quad \eta(V_{-1},V_{-1})=0.
\end{displaymath}
Given initial data for $V_{-1}$ and ${\bf F}_{-1},$ these equations are compatible.
Note that they describe a charged mass-less fluid interacting with an external electromagnetic fluid.

In general the equations for the fields appearing in the expansion $(2.3.1)$
 are obtained from the equations
$(2.2.5)$. The procedure for the higher orders is as follows:
\begin{enumerate}
\item Consider a given electromagnetic field ${\bf F}_{-1}$, solution
of the differential equations
\begin{equation}
d{\bf F}_{-1}=0,\quad d\star {\bf F}_{-1}=0
\end{equation}
for some initial value of ${\bf F}_{-1}$ on a space-like hypersurface.
Physically ${\bf F}_{-1}$ is interpreted as the external electromagnetic field.
\item Then one has to solve the equation
\begin{equation}
^{\eta}\nabla_{V_{-1}} {V}_{-1}=(\iota_{V_{-1}} {\bf F}_{-1})^{\sharp}
\end{equation}
subject to the condition
\begin{displaymath}
\eta(V_{-1}, V_{-1})=0
\end{displaymath}
and for given initial data in the space-like hypersurface $t=t_0$.
This is possible because equation (2.3.4) can be re-written as an ordinary
differential equation and one can apply standard results (see for instance the {\it appendix}).

\item Then one solves the equation for $\rho_1$, which is
\begin{equation}
d\star \big(\rho_1 {V}^{\flat}_{-1}\big)=0,
\end{equation}
for given initial values in a space-like hypersurface.
\item The $2$-form ${\bf F}_{0}$ is a solution to the Maxwell
equation, that can be written as
\begin{equation}
d{\bf F}_0=0,\quad d\star {\bf F}_0=-\star \rho_1{V}^{\flat}_{-1},
\end{equation}
where one has to specify the initial values on a space-like
hypersurface.
\item $V_0$ is the solution of the equation
\begin{equation}
^{\eta}\nabla_{V_{-1}}{V}_{0}+\,^{\eta}\nabla_{V_0}{V}_{-1}=(\iota_{V_{-1}}
{\bf F}_{0}+\iota_{V_{0}} {\bf F}_{-1})^{\sharp}
\end{equation}
subject to the requirement that $\eta(V_{-1}, V_0)=0$ and for fixed initial
 values of the vector field $V_0$ in a space-like hypersurface.

\item The density $\rho_2$ is defined as the solution of
\begin{equation}
d\star\big(\rho_2 {V}^{\flat}_{-1}) + d\star\big(\rho_2
{V}^{\flat}_{-1}\big)=0,
\end{equation}
again after given the initial values of $\rho_2$ on a space-like hypersurface.
\item The equation
\begin{equation}
 d\star{\bf F}_1=-\star
\rho_2{V}^{\flat}_{-1}-\star \rho_1 {V}^{\flat}_0
\end{equation}
can be solved for ${\bf F}_1$, once the initial values for ${\bf F}_1$ are specified.

\item $V_1$ is a solution of the equation
\begin{equation}
^{\eta}\nabla_{V_{-1}}{V}_{1}+\,^{\eta}\nabla_{V_0}{V}_{-1}+\,^{\eta}\nabla_{V_1}{V}_{-1}=(\iota_{V_{-1}}
{\bf F}_{0}+\iota_{V_{0}} {\bf F_{-1}}+ \iota_{V_{1}}{\bf V}_{-1})^{\sharp}.
\end{equation}
We need to specify the initial values on a space-like hypersurface.
\end{enumerate}

Through a generalization of this procedure, the fields $(2.3.2)$ can be solved order by
order in $\epsilon$. The only non-linear differential equation to be
solved is for $V_{-1}$. Indeed it can be written as an ordinary
second order differential
 equation for the integral curves of $V_{-1}$. These properties make it easier to solve
  both the analytical and numerical treatment of the problem than to solve the original equations $(2.2.5)$,
   which are a system of non-linear and coupled partial differential equations.

The vector field $V_{-1}$ has a difficult physical interpretation,
because it corresponds to a charged cold fluid composed of mass-less
particles and at the same time interacting with an external electromagnetic
field. There is no known
 classical physical system (in vacuum) with such characteristics (in quantum physics,
 the low energy limit of graphene admits states which are mass-less and interact
 with the electromagnetic field [56]).

 \section{Statement of the main problem considered in this thesis and out-line of the strategy to solve it}

It was claimed that the model introduced in [6] is able
 to provide a consistent treatment of the back reaction and self-force problems that
 appear in classical electrodynamics, for some situations which are of
  practical interest like ultra-relativistic plasmas. This claim is based
 on the assumption that the charged cold fluid model is an acceptable description of the dynamics of
  bunches of particles in the ultra-relativistic regime.

  On the other hand, fluid models have been used intensively in the
 description of the dynamics of plasmas [8, 37-39]. However, these
 usual models are based on assumptions on the moments of the
 distribution function, which are difficult to check in experimental conditions.

These motivate the main problem considered in this thesis:

{\it Is it mathematical justified to use the charged cold fluid model $(2.2.5)$
under the conditions present in the currently used particle accelerators?}

We will estimate the value of the
 differential operators appearing in the equation  $^{\eta}\nabla_V {V}(x)=(\iota_V{ \bf F})^{\sharp}$,
  where $V$ is the mean field $(2.1.6)$ for a given distribution function $f$.
 Then we will show in {\it chapter 5} that the
    differential expression for the charged cold fluid model equation is bounded and controlled
     by powers of the diameter $\alpha$ of the distribution function $f(x,y)$, powers of the
      {\it energy}\footnote{The notion of energy of a bunch of particles
 that we will use in those bounds is not trivial and will be introduced in {\it chapter 4}.}
 of the system and powers of the coordinate time evolution $t$.
      The relation is such that for narrow distribution functions
       and in the ultra-relativistic regime, the charged cold fluid model is a good
        approximation of the kinetic model.

        The strategy that we will follow is the following. Since the fluid
        model $V(x)$ is an approximate description of the system, we interpret
        $V(x)$ as an averaged quantity. On the other hand, given a
        dynamical system, we can associate a {\it non-linear connection}. That connection can
        be averaged, using a distribution function. If in addition, the
        difference between the original connection and the averaged connection
        is small, one can substitute the original one by the averaged connection
in the description of the dynamics.

        Any dynamical system described by a connection has an associated kinetic model.
        In particular, this is true for the Lorentz force equation and the averaged Lorentz
 force equation, that we will define. Since the Lorentz connection is similar to the averaged
 Lorentz connection in a sense that we will explain later,
 one can also substitute the associated kinetic models.

        Working with the averaged model has technical advantages. In particular one
 can give estimates of the value of some differential operators which appear in the
 fluid models. This  will be done for the charged cold fluid model.

\chapter{The averaged connection}

In this {\it chapter} we introduce the notions of non-linear
connections and the associated averaged connections, before applying the method
 to the connection associated with the Lorentz force equation in the next {\it chapter}.
 The construction is adapted from reference [22]. This {\it chapter} explains
 the mathematical theory that we will use in {\it chapters 4, 5 and 6}.

\section{Non-Linear connection associated with a second order differential
equation}
\subsection{Second order differential equations and the associated non-linear Berwald-type connection}

Let {\bf M} be an $n$-dimensional smooth manifold. A {\it natural
coordinate system} on the tangent bundle $\pi:{\bf
TM}\longrightarrow {\bf M}$ is constructed in the following way. Let $(x,{\bf U})$ be a local
coordinate system on ${\bf M}$ ,
where ${\bf U}\subset {\bf M}$ is an open sub-set of ${\bf M}$ and $x:{\bf
U}\rightarrow {\bf R}^n$ a local coordinate system. An arbitrary
tangent vector at the point $p\in {\bf U}$ is of the form
$X_p:=X=X^k\frac{\partial}{\partial x^k}|_p$. The local
coordinates associated with the tangent vector $X_p\in {\bf T}_x{\bf
M}\subset{\bf TM}$ are $(x^k,y^k)$. We will identify the point $x\in {\bf M}$ with its coordinates,
 by notational convenience. {\bf N} is a sub-bundle of the tangent bundle {\bf TM}.
From the imbedding $e:{\bf N}\hookrightarrow {\bf TM}$,
 $e({\bf N})$ acquires the induced differential structure from {\bf TM}; $e({\bf N})$ is denoted by {\bf N}.

We recall the following notion of connection [20, pg 314]. Let $\pi:{\bf N}\longrightarrow {\bf M}$
 be a bundle over {\bf M} and consider
the differential function $d\pi:{\bf TN}\longrightarrow {\bf TM}$.
Then the vertical bundle is $\mathcal{V}=ker (d\pi)\subset {\bf TN}$.
\begin{definicion}
A connection in the sense of Ehresmann is a distribution $\mathcal{H}\subset {\bf TN}$ such that
\begin{enumerate}
\item There is a decomposition
at each point $u\in{\bf N}$, ${\bf T}_u {\bf N}=\mathcal{H}_u\,\oplus \mathcal{V}_u.$

\item The horizontal lift exists for any curve $t\mapsto \sigma(t)\in
{\bf M},\,\,\,t_1 \leq t \leq t_2$ and is defined for each $\xi \in {\bf T}_u{\bf M}$ and $ u\in\pi^{-1}(x)$.
\end{enumerate}
\end{definicion}

\paragraph{}
Let us consider a set of $n$ second order differential
equations, with $n$ the dimension of {\bf M}. The solutions are parameterized curves on {\bf M}. Assume
that the system of differential equations describes the
flow of a vector field $ ^G\chi\in \Gamma{\bf TN}$. In
particular, the system of differential equations has the following form
\begin{equation}
\frac{d^2 x^i}{dt^2} - G^i(x,\frac{dx}{dt})=0,\quad i=1,...,n.
\end{equation}
This system of differential equations
is equivalent to the following system of first order differential
equations on {\bf N},
\begin{equation}
\left\{
\begin{array}{l}
\frac{d y^i}{dt} - G^i(x,y)=0,\\
\frac{d x^i}{dt}=y^i,\quad \quad\quad \quad
i=1,...,n.\\
\end{array} \right.
\end{equation}
The coefficients $G^i(x,y)$ are called {\it spray coefficients} if
they are homogeneous functions of degree one on the coordinate $y$; in
the general case where they are not homogeneous those coefficients are called semi-spray coefficients.
$G^i(x,y)$ transform
under a change of natural local coordinates on {\bf N}, induced from changes
of coordinates on {\bf M}, in such a way that the system of
differential equations (3.1.1) is {\it covariant}. Explicitly, if the
change of local natural coordinates on the manifold {\bf N} is
 \begin{displaymath}
\left\{
\begin{array}{l}
\tilde{x}^i =\tilde{x}^i(x),\\
\tilde{y}^i=\frac{\partial \tilde{x}^i}{\partial x^j} y^j,\\
\end{array} \right.
\end{displaymath}
then the associated co-frame transforms as
 \begin{displaymath}
\left\{
\begin{array}{l}
d\tilde{x}^i =\frac{\partial \tilde{x}^i}{\partial x^j}dx^j,\\
d\tilde{y}^i=\frac{\partial^2 \tilde{x}^i}{\partial x^k \partial x^j}y^k
dx^j + \frac{\partial \tilde{x}^i}{\partial x^j} dy^j.\\
\end{array} \right.
\end{displaymath}
The induced transformation in the associated system of differential
equations is
\begin{displaymath}
\frac{d{y}^i}{dt}- {G}^i({x},{y})=0\,\,\Rightarrow\,\,
\frac{d\tilde{y}^i}{dt}- \tilde{G}^i(\tilde{x},\tilde{y})=0,
\end{displaymath}
where the coefficients $\tilde{G}^i(x,y)$ are
\begin{displaymath}
\tilde{G}^i(\tilde{x},\tilde{y})=\,\sum_{j,k} \big(\frac{\partial \tilde{x}^l}{\partial
x^j}\big)y^j\big(\frac{\partial \tilde{x}^s}{\partial
x^k}\big)y^k\frac{\partial^2 \tilde{x}^i}{\partial x^l \partial
x^s} - \sum_j \big(\frac{\partial \tilde{x}^i}{\partial{x}^j}\big){G}^j({x},{y}).
\end{displaymath}

The vertical distribution $\mathcal{V}$ admits a local {\it
holonomic} basis given by
\begin{equation}
\{\frac{\partial}{\partial y^1},...,\frac{\partial}{\partial
y^n}\},\quad i,j=1,...,n.
\end{equation}
Using these spray coefficients it is possible
 to define a horizontal $n$-dimensional
distribution of the fiber bundle ${\bf TN}\longrightarrow {\bf N}$. The basis for the distribution is
\begin{equation}
\{\frac{\delta}{\delta x^1},...,
\frac{\delta}{\delta x^n}\},\quad\quad \frac{\delta}{\delta x^k}:=\frac{\partial}{\partial
x^k}-\frac{\partial{G^i}}{\partial y^k} \frac{\partial}{\partial
y^i},\quad i,j=1,...,n.
\end{equation}
It generates a supplementary distribution to the vertical
distribution.
The non-linear connection coefficients $N^i\,_k (x,y)$ are defined by the
relation
\begin{displaymath}
G^i (x,y):=y^k N^i\, _k (x,y).
\end{displaymath}
For a spray, the connection coefficients of the non-linear connection are:
\begin{displaymath}
N^i\,_k (x,y)=\frac{\partial G^i(x,y)}{\partial y^k}.
\end{displaymath}
Since the spray coefficients $G^i$ are transformed under the a change in natural coordinates in a well defined
way, the non-linear connection coefficients $N^i\,_j(x,y)$ are also transformed in a characteristic form [24, 34],
\begin{displaymath}
\big(\tilde{x}^i =\tilde{x}^i (x),\,\,\tilde{y}^i=\frac{\partial
\tilde{x}^i}{\partial x^j}y^j\big)\,\,\Rightarrow \,\,\tilde{N}^i\,_m (x,y)\,
\frac{\partial \tilde{x}^m}{\partial x^j}(x)=\,N^m_{j}(x,y)\,
\frac{\partial \tilde{x}^i}{\partial x^m}(x)+\,\frac{\partial^2 \tilde{x}^i}{\partial x^k\partial x^j}(x) y^k.
\end{displaymath}

 Given a spray $G^i(x,y)$, we define the connection coefficients such that the only
  non-zero coefficients correspond to the covariant derivative of
 horizontal sections of $\Gamma{\bf TN}$ along horizontal sections of ${\bf TN}$ and such that they are
 given by the Hessian of the spray:
\begin{displaymath}
\Gamma^i \,_{jk}(x,y):=\frac{1}{2}\, \frac{\partial^2 G^i(x,y)}{\partial y^j \partial y^k}.
\end{displaymath}
All the other coefficients are zero. This type of connection resembles the so-called Berwald
connection used in Finsler geometry [45]. From the point of view of the geometry of sprays, it is a natural
connection. Note that, while $\Gamma^i \,_{jk}(x,y)$ can be
associated with a linear connection on ${\bf TN}$, the connection
coefficients $N^i\,_j(x,y)$ cannot (this is why they are called non-linear connection coefficients);
${ N}^i\,_j(x,y)$
 determines a connection which is non-linear in the direction of the derivation.

Given the non-linear connection, one can define the horizontal lift of the tangent vectors;
the horizontal lift of $X =X^i\partial_i \,\in {\bf T}_x{\bf M}$ to the space
${\bf T}_u{\bf N}$ is defined by  $h(X)=X^i\frac{\delta}{\delta x^i}$.
This lift is defined here using local coordinates. However, an intrinsic definition can be found in [13].
After introducing this lift, one can define the horizontal lift of vector fields and
tensor fields of the corresponding bundles.

\subsection{The pull-back bundle}

Let us consider the product ${\bf N}\times{\bf TM}$ and the canonical projections
\begin{displaymath}
\pi _1:\pi ^* {\bf TM}\longrightarrow {\bf N},\quad
(u,\xi)\longrightarrow u,
\end{displaymath}
\begin{displaymath}
\pi _2 :\pi ^* {\bf TM}\longrightarrow {\bf TM},\quad
(u,\xi)\longrightarrow \xi.
\end{displaymath}
The pull-back bundle $\pi^* {\bf TM}\longrightarrow {\bf N}$ of the bundle
{\bf TM} is the {\it minimal} sub-bundle of the cartesian product
${\bf N}\times{\bf TM} $ such that the following equivalence
relation holds: for every $u\in {\bf N} $ and $(u,\xi) \in \pi^{-1}
_1 (u)$, $ (u,\xi)\in {\bf \pi^* TM}$ {iff} $\pi \circ\pi
_2(u,\xi)=\pi(u)$;
The pull-back bundle $\pi^*{\bf TM}\longrightarrow {\bf N}$ is such
that the following diagram commutes,
\begin{displaymath}
\xymatrix{\pi^*{\bf TM} \ar[d]_{\pi_1} \ar[r]^{\pi_2} &
{\bf TM} \ar[d]^{\pi}\\
{\bf N} \ar[r]^{\pi} & {\bf M}.}
\end{displaymath}
$\pi^* {\bf TM}\longrightarrow {\bf N}$ is a real vector bundle
with fibers diffeomorphic to  ${\bf T}_x{\bf M}$. For instance, let $\{e_i, i=0,...,n-1\}$
be a local frame for the sections of the tangent bundle ${\bf TM}$. Then $\{\pi^*e_i, i=0,...,n-1\}$
 is a local frame for the sections of the pullback bundle $\pi^*{\bf TM}$.
 Let $\xi^i(x,y)\pi^*_{(x,y)}e_i(x)$ be an arbitrary element in the fiber over
 $(x,y)\in {\bf N}$; the element $\pi^*_{(x,y)}e_i(x)$ is the unique element
 in $\pi^*{\bf TM}$ such that $(\pi\circ\pi_1)(\pi^*_{(x,y)}e_i(x))=(\pi_2\circ \pi)(\pi^*_{(x,y)}e_i(x))$
 and that $\pi_1\circ(\pi^*_{(x,y)}e_i(x))=e_i(x)$.

Another way to visualize this pull-back bundle is the following.
Let us consider the bundle $\pi:{\bf N}\longrightarrow {\bf M}$
and a fiber $\pi^{-1}(x)\subset {\bf N}$. On each point $u\in \pi^{-1}(x)$
 we attach a copy of the vector space ${\bf T}_x{\bf M}$. This assignment
 is done by the definition of $\pi^*$ on a local frame:
$\pi^*:\{e_1(x),...,e_n(x)\}\longrightarrow \{\pi^*|_u e_1(x),...,\pi^*|_u e_n(x)\}$
 and taking linear combinations of the elements of this local frame.
 When we consider sections of the bundle $\Gamma\pi^*{\bf TM}$ these
 linear combinations are $u$-dependent, instead of $x$-dependent.

Similarly, other pull-back bundles can be constructed from other tensor bundles
 over {\bf M}, for instance $\pi^* {\bf T^*M}\longrightarrow {\bf N}$
and $\pi^*{\bf T}^{(p,q)}{\bf M}\longrightarrow {\bf N}$, with ${\bf N}\subset {\bf TM}$
 a sub-bundle, ${\bf T}^*{\bf M}$ the vector bundle of $1$-form over
 {\bf M} and ${\bf T}^{(p,q)}{\bf M}$ the bundle of $(p,q)$-tensors over {\bf M}.

Given a non-linear connection on the bundle ${\bf TN}\longrightarrow
{\bf N}$, there are several related linear connections on the
pull-back bundle $\pi^*{\bf TM}\longrightarrow {\bf N}$.

Let $\chi$ a semi-spray defined on {\bf N}. We stipulate the following connection on $\pi^*{\bf TM}$,
 defined by the conditions
\begin{equation}
\nabla_{\frac{\delta}{\delta x^j}} \pi^* Z:= \,^{\chi}\Gamma(x,y)^i\,_{jk}\,Z^k\,\pi^*e_i,
\quad\quad \nabla_{V} \pi^* Z:=0,\quad V\in \mathcal{V}.
\end{equation}
Here $\{\pi^*e_i, \,i=0,...,n-1\,\}$ is a local frame for sections $\Gamma(\pi^*{\bf TM})$.
This connection can be generalized to general tensor bundles over ${\bf M}$.

\section{The average operator associated with a family of automorphisms}

\subsection{Average of a family of automorphisms}

The averaged connection was introduced in the context of positive
definite Finsler geometry in [22]. However, in this thesis
we need to formulate the theory for arbitrary linear
connections on the bundle $\pi^* {\bf T M}\rightarrow {\bf N}$,
where ${\bf N}\longrightarrow {\bf M}$ is a sub-bundle of
the tangent bundle ${\bf TM}\longrightarrow {\bf M}$.

 Let $\pi^* ,\pi _1 ,\pi _2 $ be the canonical
projections of the pull-back bundle $\pi^* {\bf T}^{(p,q)}{\bf
M}\rightarrow {\bf N}$, ${\bf T}^{(p,q)}{\bf M}$ being the tensor
bundle of type $(p,q)$ over ${\bf M}$, $\pi^* _u {\bf T}^{(p,q)}{\bf
M}$ the fiber over $u\in {\bf N}$ of $\pi^*{\bf T}^{(p,q)}{\bf M}$, ${\bf T}^{(p,q)}_x {\bf M}$ the
tensor space over $x\in {\bf M}$ $S_x$ a
generic element of ${\bf T}^{(p,q)}_x {\bf M}$ and $S_u$ is the
evaluation of the section $S\in\,\Gamma \big(\pi^* {\bf T}^{(p,q)}{\bf M}\big)$ at the point
$u\in{\bf N}$.

For each tensor $S_z \in {\bf
T}^{(p,q)} _z{\bf M}$ and $\, v\in
\pi^{-1}(z)$, $z\in {\bf U}\subset {\bf M}$ the following isomorphisms are defined:
\begin{displaymath}
\pi _2 | _v :\pi^*  _v {\bf T}^{(p,q)}{\bf M}\longrightarrow {\bf
T}^{(p,q)} _z{\bf M},\quad S_v\mapsto S _z
\end{displaymath}
\begin{displaymath}
\pi ^*  _v :{\bf T}^{(p,q)}_z{\bf M}\longrightarrow \pi^* _v {\bf
T}^{(p,q)}_z {\bf M} ,\quad S _z\mapsto  \pi^* _v S_z .
\end{displaymath}
To define the averaging operation we need two type of structures:
\begin{enumerate}
\item A family of non-intersecting,
oriented sub-manifolds
\begin{displaymath}
 {\bf N}_{{\bf U}}:= \bigsqcup_{x\in {\bf U}} {\bf
N}_x ,\, \quad {\bf
N}_x \subset{\bf T}_x{\bf M}.
\end{displaymath}

\item A measure at each point $x\in {\bf M}$, which is an element $f(x,y)\,\omega_x(y)\in
\bigwedge^{m} {\bf N}_x$, where $m$ is the dimension of ${\bf
N}_x$ and $f_x:{\bf N}_x\longrightarrow [0,\infty],\,f_x:=f(x,\cdot)$
 is required to have compact support on ${\bf N}_x$.
\end{enumerate}
Consider a family of endomorphisms,
$
\{A_w:\pi ^*_u {\bf TM} \longrightarrow \pi^*_u {\bf TM},\, u\in
\pi ^{-1}(x)\}.$
Let us consider the integral operations
\begin{displaymath}
\Big(\int _{{\bf N}_x}
 \pi
_2 |_u  A_u  \pi^* _u \Big)\cdot\,S(x) := \int _{{\bf N}_x}\big(
 \pi
_2 |_u  A_u  \pi^* _u \,S(x,u) \big)\,f(x,u)\,\omega_x (u),
\end{displaymath}
The volume function is defined as
 \begin{displaymath}
x\mapsto  vol({\bf N}_x):=\int_{{\bf N}_x}\omega_x (u)f(x,u).
 \end{displaymath}

\begin{definicion}
Consider a family of endomorphisms,
\begin{displaymath}
\{A_w:\pi ^*_w {\bf TM} \longrightarrow \pi^*_w {\bf TM},\, w\in
\pi ^{-1}(x)\}.
\end{displaymath}
The average endomorphism of this family is the endomorphism:
\begin{displaymath}
<A>_x : {\bf T }_x {\bf M} \longrightarrow  {\bf T}_x{\bf M}
\end{displaymath}
\begin{displaymath}
 S_x \mapsto\frac{1}{vol({\bf \Sigma}_x)}\Big(\int _{{\bf \Sigma}_x}
 \pi
_2 |_u  A_u  \pi^* _u \Big)\cdot\,S_x ,
\end{displaymath}
\begin{displaymath}
\quad u\in {\pi^{-1}(x)},\, S_x
\in {\bf \Gamma }_x {\bf M}.
\end{displaymath}
\end{definicion}
We denote the averaged endomorphisms by symbols between brackets.

 {\bf Remark}. There is a similar notion which applies to families of homomorphisms,
 instead of endomorphisms between different vector bundles.

 The averaging operation has the following effect. Let us consider an
 arbitrary tensor $S$ of a tangent space ${\bf T}^{(p,q)}_x{\bf M}$.
 Then the action of the integrand on $S$ is obtained as follows:
 \begin{enumerate}
\item First, $\pi^*_u S(x)$ {\it moves} $S$ from the fiber $\pi^{-1}(x)$ to
 the fiber $\{\pi^{-1}_1(u),\,u\in\pi^{-1}(x)\subset {\bf N}\}$ of the bundle $\pi^*{\bf T}^{(p,q)}{\bf M}$.

\item On this image, the operator $A_u$ acts: $A_u:\pi^{-1}_1(u)\longrightarrow \pi^{-1}_1(u)$.

\item The second projection again changes the fiber from $\pi^{-1}_1 (u)$ to the fiber $\pi^{-1}(x)$.
However, repeating this procedure for each $u\in {\bf U}$, being {\bf U} an open set. The
result is not an element of
${\bf T}^{(p,q)}_x{\bf M}$, since there is a dependence on $u\in {\bf
\Sigma}_x$.

\item The integration of this variable provides the desired element, eliminating the dependence on $u$.

\end{enumerate}
From this short discussion we observe that the geometric
interpretation of the average operation is quite subtle. We are actually seeking an intrinsic definition,
 besides the general one in [50].

One can prove the following fact. The averaging operator
acting on a element $S^i e_i$ is the following:
\begin{displaymath}
S^i(x) e_i(x) \mapsto \big(\int_{{\bf \Sigma}_x}\omega_x(u) [A_u ]^i\,_j
S^j(x) \big)e_i(x),
\end{displaymath}
where $[A_u ]^i\,_j$ is the coordinate representation on a given
basis of the linear operator $A$ at the point $u\in \pi^{-1}(x)$.

\subsection{Examples of geometric structures which provide an averaging procedure}

\begin{enumerate}
\item {\it Lorentzian structures} [23]. The geometric data is a Lorentzian
metric $\eta$ defined on {\bf M}. The disjoint union of the family of sub-manifolds ${\bf
\Sigma}_x\subset {\bf T}_x{\bf M}$ defines the fibre bundle $\pi:{\bf
\Sigma}\longrightarrow {\bf M}$,
 which we called the unit hyperboloid bundle over $x\in {\bf M}$,
\begin{displaymath}
{\bf \Sigma}:=\bigsqcup_{x\in {\bf M}}\{{\bf \Sigma}_x \subset{\bf T}_x{\bf M}\,\},\quad\quad {\bf
\Sigma}_x:=\{y\in {\bf T}_x{\bf M}\, | \,\eta(y,y)=1\}
\end{displaymath}
The manifold ${\bf \Sigma}_x$ is non-compact and oriented. The measure on ${\bf\Sigma}_x$ is given by
the following $(n-1)$-form
\begin{displaymath}
f(x,y)\omega_x(y):=f(x,y)\sqrt{\eta}\,\frac{1}{y^0}\,dy^1\wedge\cdot\cdot\cdot\wedge
dy^{n-1},\quad\quad y^0=y^0(x^0,x^1,...,x^{n-1},y^1,...,y^{n-1}),
\end{displaymath}
The function $y^0$ defines the parameterized hypersurface ${\bf
\Sigma}_x\subset {\bf T}_x{\bf M}$, since $y^0$ can be solved from
the condition $\eta_{ij(x)}y^i y^j=1$. This equation can be
expanded
\begin{displaymath}
\eta_{00}y^0 y^0\,+2\sum^{n-1}_{a=1}\eta_{0a}y^0y^a
\,+(\sum^{n-1}_{a,b=1}\eta_{ab}y^ay^b)=1.
\end{displaymath}
There are two type of solutions for $y^0$:
\begin{enumerate}
\item If $\eta_{00}\neq 0$, one obtains a two-fold hyperboloid
\begin{displaymath}
 y^0=\frac{1}{\eta_{00}}\Big(-\sum^{n-1}_{a=1}\eta_{0a}y^a\pm
\sqrt{(\sum^{n-1}_{a=1}\eta_{0a}y^a)^2\,-\eta_{00}((\sum^{n-1}_{a,b=1}\eta_{ab}y^ay^b)-1)}\,\Big).
\end{displaymath}

\item If $\eta_{00}=0$, the solution for $y^0$ is:
\begin{displaymath}
y^0=\frac{1-(\sum^{n-1}_{a,b=1}\eta_{ab}y^ay^b)}{2\sum^{n-1}_{a=1}\eta_{0a}y^a}.
\end{displaymath}
\end{enumerate}

Note that the Lorentzian metric $\eta$ does not determine the manifolds $\{{\bf
\Sigma}_x,\,x\in{\bf M}\}.$ For instance, one can consider $\widetilde{{\bf \Sigma}}$
 to be the collection of null cones over {\bf M}:
\begin{displaymath}
{{\bf \Sigma}}:=\bigsqcup_{x\in {\bf M}}\{{{\bf NC}}_x \subset{\bf
T}_x{\bf M}\},\, {{\bf NC}}_x:=\{y\in {\bf T}_x{\bf
M}\setminus\{0\}\, | \,\eta(y,y)=0\}.
\end{displaymath}
$\pi:{\bf NC}\longrightarrow {\bf M}$ is the null cone bundle over {\bf M} and ${\bf NC}_x$ is
the null cone over $x$; on the other hand, $e:{\bf NC}\hookrightarrow {\bf TM}$ is a
sub-bundle of ${\bf TM}\longrightarrow {\bf M}$.

\item {\it Finsler structures} [24, 45]. In this case, the Finsler function
 ${F}(x,y)$ defines the fundamental tensor $g_{ij}(x,y)=\frac{1}{2}\frac{\partial^2 F^2(x,y)}
 {\partial y^j\partial y^k}$ which is positive definite, homogeneous of degree
 zero on $y$, smooth and lives on the sub-bundle ${\bf N}:={\bf TM}\setminus \{0\}$. The bundle
${\bf \Sigma}$ is defined as the disjoint union,
\begin{displaymath}
{\bf \Sigma}:=\bigsqcup_{x\in {\bf M}}\{{\bf I}_x \subset{\bf
T}_x{\bf M}\},\,\,\, {\bf I}_x:=\{y\in {\bf T}_x{\bf M}\, |
\,F(x,y)=1\,\}.
\end{displaymath}
${\bf \Sigma}$ is the indicatrix bundle over {\bf M}. The manifold
${\bf I}_x$ is compact and strictly convex for each $x\in{\bf M}$
and is the indicatrix at $x$. The volume form is
\begin{displaymath}
\omega_x(u)=dvol(x,y):=\sqrt{g}\,\frac{1}{y^0}\,dy^1\wedge\cdot\cdot\cdot\wedge
dy^{n-1},\quad \quad y^0=y^0(y^1,...,y^{n-1}),
\end{displaymath}
where the function $y^0$ is a solution of the implicit equation
$F(x,y)=1$. We can see that locally this equation has a solution
using the implicit function theorem and the homogeneous properties
of the function F (in particular, using Euler's theorem of
homogeneous functions). If we take the derivative respect to $y^0$
of the function $\phi(x,y)=F(x,y)-1$ and we put it equal to zero, we then get
the condition of vanishing jacobian:
\begin{displaymath}
\frac{\partial}{\partial y^0}(F(x,y)-1)=\frac{\partial}{\partial
y^0}\big(g_{ij}(x,y)y^iy^j\,-1\big)=0.
\end{displaymath}
Using Euler's theorem one obtains:
\begin{displaymath}
0=\frac{\partial}{\partial y^0}(F(x,y)-1)=\big(2g_{0j}(x,y)y^j\,+2\frac{\partial }{\partial
y^0}(g_{ij}(x,y))y^iy^j\big)=
\end{displaymath}
\begin{displaymath}
=\big(2g_{0j}(x,y)y^j\,+\frac{\partial}{\partial y^0}(\frac{\partial^2 F^2(x,y)}
 {\partial y^j\partial y^k})y^jy^k\big).
 \end{displaymath}
Commuting the derivatives and considering Euler's theorem for homogenous functions,
 since $F$ is homogeneous on $y$, the above expression is
 \begin{displaymath}
=\big(2g_{0j}(x,y)y^j\,+y^j\frac{\partial}{\partial y^j}(\frac{\partial^2 F^2(x,y)}
 {\partial y^0\partial y^k})y^k\big)
 =2g_{0j}(x,y)y^j=0.
\end{displaymath}
Since the metric $g$ is positive definite, the only solution is
$y=0$, which is outside the indicatrix ${\bf I}_x$. Therefore, we can apply the
hypothesis of the implicit function theorem and the equation
$\phi(x,y)=0$ can be solved for $y^0$.

\item {\it Symplectic structures} [25]. In this case, there is defined on
${\bf T}^*{\bf M}$ a non-degenerate, closed $2$-form $\omega$. Due to Darboux's theorem [25, pg 246],
 there is
 a canonical local coordinate system of ${\bf T}^*{\bf M}$ such that the symplectic
 form $\omega$ can be written as
\begin{displaymath}
\omega=\sum^{n-1}_{i=0} dp_i\wedge dq^i.
\end{displaymath}
Associated with $\omega$ there is defined on the dual tangent bundle ${\bf T}^*{\bf M}$ a volume
$2n$-form
\begin{displaymath}
S=\omega\wedge\ \cdot \cdot \cdot \wedge \omega.
\end{displaymath}
Using canonical coordinates $(q,p)$, the $2n$-differential form can be written as:
\begin{displaymath}
S(p,q)= dp^0\wedge\cdot \cdot \cdot dp^{n-1} \wedge dq^0\wedge \cdot \cdot \cdot dq^{n-1}.
\end{displaymath}

 Let us assume the
existence of a nowhere zero vector field $V$ on {\bf TM} (therefore the Euler
characteristic of {\bf TM} must be different from zero). Then we can
construct the $(2n-1)$-form
\begin{displaymath}
\omega_q(p) =\iota_V S,\quad V\in\Gamma{\bf T}({\bf T}^*{\bf M})
\end{displaymath}
$\iota_V S$ is a non-degenerate $(2n-1)$ differential form
 whose value on $V$ is zero, since $\iota_Z \iota_Z S=0$ for any vector $Z$.
Let us chose a distribution of commuting vector fields,
 $\{X_i,\quad [X_i, X_j]=0,\quad i,j=1,...,2n-1\}$ locally supplementary to $V$
  such that $\{V,X_1,...,X_{2n-1}\}$ is a local frame of ${\bf TM}$.
  The distribution $\{ X_1,...,X_{2n-1}\}$
 is integrable and $\iota_V S$ is a volume form on the integral manifold
  ${\bf S}^{\bot}$. On the other hand ${\bf S}^{\bot}$ is a fibered manifold:
\begin{displaymath}
\pi:S^{\bot}:=\bigsqcup_{x\in {\bf M}} {\bf S^{\bot}}_x\longrightarrow
{\bf M}.
\end{displaymath}
Therefore we can define an averaging operation.

An interesting thing about this example is that we can only construct
{\it local averaging procedures}. The overlapping of open sets where the averaging operation is applied
non-trivial and in general one needs more structures to define consistently the averaging procedure globally.

\item {\it Hermitian Vector Bundles} [25]. The construction is
similar to the one in the {\it Finslerian case}. The sub-manifolds ${\bf
\Sigma}_x$ are defined as:
\begin{displaymath}
{\bf \Sigma}_x:= \{y\in{\bf T}_x{\bf M}\,| \,H(y,y)=1\},
\end{displaymath}
where $H$ is the hermitian structure on {\bf M}. Therefore
\begin{displaymath}
{\bf \Sigma}:=\bigsqcup_{x\in {\bf M}} {\bf \Sigma}_x.
\end{displaymath}
Let us assume that the hermitian structure is of the form $H=\eta+ \imath
\omega$, where $\eta$ is a Riemannian structure and $\omega$ is a
complex structure. To define the measure and the volume form we
can use either the complex structure $\omega$ or the Riemannian metric $\eta$.
\end{enumerate}

\subsection{Average operator acting on sections}

The averaging operation can be extended to a family of operators acting on sections of tensor bundles.
This is especially important for the next {\it
section}. Let $\pi^* ,\pi _1 ,\pi _2 $, $\pi^* {\bf
T}^{(p,q)}{\bf M}$ and ${\bf T}^{(p,q)}{\bf M}$ be as before. Then let us consider
the sections $S\in \Gamma({\bf T}^{(p,q)}{\bf M})$ and
$\pi^*S\in \Gamma( \pi^* {\bf T}^{(p,q)}{\bf M})$ and
the isomorphisms
\begin{displaymath}
\pi _2|_v :\Gamma(\pi^*  {\bf T}^{(p,q)}{\bf M})\longrightarrow
\Gamma({\bf T}^{(p,q)} {\bf M}),\quad S_v\mapsto S _z,
\end{displaymath}
\begin{displaymath}
\pi ^* :\Gamma ({\bf T}^{(p,q)}{\bf M})\longrightarrow
\Gamma(\pi^*{\bf T}^{(p,q)}{\bf M}) ,\quad S _z\mapsto \pi^*
_v S_z .
\end{displaymath}
Both isomorphisms are defined pointwise.
\begin{definicion}
Consider the family of fiber preserving endomorphisms
\begin{displaymath}
\{A({\bf W}):\Gamma (\pi^* {\bf TM}) \longrightarrow \Gamma \Big(\pi^* {\bf
TM}\Big),\, {\bf W}\in \pi ^{-1}({\bf U}),\,{\bf U}\in{\bf M}\}.
\end{displaymath}
The
averaged operator of this family is the map
\begin{displaymath}
<A>: \Gamma({\bf T_U }{\bf M}) \longrightarrow  \Gamma({\bf T_U}{\bf
M})
\end{displaymath}
such that at each point $x\in {\bf U}$ it is given by:
\begin{displaymath}
(<A> \cdot S)(x):=\frac{1}{vol({\bf N}_x)}\Big(\int _{{\bf
N}_x}
 \pi
_2 |_u  \big(A  \pi^* \cdot S \big)(u)\Big),
\end{displaymath}
\begin{displaymath}
\quad u\in {\pi^{-1}(x)},\, S \in {\Gamma } {\bf TM},
\end{displaymath}
where $\big(A  \pi^* \cdot S \big)(u)$ is the evaluation of the section $A  (\pi^* \cdot S )$ at $u$.
\end{definicion}
A similar definition holds if the operators act on cartesian products
of $\Gamma \big(\pi^*{\bf T}^{(p,q)}{\bf M}\big)$.

\section{Averaged connection of a linear connection on $\pi^*{\bf TM}$}

We adopt a differential volume form $f(x,y)\,\omega_x(y)$ such that $(d\omega_x(y))|_{{\bf N}_x}=0$.
Therefore we denote $\omega_x(y)|_{{\bf N}_x}=dvol(x,y)$.
Let us assume that a non-linear connection is defined on ${\bf \Sigma}$,
with ${\bf \Sigma}\longrightarrow {\bf M}$ a vector bundle.

\begin{definicion}
Let ${\bf M}$ be a $n$-dimensional smooth manifold, $\pi (u)=x$ and
consider a differentiable real function $f\in \mathcal{F}({\bf M})$.
Then $\pi^* f\in \mathcal{F}({\bf
\Sigma})$ is defined by the condition
\begin{equation}
\pi^* _u f=f(x).
\end{equation}
The horizontal lift of the tangent vector $ X^i\frac{{\partial}}{{\partial}x^i}|_x\in {\bf T}_x{\bf M}$ is
\begin{align}
h:\Gamma{\bf T}{\bf M} &\longrightarrow \Gamma {\bf TN}\\
  X^i\frac{{\partial}}{{\partial}x^i}|_x &\mapsto X^i\frac{{\delta}}{{\delta}x^i}|_{u},\,\,\,
u\in \pi^{-1}(x).
\end{align}
\end{definicion}
\begin{proposicion}
Let ${\bf M}$ be a $n$-dimensional manifold and assume that {\bf N} is endowed with a
non-linear connection, $u\in \pi^{-1}(x)\subset {\bf N}$, with $x\in
{\bf M}$. Let us consider a linear connection $\nabla$ defined on
the vector bundle $\pi^*{\bf TM}\longrightarrow {\bf N}$. Then a linear covariant derivative along $X$
$<{{\nabla}}>_X $ is defined on ${\bf M}$, and is determined by the following conditions:
\begin{enumerate}
\item $\forall  X\in {\bf T}_x {\bf M}$ and $Y\in \Gamma {\bf TM}$, the
covariant derivative of $Y$ in the direction $X$, is given by the
following averaging operations:
\begin{equation}
<\nabla>_X  Y:= <\pi _2|_u
{\nabla}_{{h}_u (X)} {\pi}^* _v Y\,>_u ,\,\, \forall v\in {\bf U}_u,
\end{equation}
where ${\bf U}_u$ is an open neighborhood of $u\in \pi^{-1}(x)$.

\item For every smooth function $f\in {\bf \mathcal{F}}{\bf M}$
the covariant derivative is given by the following average:
\begin{equation}
<\nabla>_X f:=<\pi _2 |_u \nabla _{h_u (X)} \pi^* _v f>_u
,\,\forall v\in {\bf U}_u.
\end{equation}
\end{enumerate}
\end{proposicion}
{\bf Proof}: it is shown in reference [22, {\it section 4}] or in the {\it appendix}.\hfill$\Box$

For the physical examples that we are interested, the notion of
volume that we use is obtained by isometric embedding of the ambient
 Lorentzian structure $\eta$  on the unit hyperboloid
times a positive weight function $f$. The
function $f_x:=f(x,\cdot)$ will be required later to be at least $L^1({\bf \Sigma}_x)$
and with compact support ${\bf \Sigma}_x$. This implies that the volume function is finite,
\begin{displaymath}
vol({\bf \Sigma}_x):=\int_{{\bf \Sigma}_x} f(x,y)\,dvol(x,y) <\infty.
\end{displaymath}
The manifold ${\bf N}_x$ is oriented. In particular, the
integration is performed in the unit tangent hyperboloid, which is
\begin{displaymath}
{\bf N}_x
:=\{y\in {\bf T} _x {\bf M},\,\mid\,\eta(y,y)=1,\, y^0>0 \}.
\end{displaymath}

Note that {\it proposition (3.3.2)} also holds in the more general
case where the function $f$ is not bounded and does not have compact
support, but all the relevant integrals (in particular the volume function
 and average of the connection coefficients) are finite.
\begin{definicion}(Generalized Torsion)
Let ${\nabla}$ be a linear connection on $\pi^*{\bf
TM}\longrightarrow {\bf N}$, then the generalized torsion tensor
acting on the vector fields $X,Y \in {\bf TM}$ is defined as
\begin{displaymath}
Tor(\nabla):{\Gamma}\pi^*{\bf TM}\times \Gamma\pi^*{\bf T M}\longrightarrow
{\Gamma}\pi^*{\bf TM}
\end{displaymath}
\begin{equation}
(\pi^*X,\pi^*Y)\longrightarrow
Tor({{\nabla}})(\pi^*X,\pi^*Y)=\nabla _{h  (X)} \pi^* Y
-\nabla _{h(Y)} \pi^*X -\pi^* [X,Y].
\end{equation}
\end{definicion}
This tensor is similar to the usual torsion tensor $Tor$,
\begin{displaymath}
Tor(\nabla):{\Gamma}{\bf TM}\times \Gamma{\bf TM}\longrightarrow
{\Gamma}{\bf TM}
\end{displaymath}
\begin{equation}
(X,Y)\longrightarrow Tor({{\nabla}})(X,Y)=\nabla _{X }Y -\nabla _{
Y} X - [X,Y].
\end{equation}
\begin{proposicion}
The averaged connection $<\nabla>$ has a torsion $Tor (<\nabla>)$ such that
\begin{equation}
Tor(<\nabla>)=<Tor(\nabla)>.
\end{equation}
\end{proposicion}
{\bf Proof:} It is shown in reference [22] and in the {\it appendix} of this thesis.\hfill$\Box$
\begin{corolario}
Let {\bf M} be an $n$-dimensional manifold and $\nabla$ a linear
connection on the bundle $\pi^* {\bf TM}\longrightarrow {\bf M}$
with $Tor(\nabla)=0$. Then $Tor({<{\nabla}>})=0$.
\end{corolario}
{\bf Proof}: It is directly shown with the proof of {\it proposition} $(3.3.4)$.\hfill$\Box$

If $Tor((<\nabla>))=0$ we say that the connection $(<\nabla>)$ is torsion free.
\begin{corolario}
Let {\bf M} be an $n$-dimensional manifold. If the connection
$\nabla$ on $\pi^*{\bf TM}$ has the connection coefficients $\Gamma^i\,_{jk}(x,y)$, then
the averaged connection $<\nabla>$ has the coefficients
\begin{equation}
<{\Gamma}^i\, _{jk}>(x)= \frac{1}{vol({\bf N}_x)} \int_{{\bf
N}_x} \Gamma^ i\,_{jk}(x,y)\,dvol(x,y).
\end{equation}
\end{corolario}
{\bf Proof:} Let $\{e_i\}$, $\{\pi^* e_i\}$, $\{h(e_i)\}$ be
local frames for the sections of the vector bundles ${\bf TM}$,
$\pi^*{\bf TM}$ and the horizontal bundle $\mathcal{H}$ respectively, such
that the covariant derivative is defined through the relations:
\begin{displaymath}
\nabla_{h(e_j)} \pi^* e_k =\Gamma^i\, _{jk} \pi^* e_i,\quad
i,j,k=0,1,2,...,n-1.
\end{displaymath}
Then let us take the covariant derivative
\begin{displaymath}
<\nabla>_{e_j}e_k=\frac{1}{vol({\bf
N}_x)}\Big(\int_{{\bf N}_x} \pi_2 (\nabla_{\iota(e_j)}
\pi^* e_k)\,dvol_x(y)\Big) =\frac{1}{vol({\bf
N}_x)}\Big(\int_{{\bf N}_x}\pi_2 \Gamma^i\, _{jk} \pi^* e_i\,
dvol_x(y)\Big)=
\end{displaymath}
\begin{displaymath}
=\frac{1}{vol({\bf N}_x)}\Big(\int_{{\bf N}_x}\Gamma^i\,
_{jk}\,dvol_x(y)\Big)
 e_i.
\end{displaymath}
The relation (3.3.8) follows from the definition of the connection coefficients of the averaged
connection,
\begin{displaymath}
<\nabla>_{e_j}e_k(x)=<{\Gamma}>^i\,_{jk}(x) e_i (x)
\end{displaymath}
 \hfill$\Box$

\chapter{Comparison of the Lorentz force equation and the averaged Lorentz force equation}

\section{Introduction}

We start this {\it chapter} considering several aspects of two different, although related topics.
 The first one is the notion of {\it (semi)-Randers space} in the category of metrics
  with indefinite signature. The second deals with a geometric interpretation
   of the Lorentz force equation and the associated {\it averaged Lorentz equation}.
    Both are related and the discussion of the first topic helps to understanding the second.

The discussion of the above leads to a framework where the results on the averaged Lorentz connection can be formulated properly. We also introduce a metric structure in the space of connections on some pull-back bundles. Using this metric structure, it is possible to compare geodesics of the Lorentz connection and averaged Lorentz connection, which is the main result of this {\it chapter}.
\paragraph{}
The notion of Randers space, which introduces a non-reversible space-time structure, can be traced back to the
original work by G. Randers [26]. The main motivation was the investigation
  of a geometric structure encoding the time asymmetry.
   One of the results of that study was a unifying theory of gravitation
    and electrodynamics of charged point particles,
     encoding both physical interactions in an unifying space-time metric structure.

However, one can consider that the non-degeneracy of the associated {\it metric tensor} $g(x,y)$ was not discussed in detail, in particular
when the issue of the gauge invariance associated with the electromagnetic potential is also considered. From a physical point of view, gauge invariance is a natural requirement. The combination of this requirement with the non-degeneracy
  criterion is non-trivial. Combining both has lead us to define Randers spaces in the context of pre-sheaf theory (this relation between Randers spaces and pre-sheave theory is only slightly treated in this thesis, since the details are still under construction).
   We hope that the identification of the appropriate formalism could provide
    a tool to formulate problems on Randers spaces in a consistent way.

 There are other difficulties associated with the signature of the tensor $g(x,y)$.
  Indeed, while for positive definite Finsler metrics
there is a satisfactory treatment (for instance [24, {\it chapter 11}]), for
indefinite signatures, the theory of Finsler spaces is less universally accepted and
 several proposals are currently being used in the literature.

\paragraph{}
There are two general formalisms for
 indefinite Finsler spaces (that we call
semi-Finsler structures): Asanov's formalism [27] and Beem's formalism [28],
[29]. We will argue why both treatments and the corresponding physical interpretations
 are unsatisfactory, in particular when we try to apply them to Randers-type spaces. We can  see
  the main problem with Asanov's definition when one considers gauge invariance
   issues related to the structure of Randers-type metrics. Also when one considers the possibility of light-like geodesics. The major problem with
   Beem's formalism is that there is no natural definition of Randers-type metric
   in that formalism. This is because Beem's formalism is based on
   homogeneous functions of degree two in the velocity variables $y$,
   while Randers-type functions are by definition homogeneous of degree one in $y$.

In {\it section 4.3} we will provide a definition of semi-Randers
space which is gauge invariant [14]. It has the advantage that all the notions involved are obtained directly
  from the Lorentz force equation and that it is a gauge invariant definition. However, this definition does not correspond to
 a  Finsler or Lagrange structure. Indeed,
  we show that even being possible, there are severe practical difficulties to find a Lagrangian definition of semi-Randers
   space which is at the same time gauge invariant and globally defined in the tangent space {\bf TM}. This happens even in the absence of topological obstructions like the existence of {\it monopoles} for the $1$-form $A$.
    Due to these difficulties we adopted a non-Lagrangian point of view in defining semi-Randers spaces.

In {\it section 4.4} we will propose a geometric description of the dynamics of one charged point
particle interacting with an external electromagnetic field. This interpretation is natural
in the framework for Randers-type space discussed in {\it section 4.3}. All relevant geometric
data is extracted from the semi-Riemannian metric $\eta$ and from the Lorentz force equation,
which in an arbitrary local coordinate system reads
\begin{equation}
\frac{d^2 \sigma^i}{d\tau^2} +\, ^{\eta}\Gamma^i\,_{jk} \frac{d
\sigma^j}{d\tau}\frac{d \sigma^k}{d\tau} +\eta^{ij}(dA)_{jk} \frac{d
\sigma^k}{d\tau}\sqrt{\eta(\frac{d \sigma}{d\tau},\frac{d
\sigma}{d\tau})}=0,\quad i,j,k=0,1,2,3,
\end{equation}
where $\sigma: {\bf I}\longrightarrow {\bf M}$ is a solution curve on {\bf M},
 $^{\eta}\Gamma^i\,_{jk}$ are the coefficients of the Levi-Civita connection
  $^{\eta}\nabla$ of $\eta$, $dA$ is the exterior derivative of the $1$-form $A$
   and the parameter $\tau$ is the proper-time of $\eta$ along the curve $\sigma$.
Then we interpret the equations $(4.1.1)$ as the auto-parallel
 condition of a linear connection (in a convenient bundle). We called it the Lorentz connection. This interpretation
  does not make any additional assumption beyond the information already
   contained in the system $(4.1.1)$ and the space-time metric $\eta$, except for
    some additional constrains on the generalized torsion, necessary to determine the connection coefficients completely.

     In {\it section 4.5} we obtain the averaged Lorentz connection
      associated with the Lorentz connection, applying the averaging method discussed in {\it chapter 3}.

In {\it section 4.6} we will compare the solutions of the original system (4.1.1) and those of the auto-parallel
 curves of the averaged Lorentz connection. The result is that for
the same initial conditions, in the {\it ultra-relativistic limit} and
for narrow $one$-particle probability distribution functions, the
solutions of both differential equations remain similar, even after a long
time evolution, since there is a competition between time evolution and other factors.
Therefore the original Lorentz force equation can be approximated by the averaged Lorentz force equation.

{\bf Remark}. The natural object extracted from equation $(4.1.1)$ is what
        we call {\it almost connection} (see the {\it appendix} for a formalization of the notion). In spite of
         this subtlety, we use through almost all the thesis the name connection
          (strictly speaking projective connection), since most of the calculations that
           we perform are also suitable for almost-connections (or {\it projective almost-connections}).

\subsection{On the physical interpretation of the formalism}
In the next sections we present a formal theory. However, the way the results are constructed
depends on the physical problems which motivated them. The main problem was to model the dynamics of
a bunch of particles in an accelerator machine, under the action of an external an
electromagnetic field. In this {\it chapter} we will consider the point particle dynamics point of view, which is related with the system of differential equations $(4.1.1)$.

There are some hypotheses in the results that we consider which are related with the main problem, although it is not explicitly mentioned:
\begin{enumerate}
\item Bound conditions. For instance, all the parameter and variables that will appear in our results are assumed bounded in compact domains {\bf K} of the space-time {\bf M}. The reason for this assumption is that we are trying to model systems like a bunch of particles in accelerator machines. The evolution of a bunch starts at a given instant such that $t=0$ with the {\it injection} and {\it separation} process of different bunches and the final time $t=T$, where the bunch reaches the target. Every coordinate and parameter of the model is bounded in ${\bf K}\subset {\bf M}$. The external electromagnetic fields are also bounded.

\item Ultra-relativistic regime. This is because the type of systems that we are describing are ultra-relativistic. We will define an energy function $E(x)$ which resembles the energy function used in accelerator physics. The ultra-relativistic regime happens when  $E(x(t))>>1$,where the mass of the specie of particle composing the bunch is set equal to $1$ and $c=1$.

\item Narrow distributions. This is one of the characteristics of the bunches in an accelerator machine. The narrowness of the distribution function is defined through the diameter $\alpha$ of the distribution function in the velocity space. The narrowness condition means that this diameter is small compared to the rest mass of the charged particles, $\alpha<<1$.

\item Adiabatic evolution. It is true that the change in energy is very slow compared to the energy itself, once the ultra-relativistic regimen has been reached. This is expressed by the condition $\frac{d log E}{dt}<<1$.
\end{enumerate}
The following reasons show why we have adopted the system of differential equations (4.1.1) as starting point for our geometrization of the electrodynamics of point particles are:
\begin{enumerate}

\item It seems that there is not a satisfactory and simple geometrization metric formalism for the interaction of a charged particle with an interacting external electromagnetic field (this is the main conclusion of sections 4.1-4.3).

\item The system of differential equations (4.1.1) is simple, contains all the symmetries that we are interested in and describes all the phenomenology of the dynamics of the charged point classical particles.

\item There exists an standard theory of geometric differential equations and its associated non-linear connections.

\item This geometric theory of differential equations provides the framework to apply the geometric averaging procedure described in {\it chapter 3}.

\end{enumerate}

\section{Criticism of the notion of semi-Randers space as space-time structure}

\subsection{Randers spaces as space-time structures}

Before moving to the more specific problem of defining semi-Randers spaces, let us discuss
 the notion of semi-Finsler structure.
Let ${\bf M}$ be a $\mathcal{C}^{\infty}$ $n$-dimensional manifold, {\bf TM}
its tangent bundle manifold with ${\bf TM}\supset {\bf N}$ and with projection $\pi:{\bf
N} \longrightarrow {\bf M}$, the restriction to ${\bf N}$ of the canonical projection $\pi:{\bf TM}\longrightarrow {\bf M}$. Therefore, $({\bf N},\pi)$ is a sub-bundle of {\bf TM}.

Let us consider the following two standard definitions of semi-Finsler
structures currently being used in the literature:
\begin{enumerate}
\item {\it Asanov's definition} [27],
\begin{definicion}

A semi-Finsler structure $F$ defined on the $n$-dimensional
manifold ${\bf M}$ is a positive, real function  $F:{\bf
N}\longrightarrow ]0, \infty [$ such that:
\begin{enumerate}
\item It is smooth in ${\bf N}$,

\item It is positive homogeneous of degree $1$ in $y$,
$
F(x,{\lambda}y)=\lambda F(x,y),\,\,\,\forall \lambda
>0,$
\item The vertical Hessian matrix
 \begin{equation}
g_{i j}(x,y): =\frac{1}{2}\frac{{\partial}^2 F^2
(x,y)}{{\partial}y^{i} {\partial}y^{j} }
\end{equation}
is non-degenerate on ${\bf N}$.
\end{enumerate}
$g_{ij}(x,y)$ is the matrix of the fundamental tensor.
The set $
{\bf N}_x$ is the {\it admissible set} of tangent vectors at $x$;
 the disjoint union ${\bf N}=\bigsqcup_{x\in {\bf M}}
{\bf N}_x$ is the admissible set of vectors over {\bf M}.
\end{definicion}
In the particular case when the manifold is $4$-dimensional and
$(g_{ij})$ has signature $(+,-,-,-)$, the pair $({\bf M},F)$ is a Finslerian
space-time.

\item {\it Beem's definition} [28], [29]
\begin{definicion}

A semi-Finsler structure defined on the $n$-dimensional manifold
${\bf M}$ is a real function  $L:{\bf TM}\longrightarrow {\bf R}$ such
that

\begin{enumerate}
\item It is smooth in the slit tangent bundle $\tilde{{\bf N}}:={\bf
TM}\setminus \{0\}$

\item It is positive homogeneous of degree $2$ in $y$,
$
L(x,{\lambda}y)=\lambda^2\, L(x,y),\,\,\,\forall \lambda
>0,$
\item The Hessian matrix
 \begin{equation}
g_{ij}(x,y): =\frac{1}{2}\frac{{\partial}^2 L
(x,y)}{{\partial}y^{i} {\partial}y^{j} }
\end{equation}
is non-degenerate on $\tilde{{\bf N}}$.
\end{enumerate}
\end{definicion}
\end{enumerate}
In the particular case when the manifold is $4$-dimensional and
$g_{ij}$ has signature $(+,-,-,-)$, the pair $({\bf M},L)$ is a Finslerian
space-time.

\subsection{Comparison of Asanov's and Beem's formalism}

Some differences between the above definitions are highlighted
below:
\begin{enumerate}

\item In Beem's framework there is a geometric definition of light-like vectors and it is possible to derive
Finslerian geodesics, including light-like geodesics, from a
variational principle [30]. By construction, in Asanov's
formalism it is not possible to do that in an invariant way, because
light-like vectors are excluded in the formalism from the beginning,
since nothing is said about how to extend the function $F^2$ from ${\bf {N}}$ to {\bf TM}.

\item Let $\Theta({\bf M})$ be the set of all piecewise smooth paths $\sigma:{\bf I}\longrightarrow {\bf M}$.
 In Asanov's framework, given a parameterized
path $\sigma:{\bf I}\longrightarrow {\bf M},\, {\bf I}\subset{\bf
R}$ on the semi-Finsler manifold $({\bf M},F)$ such that $\dot{\sigma}\in {\bf N}$
for all $t\in {\bf I}$, the length
functional acting on $\sigma$ is given by the following expression:
\begin{displaymath}
\mathcal{E}_A:\Theta({\bf M})\longrightarrow {\bf R}
\end{displaymath}
\begin{equation}
\sigma(t)\mapsto \mathcal{E}_A(\sigma):=\int^{t_{max}}
_{t_{min}} F (\sigma (t),\dot{\sigma}(t)) dt,\quad {\bf
I}=[t_{min},t_{max}].
\end{equation}
Due to the homogeneous condition of the Finsler function $F$,
$\mathcal{E}_A$ is a re-parametrization invariant functional. On the
other hand, if we consider Beem's definition, the energy
functional is given by the following expression [30]:
\begin{displaymath}
\mathcal{E}_B:\Theta({\bf M})\longrightarrow {\bf R}
\end{displaymath}
\begin{equation}
\sigma(t)\longrightarrow \mathcal{E}_B(\sigma):=\int^{t_{max}}
_{t_{min}} {L (\sigma(t),\dot{\sigma}(t))} dt,\quad {\bf
I}=[t_{min},t_{max}].
\end{equation}
Formulated in this way, Beem's energy functional is not
re-parametrization invariant, because the fundamental function $L$ is homogeneous of
degree two in $y$.

\item A third difference emerges when we consider the category of Randers-type spaces:
\begin{definicion}(semi-Randers Space as semi-Finsler Space)

In Asanov's framework, a semi-Randers space is characterized by a semi-Finsler function
of the form:
\begin{equation}
F(x,y)=\sqrt{\eta_{ij}(x) y^i y^j}+A_i(x)y^i,
\end{equation}
where ${\eta}_{ij}(x)dx^i\otimes dx^j$ is a
semi-Riemannian metric defined on {\bf M} and $A(x,y):=A _i (x)y^i$ is the result of the action of the $1$-form $A(x)=A_i (x)dx^i$ on $y\in{\bf T}{\bf M}$.
\end{definicion}
In the positive definite case and when $\eta$ is a Riemannian metric,
the requirement that $g_{ij}$ is non-degenerate implies that the
$1$-form $(A _1,...,A _n )$ is bounded by ${\eta}$:
\begin{displaymath}
A _i A _j {{\eta}}^{ij}< 1,\,\quad
{\eta}^{ik}{\eta}_{kj}=\delta^i\, _j.
\end{displaymath}
The indefinite case is quite different, since there is not a natural Riemannian
 metric that induces a norm in the space of homomorphisms. Therefore, the
  criterion for non-degeneracy is not trivial and further structure is required.

 Secondly, for both the positive definite and indefinite metric, only
   the variation of the length functional (4.2.3) (for fixed initial and final point variations)
 is invariant under
 the gauge transformation $A\mapsto A+d\lambda$ (not directly the integrand itself); the Finsler function $(4.2.5)$ is not gauge invariant as well.
Even in the case that we could define a metric and norm, transforming the $1$-form $A$
 by a gauge transformation can change the norm and therefore the hessian $(g_{ij})$ can become degenerate.

On the other hand, the notion of Randers space in Beem's formalism is even more problematic.
 In this case there is not a formulation of semi-Randers spaces (because eq. (4.2.5) is positive homogeneous of degree one in $y$). This suggests that a proper formulation of the notion of semi-Randers space requires going beyond metric structures.

\item It is interesting to have a definition of
semi-Randers structure capable of taking light-like trajectories for charged particles into account. As we have seen, Asanov's treatment is not able to consider light-like vectors. On the other hand, Beem's formalism is not capable of considering this problem, since there is not a known Randers-type structure in Beem's formalism. However, the
asymptotic expansion of the ultra-relativistic charged cold fluid model presented
in [6] is an example where those light-like trajectories appear naturally.
In that model, the leading order contribution to the mean velocity field of the charged cold fluid is a
light-like velocity vector field, interacting with the external electromagnetic field; perturbative corrections change the velocity vector field to a time-like vector field.
\end{enumerate}

The above observations make it reasonable to introduce a
non-metric interpretation for semi-Randers spaces. The option that we have adopted
has been to formalize a geometric structure from the geometric and physical data that
we have: the Lorentz force equation and the Lorentzian metric $\eta$. This will lead us to solve some of the problems mentioned before. It provides a rigorous framework to discuss further developments.

\section{Non-Lagrangian notion of semi-Randers space}
\subsection{Non-Lagrangian notion of semi-Randers space}

Let us assume the existence of a smooth semi-Riemannian
 structure $\eta$ on the manifold {\bf M}. This implies that the function
\begin{displaymath}
\eta:{\bf TM}\times{\bf TM}\longrightarrow {\bf R}
\end{displaymath}
\begin{displaymath}
(X,Y)\mapsto \eta_{ij}(x)X^iY^j,\quad X, Y\in\,{\bf T}_x{\bf M}
\end{displaymath}
is smooth in the variables $x, X^i, Y^j$. Since we will use the
square root $\sqrt{\eta_{ij}(x)\,X^iY^j}$, we also require that
$\sqrt{\eta_{ij}(x)\,X^iX^j}$ is smooth in $^{\eta}{\bf N}:=
\bigcup_{x\in {\bf TM}}\, \{X\in\, {\bf T}_x{\bf M},\,\,\eta_{ij}(x)\,X^iX^j>0\,\}$.
The null-cone is $^{\eta}{\bf NC}:=\bigsqcup_{x\in {\bf M}}\,\{y\in {\bf T}_x{\bf M}\,|\,\eta(y,y)=0\}$.
We propose a notion of semi-Randers space based on the following
\begin{definicion}
A semi-Randers space consists of a triplet $({\bf M},\eta
,{\bf F})$, where ${\bf M}$ is a space-time manifold, $\eta$ is a semi-Riemannian
metric continuous on {\bf TM} and smooth on ${\bf TM}\setminus \,^{\eta}{\bf NC}$ and a $2$-form
 ${\bf F}\in { \bigwedge}^2 {\bf M}$ such that $d{\bf F}=0$.
\end{definicion}
${\bf F}$ is in the second de Rham cohomology group $H^2( {\bf M})$. Due to
Poincar\'{e}'s lemma, there is a locally smooth $1$-form $A$ such that
$dA={\bf F}.$ Any pair of locally smooth $1$-forms $\tilde{A}$ and $A$ such that $\tilde{A}=A+d\lambda$,
with $\lambda$ a locally smooth real function defined on the given open neighborhood,
 are {\it equivalent} and produce under exterior derivative the same cohomology class
  $[{\bf F}]\in\,H^2( {\bf M})$ that contains the element ${\bf F}$:
  $d(d\lambda +A)=dA={\bf F}$. Note that we are speaking of locally smooth
  $1$-forms ${A}$ and of globally smooth $2$-forms ${\bf F}$. Therefore,
instead of giving ${\bf F}$, one can consider the equivalence class of $1$-forms $A$,
\begin{displaymath}
[A]:=\{\tilde{A}=A+d\lambda,\,\, dA={\bf F}\, \,\textrm{in the intersection of
 the opens sets where $A$ and $\lambda$ are defined}\,\},
\end{displaymath}
with $A$ a locally smooth $1$-form defined on the open set ${\bf U}\subset{\bf M}$,
 $\tilde{A}$ a locally smooth $1$-form defined on the open set $\tilde{{\bf U}}\subset{\bf M}$
  and $\lambda$ a locally smooth function defined on ${\bf U}\cap \tilde{{\bf U}}$.
 Then if two locally smooth forms $1$-forms $^{\mu}A$ and $^{\nu}A$, representatives of $[A]$,
  are defined on $^{\mu}{\bf U}$ and $^\nu{\bf U}$ respectively, one has that
   that $d(\,^{\mu}A-\,^{\nu}A)=0$. For each point of the open
    neighborhood $^{\mu \nu}{\bf U}=\,^{\mu}{\bf U}\cap\,^{\nu}{\bf U}$, there
     is a locally smooth function defined in an open neighborhood of $x$
      ${\bf U}(x)\subset\,^{\mu \nu} {\bf U}$  such that $(\,^{\mu}A-\,^{\nu}A)=d(\,^{\mu \nu}\lambda)$
       (a consequence of the Poincar\'{e} lemma).

Based on these arguments, we give an alternative definition of semi-Randers space:
\begin{definicion}
A semi-Randers space consists of a triplet $({\bf M},\eta
,[A])$, where ${\bf M}$ is a space-time manifold, $\eta$ is a semi-Riemannian
 metric continuous on {\bf TM} and smooth on ${\bf TM}\setminus \,^{\eta}{\bf NC}$ and
  the class of locally smooth $1$-forms $A$ is defined such that $dA={\bf F}$ for any $A\in [A]$.
\end{definicion}
\begin{proposicion}
These  definitions of semi-Randers space are equivalent.
\end{proposicion}
{\bf Proof}. We proved already one of the directions of the equivalence. To show the other direction,
 one needs to construct locally $1$-forms which produce the required $2$-form
 ${\bf F}$ under exterior differentiation. This is achieve by the
 Poincar\'{e} lemma in a star-shaped domain [32, pg 155-156]. The formula for the $1$-form $A$ is
 \begin{displaymath}
 A(x)=\big(\,\int^1_0\, t\sum^{n-1}_{k=0} x^k\,{\bf F}_{kj}(tx)\,dt\big)dx^j.
 \end{displaymath}\hfill$\Box$

  We will adopt {\it definition 4.3.2}, since it has the advantage that it allows a discussion of
  some local issues related with the inverse variational problem of
   the Lorentz force equation. Essentially, this is the reason that even if the topology of ${\bf M}$ is
    trivial, the $1$-forms $A$ are only locally smooth.

As we have learned from the discussion above, a proper treatment of semi-Randers spaces
 combined with gauge invariance requires consideration locally smooth potentials.
  There are also locally smooth functions and local{\it compatibility conditions}.
  These kind of structures are formalized by the notion of pre-sheaf structure
   (and the related notion of sheaf structure) [32, 33].

Let us denote the set of locally smooth functions over {\bf M} by $\bigwedge^p_{loc} {\bf M}$. This is a pre-sheaf structure. The pre-sheave of locally smooth functions on open sets of {\bf M} is denoted by $\mathcal{F}_{loc}({\bf M})$. Given a Lorentz semi-Randers structure $({\bf M}, \eta, [A])$, for each of the representatives $A\in [A] \in \bigwedge^1_{loc} {\bf M}$, there is on {\bf M} a function $F_A$ defined by
the following expression:
\begin{equation}
F_A(x,y)=\left\{
\begin{array}{l l}
\sqrt{\eta_{ij}(x)y^i y^j}+A_i(x)y^i &\quad \textrm{for  } \,\,\eta_{ij}(x)y^iy^j\geq 0, \\
 \sqrt{-\eta_{ij}(x)y^i y^j}+A_i(x)y^i &\quad \textrm{for  } \,\,\eta_{ij}(x)y^iy^j\leq 0. \\
\end{array} \right.
\end{equation}
The following properties follow easily from definition $(4.3.1)$ and from the definition ${F}_A$:
\begin{proposicion}
Let $({\bf M,\eta, [A])}$ be a semi-Randers space with $\eta$ a semi-Riemannian metric,
 $A\in [A]$ and $F_A$ given by equation $(4.3.1)$. Then
\begin{enumerate}
\item On the null cone ${\bf NC}_x:=\{y\in{\bf T}_x{\bf M}\,|\,
\eta_{ij}(x)\,y^i y^j=0\}$ $F_A$ is of class $\mathcal{C}^0$, for $\eta
\in \mathcal{C}^0$ and $A\in \bigwedge^p_{loc} {\bf M}$.

\item The subset where $\eta_x(y,y)\neq 0$ is an open subset of ${\bf T}_x{\bf M}$
 and $F_A$ is smooth on ${\bf T}_x{\bf M}\setminus {\bf NC}_x$, for $\eta$ smooth and $A\in \bigwedge^1_{loc} {\bf M}$.

\item The function $F_A$ is positive homogeneous of degree $1$ in $y$.
\end{enumerate}
\end{proposicion}
{\bf Remark}. There is no constraint on the non-degeneracy of the fundamental tensor $g_{ij}$.
Therefore, it is not required that any representative $A$ of $[A]$
 be bounded by $1$, as of is the case for positive definite Randers spaces [24, {\it chapter 11}].

\subsection{Variational principle on semi-Randers spaces}

Let $^t\Theta({\bf M})$ be the set of piecewise smooth curves on {\bf M} with time-like tangent vector field.
  The functional acting on $\sigma$ is given by the integral:
  \begin{displaymath}
  \mathcal{E}_{F_A}: \Theta({\bf M})\longrightarrow {\bf R}
  \end{displaymath}
\begin{equation}
\sigma\mapsto  \mathcal{E}_{F_A}(\sigma):=\int _{\sigma}
F_A(\sigma(\tau),\dot{\sigma}(\tau))\,d\tau
\end{equation}
 with $\tau$ the proper time associated with $\eta$ along the curve $\sigma$.
The functional is gauge invariant up to a constant: if we choose another representative
 $\tilde{A}=A+d\lambda$, then $\mathcal{E}_{F_A}(\sigma)=
 \mathcal{E}_{F_{\tilde{A}}}(\sigma)+ constant$, the constant coming from the boundary terms of the integral.
  Therefore, the variation of the functional is well defined on a given
 semi-Randers space $({\bf M},\eta, [A])$, for fixed initial and final points variations.

In order to guarantee the construction of the first
variation formula and the existence and uniqueness of the corresponding solution it
 is necessary that the vertical Hessian $g_{ij}$ be non-degenerate [31]. However,
 given a representative $A\in [A]$, one can not guarantee that the Hessian of $F_A$
   is non-degenerate. Due to the possibility of doing gauge transformations in
    the representative $A(x)\mapsto A(x)+d\lambda(x)$ we have
\begin{proposicion}
Let $({\bf M,\eta, [A])}$ be a semi-Randers space.
Assume that the image of the curve $\sigma$ on the manifold ${\bf M}$ is a compact subset.
Then
\begin{enumerate}
\item There is a representative $\bar{A}\in [A]$ such that the Hessian of the
 functional $F_{\bar{A}}$ is non-degenerate.

\item The functional $(4.3.2)$ is well defined on the Randers space $({\bf M,\eta, [A])}$,
 except for a constant depending on the representative $\bar{A}\in [A]$.

\item If the geodesic curves are parameterized by the proper time associated
with the Lorentzian metric  $\eta$, the Euler-Lagrange equation of the functional $F_A$ is the Lorentz force equation.
\end{enumerate}
\end{proposicion}
{\bf Proof}: There are several steps in the proof:
\begin{enumerate}
\item Using the gauge invariance of $\mathcal{E}_{F}(\sigma)$
 up to a constant, we can obtain locally an element $\bar{A}\in [A]$ such
  that $|\bar{A}_i \bar{A}_j\,\eta^{ij}| <1$ in an open neighborhood in the following way.
  Consider that we start with a $1$-form $A$ which is
    not bounded by $1$. The  $1$-form
    $\bar{A}(x)=A(x)+d\lambda(x)$ is also a representative of $[A]$. The requirement that the hessian of $\bar{A}$
     is non-degenerate is, using a generalization of the condition [24, pg 289]
\begin{displaymath}
0<\,\Big|2+\frac{\bar{A}_i(x)y^i+\,\sqrt{\eta_{ij}(x)y^iy^j}\,(\eta^{ij}\,\bar{A}_i(x)
 \bar{A}_j(x))}{\sqrt{\eta_{ij}(x)y^iy^j}+\,\bar{A}_i(x)y^i}\Big|
\end{displaymath}
This condition is obtained in [24] relating the determinant of the
 metric $\eta_{ij}$ and the determinant of $g_{ij}$, the fundamental tensor of a Randers-type metric structure. Although the authors are considering positive definite metrics, the result is valid for arbitrary signatures of the metric $\eta_{ij}$, for $y\in {\bf N}_x$

On the unit tangent hyperboloid ${\bf \Sigma}_x$ this condition reads
\begin{displaymath}
0<\Big|2+\frac{\bar{A}_i(x)y^i+\,\eta^{ij}(x)\,\bar{A}_i(x) \bar{A}_j(x)}{1+\,\bar{A}_i(x)y^i}\Big|.
\end{displaymath}
Let us assume (if it is negative, the treatment is similar) that
\begin{displaymath}
\epsilon^2(x,y):=2+\frac{\bar{A}_i(x)y^i+\,\eta^{ij}\,\bar{A}_i(x) \bar{A}_j(x)}{1+\,\bar{A}_i(x)y^i}>0.
\end{displaymath}
To write down this condition, one needs that ${1+\,\bar{A}_i(x)y^i}\neq 0$;
 the region where this does not hold is the intersection of the hyperplane
\begin{displaymath}
{\bf P}_x:=\{y\in\,{\bf T}_x{\bf M}\,\,|\,\,{1+\,\bar{A}_i(x)y^i}=0\,\}
\end{displaymath}
with the unit hyperboloid ${\bf \Sigma}_x$. The intersection is such that
${\bf P}_x\cap \,{\bf \Sigma}_x\,\subset \{y\in {\bf \Sigma}_x\,\quad|\quad\,F(x,y)=0\}$.

Let us write in detail the above condition of positiveness for the potential $\bar{A}_i=A_i(x)\,+\partial_i \lambda(x)$:
\begin{displaymath}
\epsilon^2(x,y)=2+\frac{\big({A}_i(x)+\,\partial_i \lambda(x)\big)y^i+\,\eta^{ij}\,\big({A}_i(x)\,+\partial_i \lambda(x)\big) \big({A}_j(x)\,+\partial_j \lambda(x)\big)}{1+\,\big({A}_i(x)\,+\partial_i \lambda (x) \big)y^i}.
\end{displaymath}

 If $y\in {\bf \Sigma}_x$, then for $\beta\neq 1$, $\beta y$ is not in ${\bf \Sigma}_x$. The situation is different if $y\in {\bf NC}_x$. Then $\beta y\in {\bf NC}_x$, even if $\beta\neq 1$. Now we make the approximation ${\bf \Sigma}_x\longrightarrow {\bf NC}_x$ in the asymptotic limit $y^0\longrightarrow \infty$. In this approximation, one can perform limits in the expression for $\epsilon^2(x,y)$. In particular, one can consider $\epsilon^2 (x,\beta y)$ for large $\beta$:
 \begin{displaymath}
 \lim_{\beta\rightarrow \infty} \epsilon^2 (x,\beta y)=\,2+\, \lim_{\beta\rightarrow \infty}\frac{\big({A}_i(x)+\,\partial_i \lambda(x)\big)\beta y^i+\,\eta^{ij}\,\big({A}_i(x)\,+\partial_i \lambda(x)\big) \big({A}_j(x)\,+\partial_j \lambda(x)\big)}{1+\,\big({A}_i(x)\,+\partial_i \lambda (x) \big)\beta y^i}=
 \end{displaymath}
\begin{displaymath}
=\,2+\, \lim_{\beta\rightarrow \infty}\frac{\big({A}_i(x)+\,\partial_i \lambda(x)\big)\beta y^i}{1+\,\big({A}_i(x)\,+\partial_i \lambda (x)  \big)\beta y^i}=3.
\end{displaymath}
That this limit is well defined for large enough $y^0$ can be seen in the following way. Let us assume that $1+\,\big({A}_i(x)\,+\partial_i \lambda (x) \big) y^i=0.$ Then we consider the expression $1+\,\big({A}_i(x)\,+\partial_i \lambda (x) \big)\beta y^i$ for $\beta >>1$. It is impossible that the second expression is zero except if $\beta=1$ and $1+\,\big({A}_i(x)\,+\partial_i \lambda (x)  \big)\beta y^i=0$.

One should prove that $\epsilon^2(x,y)$ is positive for any $y\in {\bf \Sigma}_x$. This is achieved because $\epsilon^2(x,y)$ is invariant. Therefore, we can change to a coordinate system where $y^0$ is arbitrary and the value of the bound does not change.

\item If the curve $\sigma$ is such that its image $\sigma({\bf I})\subset {\bf M}$ is covered by several open sets, for instance $^{\mu}{\bf U}$ and $^{\nu}{\bf U}$, $\mathcal{F}_A(\sigma)$ is evaluated using both representatives. In principle, if we consider two representatives $^{\mu}\bar{A}$ and $^{\nu}\bar{A}$, there is a contribution coming from {\it boundary terms} coming form the evaluation of $d\lambda$ on those points
   in the intersection $^{\mu}{\bf U}\cap\,^{\nu}{\bf U}$.
    This contribution do not appear in the first variation of the functional.
     Therefore, the first variation of (4.3.2) exists and does not
     depend on the representative $A$, for fixed initial and final point variations of $\sigma$. The corresponding extremal curves exist and they
      are unique. They correspond to the solutions of the Lorentz force equation.

\item If the image $\sigma({\bf I})\subset {\bf M}$ is a compact set, the number of sets $^{\mu}{\bf U}$ that we need is finite. Then we obtain a globally defined section of $\bigwedge^1_{loc}{\bf M}$.

\item Once a locally differentiable $1$-form $A$ with the required properties is obtained globally
 over the variation $Var(\sigma)$ of $\sigma:{\bf I}\longrightarrow {\bf M}$, one can follow the standard proof of the deduction of the
 Lorentz force equation from the variation of a functional [26], [2, pg 47-52]. The variation $Var(\sigma)$ must be constructed such that the $1$-form $A$ is globally defined on $Var(\sigma)$. \hfill$\Box$

\end{enumerate}
{\bf Remarks}
\begin{enumerate}
\item It is important to notice that the non-degeneracy of the metric
 $g_{ij}$ does not necessarily imply that it has the same signature as the semi-Riemannian
 metric $\eta$. Further investigations are required to determine the criteria
for conservation of signature.

\item If the curve $\sigma$ is parameterized with respect to a parameter such that the
Finslerian arc-length $F ({\sigma}, \dot{\sigma})$ is constant
along the geodesic, the geodesic equations have a complicated
form (see for instance [24, pg. 296]) and they are not invariant under arbitrary gauge
transformations of the $1$-form $A\longrightarrow A+d\lambda,\, \lambda\in
\mathcal{F}_{loc}({\bf M})$.

\item As we have mentioned before, the fact that we are speaking of local data natural leads us to consider
  notions from pre-sheaf and sheaf theory [32, {\it chapter 6}, 33, {\it chapter 2}] as the basic ingredient in the definition
  of Randers spaces. Sheaf theory is a theory that allows us to treat local objects, for example germs of locally smooth functions or locally
    smooth differential forms [33, {\it chapter 2}]. Hence, we think this is a natural framework to study the geometry and variational properties of semi-Randers spaces.
\end{enumerate}

With definition $(4.3.2)$ at hand, the problem of how to introduce the gauge symmetry in
 a Randers geometry is solved. The price to pay is
 \begin{enumerate}
\item  The class $[A]$ and the Riemannian
  metric $\eta$ are unrelated geometric objects. This is in contradiction with the
   spirit of Randers spaces as a space-time asymmetric structure.

\item One needs to consider locally smooth $1$-forms $A$ instead of
 globally defined $1$-forms. This happens even if the topology of {\bf M} is trivial or the cohomology class of $A$ is trivial.
\end{enumerate}
 We have
   seen that given a $1$-form $A\in[A]$ one has to {\it work hard} to find another
    representative $\bar{A}\in [A]$ such that Douglas's theorem [31] holds.
    Douglas's theorem states the condition under which a system of ordinary
     differential equations can be interpreted as the Euler-Lagrange equation
      coming from a Lagrangian. One of the requirements is that the vertical
       hessian of the Lagrangian must be non-degenerate, $det(g_{ij})\neq 0$.

On the other hand, light-like trajectories cannot be considered in this
formalism for semi-Randers spaces in a natural way. Beem's formalism allows the
 treatment of light-like geodesics as extremal curves of an energy functional
  for some examples of indefinite Finsler space-times.
  However, it is not known if a semi-Randers function exists in Beem's formalism.

  In conclusion, the advantage of the definition given here over Asanov's definition of semi-Randers
  space is that it is consistent with gauge invariance. Nevertheless,
   our formalism is still not completely satisfactory, since it can not be used for llight-like trajectories.

\section{Geometric formulation of the Lorentz force equation}
\subsection{The non-linear connection associated with the Lorentz force equation}

Let us consider a semi-Randers space $({\bf M},\eta, [A])$
and the bundle ${\bf N}\longrightarrow {\bf M}$,
${\bf N}:=\bigsqcup_{x\in {\bf M}}\,\{y\in {\bf T}_x{\bf M},\,\, \eta(y,y)\geq 0\}\,\subset {\bf TM}$.
Then the following diagram
\begin{displaymath}
\xymatrix{ &
{\bf TM} \ar[d]^{\pi}\\
{\bf N}\ar[ur]^{e} \ar[r]^{{\pi}} & {\bf M}.}
\end{displaymath}
where $e$ is the following natural embedding
\begin{displaymath}
e:{\bf N}\longrightarrow {\bf TM}
\end{displaymath}
\begin{displaymath}
(x,y)\mapsto (x,y),\,x\in {\bf M},\, y\in {\bf N}_x
\end{displaymath}
We will consider the differential map
   $d{\pi}:{\bf TN}\longrightarrow {\bf TM}$. Recall that the vertical
    bundle is defined as the kernel $\mathcal{V}:=ker (d{\pi})$;
     at each point one has $ker(d{\pi}|_u):=\mathcal{V}_u$, with $u\in {\bf N}$.

The system of second order differential equations $(4.1.1)$ determines
a special type of vector field on ${\bf N}$ called spray. It is
well known that a spray defines an Ehresmann connection on {\bf TN}[34].
Let us denote by $\eta(Z,Y):=\eta_{ij}(x)Z^i\, Y^j$.
\begin{definicion}
Let $({\bf M}, \eta,[A])$ be a semi-Randers space. For each tangent vector
 $y\in {\bf T}_x{\bf M}$ with $\eta(y,y)>0$, the following functions are well defined,
\begin{displaymath}
^L\Gamma^i\,_{jk}(x,y)=\, ^{\eta}\Gamma^i\,_{jk} (x)+
\frac{1}{2{\sqrt{\eta(y,y)}}}({\bf F}^i\,_{j}(x)y^m\eta_{mk}+ {\bf
F}^i\,_{k}(x)y^m\eta_{mj})+
\end{displaymath}
\begin{equation} +{\bf F}^i\,_m (x)
\frac{y^m}{2\sqrt{\eta(y,y)}}(\eta_{jk}-\frac{1}{\eta(y,y)}\eta_{js}
\eta_{kl}y^s y^l),
\end{equation}
$^{\eta}{\Gamma}^i\,_{jk}(x),\,
(i,j,k= 0,1,2,...,n)$ are the connection  coefficients of the Levi-Civita
 connection $^{\eta}\nabla$ in a local frame, ${\bf F}_{ij}:=\partial_i A_j -\partial_j A_i$
  and ${\bf F}^i\,_j=\eta^{ik}{\bf F}_{kj}$, for any representative $A\in [A]$.
\end{definicion}
The unit hyperboloid sub-bundle ${\bf \Sigma}$ acquires an induced connection,
whose connection coefficients are
\begin{displaymath}
^L\Gamma^i\,_{jk}(x,y)|_{\bf \Sigma} \,=\, ^{\eta}\Gamma^i\,_{jk} (x)\,+
\frac{1}{2}({\bf F}^i\,_{j}(x)y^m\eta_{mk}+ {\bf
F}^i\,_{k}(x)y^m\eta_{mj})+{\bf F}^i\,_m (x)
\frac{y^m}{2}(\eta_{jk}-\eta_{js}
\eta_{kl}y^s y^l).
\end{displaymath}
The structure of these functions is clear: $^{\eta}\Gamma^i\,_{jk}(x)$ are the connection
coefficients of the Lorentzian metric $\eta$; the other two terms are tensorial.
Indeed, one can define the following expressions:
\begin{displaymath}
L^i\,_{jk}(x,y)=\frac{1}{2{\sqrt{\eta(y,y)}}}({\bf F}^i\,_{j}(x)y^m\eta_{mk}+ {\bf
F}^i\,_{k}(x)y^m\eta_{mj}),
\end{displaymath}
\begin{displaymath}
T^i\,_{jk}(x,y)={\bf F}^i\,_m (x)
\frac{y^m}{2\sqrt{\eta(y,y)}}\Big(\eta_{jk}-\frac{1}{\eta(y,y)}\eta_{js}
\eta_{kl}y^s y^l\Big).
\end{displaymath}
Therefore,
\begin{displaymath}
^L\Gamma^i\,_{jk} = \,^{\eta}\Gamma^i\,_{jk} +\, T^i\,_{jk}+\,L^i\,_{jk}.
\end{displaymath}
With the functions $L^i _{jk}$ and $T^i_{jk}$, one can construct the following maps:
\begin{displaymath}
L_u:{\bf T}_u{\bf N}\times {\bf T}_u{\bf N}\longrightarrow {\bf T}_u{\bf N}
\end{displaymath}
\begin{displaymath}
(X,Y)\mapsto \,L^i\,_{jk}(x,y)X^jY^k \frac{\delta}{\delta x^i}.
\end{displaymath}
Recall that the section $\frac{\delta}{\delta x^i}$ is the horizontal lift of the section $\partial_i$.

The second operator that we define is
\begin{displaymath}
T_u:{\bf T}_u{\bf N}\times {\bf T}_u{\bf N}\longrightarrow {\bf T}_u{\bf N}
\end{displaymath}
\begin{displaymath}
(X,Y)\mapsto \,T^i\,_{jk}(x,y)X^jY^k \frac{\delta}{\delta x^i}.
\end{displaymath}
$u=(x,y)$ and $X, Y$ are arbitrary tangent vectors $X,Y\in {\bf T}_u{\bf N}$.
This can be generalized to homomorphisms acting on vector sections.

Note the following elementary property
\begin{displaymath}
T_u(Y ,Y)=0, \,\,\, \forall y\in {\bf N}_x,\,Y=y^i\frac{\delta}{\delta x^i},\,\, u=(x,y).
\end{displaymath}
However, $T_u(\cdot,Y)\neq 0$ in general.

\subsection{The Koszul connection $^L\nabla$ on {\bf TN} associated with the Lorentz force equation}

Let $\{e_0,...,e_{n-1}\}$ be a local basis for the sections of the frame
bundle associated with the tangent bundle $\Gamma{\bf TM}$ and let
us assume that each $e_i$ is a time-like tangent vector at the point $x\in {\bf M}$
(therefore, the metric cannot be diagonal in this basis, since
$\eta$ is a semi-Riemannian metric). Then
$\{\pi^*e_0,...,\pi^*e_{n-1}\}$ is a local frame for the fiber
$\pi^{-1}_u\subset \pi^*{\bf TM},\, u\in {\bf N}$, $\{h_0,...,h_{n-1}\}$
is the local frame of the horizontal distribution
$\mathcal{H}_u\,\subset {\bf T}_u{\bf N}$ obtained by the horizontal
lift $h_i=h(e_i)$ and $\{v_0,...,v_{n-1}\}$ is a local frame for the
vertical distribution $\mathcal{V}_u\,\subset {\bf T}_u{\bf N}$.

Given the set of functions $\{^L\Gamma^i\,_{jk},\,i,j,k=0,..., n-1\}$ and
the non-linear connection associated
 with the system of differential equations, there is an associated Koszul connection on {\bf TN} [35]
  (a Koszul connection is a linear connection defined through the corresponding covariant derivative).
\begin{proposicion}
Let $({\bf M}, \eta,[A])$ be a semi-Randers space and ${\bf M}$ a $n$-dimensional manifold. There is
defined a covariant derivative $^LD$ on ${\bf TN}$ determined by the following conditions:
\begin{enumerate}

\item For each $X\in \mathcal{H}_u$ and $Z\in \Gamma \mathcal{H}$
\begin{displaymath}
^LD_X Z=\,X^k\,^L\Gamma ^i\,_{jk} (x,y) Z^j\,h_i,\quad X=X^i\,h_i|_u, \quad Z=Z^i\,h_i|_v,
\end{displaymath}
with $\{h_i\}$ a local frame for the horizontal distribution, $u=(x,y)$ and $v$
an arbitrary point of an open set ${\bf \tilde{O}}\subset {\bf N}$ containing $u$.

\item The covariant derivative of arbitrary sections $Z\in {\bf \Gamma TN}$
along vertical direction is zero:
\begin{displaymath}
^LD_V Z=0, \quad \forall\, V\in \mathcal{V},\,Z\in \Gamma {\bf TN}.
\end{displaymath}

\item The covariant derivative $^LD$ is symmetric, i.e, has zero horizontal
torsion:
\begin{displaymath}
^LD_U V -\, ^LD_V U -[U,V]=0,\quad \forall\, U,V\,\in
\mathcal{H}.
\end{displaymath}

\item For all $X\in \mathcal{H}_u$ and $Z\in \Gamma \mathcal{V}$, $^LD_X Z=0$.
\end{enumerate}
\end{proposicion}
{\bf Proof}: Let us consider the Finsler geodesic equation
associated with the semi-Randers space $F_A=\sqrt{\eta_{ij}(x)y^iy^j}+\,A_i(x)y^j$ but
parameterized using the arc-length of the Lorentzian metric $\eta$:
\begin{displaymath}
\frac{d^2 x^i}{d\tau^2}+\, ^{\eta}{\Gamma}^i\,
_{jk}\frac{dx^j}{d\tau}\frac{dx^k}{d\tau}+
\eta^{ij}(dA)_{jk}\sqrt{\eta(\frac{dx}{d\tau},\frac{dx}{d\tau})}\frac{dx^k}{d\tau}=0,
\end{displaymath}
where ${\bf F}=dA$ is the exterior differential of the $1$-form
$A$ and $\eta(X,Z)=\eta_{ij}(x)X^i\,Z^j.$ From these equations, we can
read the value of the semi-spray coefficients, which are
\begin{displaymath}
G^i(x,y)\,=\, ^{\eta}{\Gamma}^i\,_{jk}(x)\,y^j
y^k+\eta^{ij}(x)(dA)_{jk}(x)\sqrt{\eta(y,y)}y^k.
\end{displaymath}
Taking the first and second derivatives with respect to $y$, we
obtain
\begin{displaymath}
\frac{1}{2}\frac{\partial}{\partial y^j} G^i (x,y) = \, ^{\eta}\Gamma^i\,
_{lj} (x)\,y^l + \eta
^{il}(x)\,(dA)_{lm}(x)\,\frac{1}{\sqrt{\eta(y,y)}}\eta_{js}(x)y^sy^m
+\eta^{il}(x)(dA)_{lj}(x) \sqrt{\eta(y,y)}.
\end{displaymath}
\begin{displaymath}
\frac{1}{2}\frac{\partial}{\partial y^k} \frac{\partial}{\partial
y^j}G^i (x,y) = \, ^{\eta} \Gamma^i\,_{kj} (x)\, - \eta
^{il}(x)(dA)_{lm}(x)\frac{1}{2({\eta(y,y)})^{3/2}}\eta_{js}(x)\,y^sy^m\eta_{kp}(x)\,y^p
+
\end{displaymath}
\begin{displaymath}
 +\eta ^{il}(x)(dA)_{lm}(x)\frac{1}{2(\sqrt{\eta(y,y)})}\eta_{jk}(x)y^m +
\eta ^{il}(x)(dA)_{lk}(x)\frac{1}{2(\sqrt{\eta(y,y)})}\eta_{js}y^s +
\end{displaymath}
\begin{displaymath}
+\eta
^{il}(x)(dA)_{lj}(x)\frac{1}{2(\sqrt{\eta(y,y)})}\eta_{ks}(x)y^s.
\end{displaymath}
One can check that $\frac{1}{2}\frac{\partial}{\partial y^j}\frac{\partial}{\partial y^k}
G^i (x,y) =\,^L\Gamma^i\,_{jk}(x,y)$.

From the structure of these coefficients one can check (following a standard
procedure, for instance in [34]) that there is a splitting of the tangent vector spaces
${\bf T}_u{\bf N}$ for each $u\in {\bf N}$. In addition,
 one can check that the connection coefficients are symmetric, $^L\Gamma^i\,_{jk}=\,^L\Gamma^i\,_{kj}$.
 The fact that the covariant derivative along the vertical directions is zero is
 an additional hypothesis used to make the covariant derivative unique.
\hfill$\Box$

{\bf Remark 1}. In this interpretation of the Lorentz force equation, the
 role of the function $F_A$ is not fundamental, since all the information is obtained directly
  from the differential equations $(4.1.1)$.

{\bf Remark 2}. We have imposed the restriction that the covariant derivatives
 along vertical directions are zero. However, there can be covariant derivatives
  which are compatible with the equations $(4.1.1)$ but which have
   non-zero vertical covariant derivatives.
\begin{proposicion}
The following properties hold:
\begin{enumerate}
\item The Lorentz connection $^LD$ is invariant under gauge
transformations $A\longrightarrow A+d\lambda$ of the locally smooth $1$-form
$A(x)=A_i(x)dx^i$.

\item Given a point $x\in {\bf M}$, $^LD$ admits a normal coordinate system centered at $x$
which coincides with the normal coordinate system associated
with $^{\eta}\nabla$ centered at $x$ iff ${\bf F}(x)=0$.
\end{enumerate}
\end{proposicion}
{\bf Proof}:
\begin{enumerate}
\item
The first property is a consequence of the fact that all the geometric
objects appearing in the connection coefficients $^L\Gamma^i\,_{jk}$
are gauge invariant.

\item If $^LD$ is affine, for any point $x\in {\bf M}$ there is a coordinate system
where the connection coefficients are zero $^L\Gamma^i\,_{jk}(x,y) =0$.
This implies
 that $^L\Gamma^i\,_{jk} y^j y^k=0$ at the point $x\in {\bf M}$. Due to the decomposition $^L\Gamma^i\,_{jk}=\,^{\eta}\Gamma^i\,_{jk}+\,T^i\,_{jk}+L^i\,_{jk}$ this is equivalent to
\begin{displaymath}
0=(\,^{\eta}\Gamma^i\,_{jk} + T^i\,_{jk}+L^i\,_{jk})y^j
y^k,\,\,\forall y\in {\bf \Sigma}_x.
\end{displaymath}
Since the transversality condition holds $(T^i\,_{jk}y^jy^k =0)$,
one obtains
\begin{displaymath}
0=(\,^{\eta}\Gamma^i\,_{jk} +\, L^i\,_{jk})y^j y^k,\,\,\forall y\in
{\bf \Sigma}_x.
\end{displaymath}
Assume that there is a normal coordinate system centered at $x$ for $^L\nabla$ and
that this coordinate system coincides with the normal coordinate
system associated with $^{\eta}\nabla$. Then at the point $x$, one
has the relation
\begin{displaymath}
 \,(\,^{L}\Gamma^i\,_{jk}y^j y^k) =\, {\bf F}^i\,_j y^j=0,\quad  \forall y\in {\bf T}_x{\bf M}.
\end{displaymath}
This last condition is strong enough to imply ${\bf F}=0$.
\end{enumerate}
\hfill$\Box$

{\bf Remark}. In the positive definite case, it is well known that the
requirement that the Chern connection of a Randers space lives on
the manifold ${\bf M}$ is that the $1$-form $A$ must be parallel in the sense that
$^{\eta}\nabla A=0$, [24, {\it chapter 11}]. This is a stronger condition than ${\bf F}=dA=0$. The
parallel condition indicates that the structure $({\bf M}, F)$ is a
 generalization of the Berwald structure in Finsler geometry: a Berwald space
  is a Finsler space where the connection coefficients live on {\bf M}; the
closeness condition indicates that the space is Douglas [24, pg 304];
 Douglas's spaces are such that they have the same geodesics as the underlying Riemannian metric $\eta$.
One can easily move the proofs to the Lorentzian and indefinite category, if one adopt Asanov's framework.
\begin{corolario} Let $({\bf M}, \eta, [A])$ be a semi-Randers space.
 Then the Lorentz force equation can be written as
\begin{displaymath}
^LD _{\dot{\tilde{x}}} \dot{\tilde{x}}=0,
\end{displaymath}
where ${x}:{\bf I}\longrightarrow {\bf M}$ is a time-like curve parameterized
 with respect to the proper time of the Lorentzian metric $\eta$,
$\tilde{x}$ is the horizontal lift on {\bf N} and $^L\nabla$ is the
non-linear connection determined by the system of differential equations $(4.1.1)$.
\end{corolario}
{\bf Proof}: A solution of the auto-parallel condition of the Lorentz
connection defines a curve on ${\bf N}$ given by $(x,y)(\tau)=(x(\tau),\dot{x}(\tau))$.
 Projecting this curve into {\bf M} by ${\pi}$, one obtains a curve $x(\tau)$
  which is a solution of the Lorentz force equation.\hfill$\Box$

\subsection{The Lorentz connection on the pull-back bundle $\pi^*{\bf TM}$}

We introduce the third framework, which will be directly used later to define the averaged connection.
Given the non-linear connection $^LD$ on ${\bf
TN}\longrightarrow {\bf N}$, there is a natural linear connection on the
pull-back bundle $\pi^*{\bf TM}\longrightarrow {\bf N}$ that we
denote by $^L\nabla$ characterized by the following:
\begin{proposicion}
The linear connection $^L\nabla$ on the
pull-back bundle $\pi^*{\bf TM}\rightarrow {\bf N}$ is determined by the following structure equations,
\begin{enumerate}
\item $^L\nabla$ on $\pi^*{\bf TM}\rightarrow {\bf N}$ is a
symmetric connection,
\begin{equation}
^L\nabla_{\tilde{X}} \pi^* Y -\,^L\nabla_{\tilde{Y}}\pi^* X
-\pi^*[X,Y]=0,
\end{equation}
where $X,Y\in \Gamma{\bf TM},\, \tilde{X},\tilde{Y}\in \Gamma{\bf
TN}$ are horizontal lifts of $X, Y\in \Gamma{\bf TM}$ to $\Gamma {\bf TN}$, with
$\eta(X,X)>0$ and $\eta(Y,Y)>0$.

\item The covariant derivative along vertical directions of
sections of $\pi^*{\bf TM}$ are zero,
\begin{equation}
^L\nabla _{v_j} \pi^*e_k=0,\quad (j,k=0,...,n-1).
\end{equation}
\item The covariant derivative along horizontal directions is
given by the formula
\begin{equation}
^L\nabla\, _{h_j} \pi^* e_k=\, ^L{\Gamma}^i\, _{jk}(x,y)\,\pi^*
e_i,\quad (i,j,k=0,...,n-1).
\end{equation}
\item By definition the covariant derivative of a function $
f\in\mathcal{F}({\bf N})$ is given by
\begin{equation}
^L\nabla _{\hat{ X}} f:= \hat{X}(f),\quad \forall \hat{X}\in {\bf T}_u{\bf
N}.
\end{equation}
\end{enumerate}
\end{proposicion}
{\bf Proof:} One can check by direct computation that the above relations
define a {\it covariant derivative} on $\pi^*{\bf TM}$ and that they are self-consistent.
A general covariant derivative can be expressed in terms of the connection $1$-forms as
\begin{displaymath}
\omega^i\,_j:=\,^L\Gamma^i\,_{jk}dx^k +\, ^L\Upsilon^i\,_{jk}{\delta y^k},
\end{displaymath}
where we have used a local frame of $1$-forms $\{ dx^0,...,dx^{n-1},\,{\delta y^0},...,{\delta y^{n-1}}\}$.
Since the covariant derivative of sections on $\pi^*{\bf TM}$ along vertical directions is zero, one obtains
\begin{displaymath}
^L\Upsilon^i\,_{jk}{\delta y^k}=0\, \Rightarrow\, ^L\Upsilon^i\,_{jk}=0
\end{displaymath}
at each point $(x,y)\in {\bf N}$. Since the torsion tensor is zero, one has
\begin{displaymath}
\,^L\Gamma^i\,_{jk} =\,^L\Gamma^i\,_{kj}.
\end{displaymath}
Therefore, we have to provide the rule for deriving sections along horizontal directions. Since the coefficients given by
formula (4.4.1) are symmetric, this rule is consistent with the torsion-free condition.
Finally, we require that the covariant derivative to be a local operator. This
 is satisfied by (4.4.5), which guaranties that $^L\nabla$ satisfies the Leibnitz rule.
 \hfill$\Box$

\begin{corolario}
Let ${\bf M}$, ${\bf N}$, $\pi^*{\bf TM}$ and $^L\nabla$ be as
before. Then the auto-parallel curves of the linear Lorentz
connection $^L\nabla$ are in one to one correspondence with the solutions
 of the Lorentz force equation,
\begin{displaymath}
^L\nabla_{\tilde{\dot{x}}}\pi^* \dot{x}=0\Leftrightarrow\,
^LD_{\dot{x}} \dot{x}=0,\quad \dot{x}=\frac{d\sigma}{d\tau}.
\end{displaymath}
\end{corolario}
{\bf Proof}: If in some coordinate system the Lorentz connection $^L\nabla$
 has the connection coefficients $^L\Gamma^i\,_{jk}$, the auto-parallel condition is
\begin{displaymath}
0=\Big(\pi^*\big(\,^L\nabla_{\frac{d x(\tau)}{d\tau}}\,\pi^* \frac{d x(\tau)}{d\tau}\big)\Big)^i\,=\Big(\frac{d^2 x^i(\tau)}{d\tau^2}+\,^L\Gamma^i\,_{jk}(x,\frac{d x(\tau)}{d\tau})\,\frac{d x^j(\tau)}{d\tau}\frac{d x^k(\tau)}{d\tau}\Big)=
\end{displaymath}
\begin{displaymath}
=\Big(\frac{d^2 x^i(\tau)}{d\tau^2}+\,\Big(
 \, ^{\eta} \Gamma^i\,_{kj}  - \eta
^{il}(dA)_{lm}\frac{1}{2({\eta(\frac{d x(\tau)}{d\tau},\frac{d x(\tau)}{d\tau})})^{3/2}}\eta_{js}\frac{d x^s(\tau)}{d\tau}\frac{d x^m(\tau)}{d\tau}\eta_{kp}\frac{d x^p(\tau)}{d\tau}
+
\end{displaymath}
\begin{displaymath}
 +\eta ^{il}(dA)_{lm}\frac{1}{2(\sqrt{\eta(\frac{d x(\tau)}{d\tau},\frac{d x(\tau)}{d\tau})})}\eta_{jk}\frac{d x^m(\tau)}{d\tau} +
\eta ^{il}(dA)_{lk}\frac{1}{2(\sqrt{\eta(\frac{d x(\tau)}{d\tau},\frac{d x(\tau)}{d\tau})})}\eta_{js}\frac{d x^s(\tau)}{d\tau} +
\end{displaymath}
\begin{displaymath}
+\eta
^{il}(dA)_{lj}\frac{1}{2(\sqrt{\eta(\frac{d x(\tau)}{d\tau},\frac{d x(\tau)}{d\tau})})}\eta_{ks}\frac{d x^s(\tau)}{d\tau}\Big)\frac{dx^j(\tau)}{d\tau}\frac{d x^k(\tau)}{d\tau}\Big).
\end{displaymath}
Using $\eta(\frac{d x(\tau)}{d\tau},\frac{d x(\tau)}{d\tau})=1,$ the above expression simplifies to
\begin{displaymath}
0=\Big(\pi^*\big(\,^L\nabla_{\frac{d x(\tau)}{d\tau}} \frac{d x(\tau)}{d\tau}\big)\Big)^i\,=
\end{displaymath}
\begin{displaymath}
=\Big(\frac{d^2 x^i(\tau)}{d\tau^2}+\,\Big(
 \, ^{\eta} \Gamma^i\,_{kj} +
\eta ^{il}(dA)_{lk}\frac{1}{2}\eta_{js}\frac{d x^s(\tau)}{d\tau}
+\eta
^{il}(dA)_{lj}\frac{1}{2}\eta_{ks}\frac{d x^s(\tau)}{d\tau}\Big)\frac{dx^j(\tau)}{d\tau}\frac{d x^k(\tau)}{d\tau}\Big).
\end{displaymath}
This is the Lorentz force equation $(4.1.1)$.
\hfill$\Box$

We need to translate from the non-linear connection $^L\nabla$
 on ${\bf TN}$ to the linear connection on $\pi^*{\bf TM}$ because we will consider the
 averaged connection, which was defined in [22] and in {\it chapter 3}.

\section{The averaged Lorentz connection}

Given the linear connection $^L\nabla$ on the bundle $\pi^*{\bf
TM}\longrightarrow {\bf \Sigma}$ we can obtain an associated
averaged connection using the theory described in {\it section 3.3}.

Usually the
measure is given as $f(x,y)\, dvol(x,y)$. The function $f(x,y)$ must be gauge invariant and
 such that the low order moments are finite. The volume form is
\begin{displaymath}
dvol(x,y)=\sqrt{-det\,{\eta}}\,\frac{1}{y^0}\,dy^1\wedge\cdot\cdot\cdot dy^{n-1},\quad y^0=y^0(x^0,...,x^{n-1},y^1,...,y^{n-1}).
\end{displaymath}
Although the dimension of the manifold ${\bf \Sigma}$ is $2n-1$, we will use the extrinsic formalism explained in {\it chapter 2}. This in particular means that all Latin indices run from $0$ to $n-1$, if nothing else is stated.
Then one can prove the following
\begin{proposicion}
The averaged connection of the Lorentz connection $^L\nabla$ on the
pull-back bundle $\pi^*{\bf TM}\longrightarrow {\bf \Sigma}$ is an affine,
symmetric connection on  ${\bf M}$. The connection coefficients are given by the formula
\begin{displaymath}
<\,^L\Gamma^i\,_{jk} >=\, ^{\eta}\Gamma^i\,_{jk}+ ({\bf
F}^i\,_{j}<\frac{1}{2}y^m>\eta_{mk}+ {\bf
F}^i\,_{k}<\frac{1}{2}y^m>\eta_{mj})+
\end{displaymath}
\begin{equation}
+{\bf F}^i\, _m\,\frac{1}{2}\big(
<{y^m}>\,\eta_{jk}-\eta_{js}
\eta_{kl}<\,y^m y^s y^l>\,\big).
\end{equation}
Each of the integrations is equal to the $y$-integration along the
fiber,
\begin{displaymath}
{vol({\bf \Sigma}_x)}=\int _{{\bf \Sigma}_x}
f(x,y)\,dvol(x,y),\quad <y^i> :=\frac{1}{vol({\bf \Sigma}_x)}\int _{{\bf \Sigma}_x} y^i
f(x,y)\,dvol(x,y),
\end{displaymath}
\begin{displaymath}
<y^m y^s y^l>:=\frac{1}{vol({\bf \Sigma}_x)}\int _{{\bf \Sigma}_x} y^m y^s y^l
f(x,y)\,dvol(x,y).
\end{displaymath}
\end{proposicion}
{\bf Proof:} Equation $(4.5.1)$ follows easily from the definition of
the averaged connection by linearity. We only need to prove that $<y^i>$ and
 the other moments are given by the corresponding integrals and the identity
  operator $Id:\pi^*{\bf TM}\longrightarrow \pi^*{\bf T M}$, $(x,y)\mapsto (x,y)$:
\begin{displaymath}
<y^i>=\frac{1}{vol({\bf \Sigma}_x)}\int _{{\bf \Sigma}_x} \pi_2\,Id \,y^i\pi^*(y^i)
f(x,y)\,dvol(x,y)=\frac{1}{vol({\bf \Sigma}_x)}\int _{{\bf \Sigma}_x} y^i
f(x,y)\,dvol (x,y)
\end{displaymath}
and similarly for other moments. Note that since $y\in {\bf \Sigma}_x$ $\eta(y,y)=1$, the factors $\sqrt{\eta(y,y)}$ do not appear in the connection coefficients.\hfill$\Box$

{\bf Remarks}.
\begin{enumerate}

\item If we consider the bundle $\pi^*{\bf TM}\longrightarrow {\bf N}$, the coefficients of the averaged Lorentz connection are

\begin{displaymath}
<\,^L\Gamma>^i\,_{kj}= \, ^{\eta} \Gamma^i\,_{kj} (x)\, - \eta
^{il}(x)(dA)_{lm}(x)<\frac{1}{2({\eta(y,y)})^{3/2}}\eta_{js}(x)\,y^sy^m\eta_{kp}(x)\,y^p>
+
\end{displaymath}
\begin{displaymath}
 +\eta ^{il}(x)(dA)_{lm}(x)<\frac{1}{2(\sqrt{\eta(y,y)})}\eta_{jk}(x)y^m >+
\eta ^{il}(x)<(dA)_{lk}(x)\frac{1}{2(\sqrt{\eta(y,y)})}\eta_{js}y^s> +
\end{displaymath}
\begin{displaymath}
+\eta
^{il}(x)(dA)_{lj}(x)<\frac{1}{2(\sqrt{\eta(y,y)})}\eta_{ks}(x)y^s>.
\end{displaymath}
In this case, the averaged connection of higher moments of the distribution function.
\item Formula $(4.5.1)$ holds in any local natural coordinate system.

\end{enumerate}

The following proposition enumerates basic properties of the averaged Lorentz connection.
 The proof is straightforward; one  needs to check the properties of the coefficients given by the formula $(4.5.1)$:
\begin{proposicion} Let $({\bf M}, \eta, [A])$ be a semi-Randers space,
$f:{\bf \Sigma}\longrightarrow {\bf R}$ non-negative with compact support in each ${\bf \Sigma}_x$
 and $<\,^L\nabla>$ be the averaged Lorentz connection. Then
\begin{enumerate}

\item $<\,^L\nabla>$ is an affine, symmetric connection on ${\bf
M}$. Therefore, for any point $x\in {\bf M}$, there is a {\it normal coordinate
 system} such that $<\,^L\Gamma^i\,_{jk}>(x)=0$.

\item $<\,^L\nabla>$ is determined by the first, second and
 third moments of the distribution function $f(x,y)$.
\end{enumerate}
\end{proposicion}
{\bf Remark}.
While the first property is a general property of the averaged connection,
the second one is a specific property of the averaged Lorentz connection, which results because we are
considering trajectories whose velocity vectors $y$ in the unit hyperboloid ${\bf \Sigma}$.
Also note that $<\,L^\nabla>$ does not preserves the norm $\eta_{ij}(x)y^i y^j$
for arbitrary $y\in{\bf T}_x{\bf M}$: $<\,L^\nabla>_Z(\eta_{ij}(x)y^i y^j)\neq 0$ for an arbitrary $Z\in {\bf T}_x{\bf M}$.

 \section{Comparison between the geodesics of $\,^L\nabla$ and\\ $<\,^L\nabla>$}

\subsection{Basic geometry in the space of connections}

Let us consider the Lorentzian manifold $({\bf M},\eta)$ with signature $(+,-,...,-)$ and ${\bf M}$  $n$-dimensional.
Let us choose on ${\bf  M  }$ a time-like vector field $U$ normalized such that
$\eta(U,U)=1$. Then one can define the Riemannian metric $\bar{\eta}_U$ [36]
\begin{equation}
\bar{\eta}_U(X,Y):=-\eta(X,Y)+2\eta(X,U)\eta(Y,U).
\end{equation}
$\bar{\eta}_U$ determines a Riemannian metric on the
vector space ${\bf T}_x{\bf M}$ that we also denote by $\bar{\eta}_U$ and
 that in local coordinates can be expressed as
$\bar{\eta}_{U}=\bar{\eta}_{ij}(x) \,dy^i\otimes dy^j$. Note that the metric
 $\bar{\eta}_U$ on {\bf M} is in local coordinates $\bar{\eta}_{U}=\bar{\eta}_{ij}(x) \,dx^i\otimes dx^j$.
  We hope that the meaning of the symbols of the type $\bar{\eta}_{U}$ is clear from the context.

The pair $({\bf T}_x{\bf M},\bar{\eta}_U)$ is a Riemannian manifold.
The Riemannian metric $\bar{\eta}_U$ induces a distance
function $d_{\bar{\eta}_U}$ on the manifold ${\bf T}_x{\bf M}$,
\begin{displaymath}
d_{\bar{\eta}_U}:{\bf T}_x{\bf M}\times {\bf T}_x{\bf M}\longrightarrow {\bf R}
\end{displaymath}
\begin{displaymath}
 (X,Y)\mapsto inf\big\{ \int^1_{0}\sqrt{ \bar{\eta}_U(\dot{\hat{\sigma}},\dot{\hat{\sigma}}})d\tau,\,\,
  \hat{\sigma}:{\bf I}\longrightarrow {\bf T}_x{\bf M},\quad \hat{\sigma}(0)=X,\,\,\hat{\sigma}(1)=Y\big\}.
\end{displaymath}
We will say that the time-like vector field $U$ defines a field of observer [60, pg 45].

We assume that $f(x,y)$ has compact support on each unit hyperboloid ${\bf
\Sigma}_x$. Then the diameter of the distribution $f_x=f(x,\cdot):{\bf \Sigma}_x \longrightarrow {\bf R};\quad (x,y)\mapsto f(x,y)$ is
$\alpha_x:=sup\{d_{\bar{\eta}}(y_1,{y}_2)\, |\, y_1,{y}_2\in
supp(f_x )\}$. We define the parameter
${\alpha}:=sup\{{\alpha}_x,\, x\in {\bf K}\}$, with ${\bf K}\subset {\bf M}$ a compact domain of the space-time {\bf M}. We will restrict always our considerations to {\bf K}. Note that $\alpha$ (and other parameters) can depend on ${\bf K}$.

Let us fix the coordinate system $(x,y)$ on ${\bf N}$. Let us denote
 the space of linear connections on ${\bf TN}$ by $\nabla_{{\bf N}}$.
This space is a finite dimensional manifold whose points are coordinated by
 the set of functions $\{\Gamma^i\,_{jk}(x,y),\,\,i,j,k=0,1,...,n-1\}$.
 We will introduce a distance function on $\nabla_{{\bf N}}$. First, we recall the definition of the norm of an operator.
\begin{definicion}
Given a linear operator $A_x:{\bf T}_x{\bf
M}\longrightarrow {\bf T}_x{\bf M}$ in the finite dimensional normed linear space $({\bf T}_x{\bf M}, \|\cdot\|_{\bar{\eta}_Z})$, its operator norm is defined by
\begin{displaymath}
\|A\|_{{\bar{\eta}}_Z}(x):=sup\,\Big\{\,\frac{\|A(y)\|_{{\bar{\eta}}_Z}}{\|y\|_{{\bar{\eta}}_Z}}
 (x),\,y\in{\bf T}_x{\bf M}\setminus \{0\}\,\Big\}.
\end{displaymath}
\end{definicion}
The norm $\|\cdot\|_{\bar{\eta}_Z}$ is constructed from the Lorentzian metric $\eta$ using a particular local time-like vector $Z$ through the definition $(4.6.1)$. We will later specify a vector field $Z$ which will be of special interest for our purposes.
\begin{proposicion}
On the space $\nabla_{{\bf N}}$, there is a distance function.
The distance between two points $^1\nabla,\,^2\nabla\in \nabla_{{\bf N}}$ is
given by
\begin{displaymath}
d_{\bar{\eta}_Z}(\,^1\nabla,
\,^2\nabla )(x):=sup\,\Big\{\frac{\sqrt{{\bar{\eta}_Z}(x)(\,^1\nabla_X
X-\,^2\nabla_X X,\,^1\nabla_X X-\,^2\nabla_X
X)}}{\sqrt{\bar{\eta}_Z(X,X)}},\,\,
\end{displaymath}
\begin{equation}
\,
 X\in \Gamma{\bf TN},\, \quad ^1\nabla,\, ^2\nabla \in \nabla_{{\bf N}}\Big\}
\end{equation}
for a Riemannian metric (4.6.1) constructed using the local time-like vector field $Z$.
\end{proposicion}
{\bf Proof}: The function $(4.6.2)$ is symmetric and non-negative. The
distance between two arbitrary connections is zero iff
\begin{displaymath}
\sqrt{{\bar{\eta}_Z}(\,^1\nabla_X
X-\,^2\nabla_X X,\,^1\nabla_X X-\,^2\nabla_X
X)}=0
\end{displaymath}
for all $X\in \Gamma{\bf T\Sigma}$. This happens iff $^1\nabla_X X =\,^2\nabla_X X$ for
 any $X \in \Gamma T\Sigma$. The triangle inequality also holds, from the triangle inequality for $\bar{\eta}$. This reads
\begin{displaymath}
d_{\bar{\eta}_Z}(\nabla_1, \nabla_3)=sup\,\Big\{\frac{\sqrt{{\bar{\eta}_Z}(\,^1\nabla_X
X-\,^3\nabla_X X,\,^1\nabla_X X-\,^3\nabla_X
X)}}{\sqrt{\bar{\eta}_Z(X,X)}}\Big\}\leq
\end{displaymath}
\begin{displaymath}
\leq sup\,\Big\{\frac{\sqrt{{\bar{\eta}_Z}(\,^1\nabla_X
X-\,^2\nabla_X X,\,^1\nabla_X X-\,^2\nabla_X
X)}}{\sqrt{\bar{\eta}_Z(X,X)}}\Big\}+
\end{displaymath}
\begin{displaymath}
+sup\,\Big\{\frac{\sqrt{\bar{\eta}_Z(\,^2\nabla_X
X-\,^3\nabla_X X,\,^2\nabla_X X-\,^3\nabla_X
X)}}{\sqrt{\bar{\eta}_Z(X,X)}}\Big\}\leq
\end{displaymath}
\begin{displaymath}
\leq d_{\bar{\eta}_Z}(\,^1\nabla,\,^2\nabla)+d_{\bar{\eta}_Z}(\,^2\nabla,\,^3\nabla).
\end{displaymath}\hfill$\Box$

Recall that for an arbitrary $1$-form $\omega$ we denote by ${\omega}^{\sharp}:=\eta^{-1}(\omega,\cdot)$
the vector obtained by duality, using the Lorentzian metric $\eta$. Similarly, given
a vector field $X$ over {\bf M}, one can define the dual one form ${X}^{\flat}:=\eta(X,\cdot)$;
$\iota_X \omega$ is the inner product of the vector $X$ with the form $\omega$.
Also recall that the difference between two connections is a tensor. In this sense,
 one can consider the difference between connections on the pull-back bundle $\pi^*{\bf TM}$
  given by the Lorentz connection $^L\nabla$ and the pull-back connection of the averaged connection
   $\pi^*<\,^L\nabla>$. Then it makes sense to take the difference $^L\nabla\,-\pi^*<\,^L\nabla>$,
    which is a tensor along ${\pi}$. We use the {\it hat}-notation to indicate integrated variables. For instance, $<\hat{y}>$ means an integration operation, where the variable integrated is $\hat{y}$. Let us consider a local frame $\{e_0,...,e_{n-1}\}$ on {\bf TM}.
\begin{proposicion}
Let $f(x,y)$ be the one-particle probability distribution function such that each function $f_x$
 has compact and connected support
$supp(f_x)\subset{\bf \Sigma}_x$. Then
\begin{equation}
(\,^L\nabla_y y-\, \pi^*<\,^L\nabla>_y y)(x)=-({\iota_{\delta}{\bf F}})^{\sharp}(x)\cdot (\iota_{y}({\delta}))(x,y))
+\mathcal{O}_2(\delta^2(y))(x,y)\,+\mathcal{O}_3(\delta^3(y))(x,y),
\end{equation}
where $\delta(x,y)=<\hat{y}>(x)-y$ does not depend on
 the $2$-form ${\bf F}$. The tensors $\mathcal{O}_i(x,y)$ are given by the following expressions:
\begin{displaymath}
\mathcal{O}_2(\delta^2(y))(x,y)=\frac{1}{2}\,{\bf F}^i\,_{m}\Big( <\hat{y}^m>(x)\delta^s(x,y)
\delta^l(x,y)+<\hat{y}^m><\delta^s(x,\hat{y})\delta^l(x,\hat{y})>\,
\end{displaymath}
\begin{equation}
+2<\hat{y}^l>(x)<\delta^m (x,\hat{y})\delta^s(x,\hat{y})>\Big)\eta_{sj}\eta_{lk}y^jy^k\,\pi^*e_i,
\end{equation}
\begin{equation}
\mathcal{O}_3(\delta^3(y))(x,y)\,= \frac{1}{2}\,{\bf F}^i\,_{m}<\delta^m(x,\hat{y})\delta^s(x,\hat{y})\delta^l(x,\hat{y})>
\eta_{sj}\eta_{lk}y^jy^k\,\pi^*e_i.
\end{equation}
\end{proposicion}
{\bf Proof}: From the expressions for the connection coefficients,
\begin{displaymath}
\pi^*<\,^L\nabla>_y y -\, ^L\nabla_y y \,=\frac{1}{2}\Big({\bf
F}^i\,_{j}(x)(<\hat{y}^m>(x)-y^m)\eta_{mk}+ {\bf
F}^i\,_{k}(x)(<\hat{y}^m>(x)-y^m)\eta_{mj})+
\end{displaymath}
\begin{displaymath} +{\bf F}^i\,
_m(x)\big(({<\hat{y}^m>(x)-y^m)}\eta_{jk}-\eta_{js} \eta_{kl}(<\hat{y}^m \hat{y}^s
\hat{y}^l>(x)-y^m y^sy^l)\big)\Big)y^j y^k \,\pi^*e_i,
\end{displaymath}
since $\eta_{ij}y^iy^j=1$. Here $\{e_i,\,i=0,...,n-1\}$ is an arbitrary frame unless otherwise is indicated.
The difference between the two connections can be expressed in
terms of the following tensors:
\begin{displaymath}
\delta^m(x,y) =<\hat{y}^m>(x)-y^m,\,\, \,\,\,\,\,\,\,\, \delta^{msl}(x,y)y_sy_l=<\hat{y}^m
\hat{y}^s \hat{y}^l>(x)\eta_{sj}\eta_{lk}\,y^jy^k -y^m
\end{displaymath}
and is given by the following expression:
\begin{displaymath}
\Big(\pi^*<\,^L\nabla>_y y -\, ^L\nabla_y y \Big)(x)=\frac{1}{2}\Big({\bf
F}^i\,_{j}(x)(\delta^m(x,y))\eta_{mk}+ {\bf
F}^i\,_{k}(x)(\delta^m(x,y))\eta_{mj}+
\end{displaymath}
\begin{displaymath}
+{\bf F}^i\,
_m(x)\big(\delta^m(x,y)\eta_{jk}-\eta_{js}
\eta_{kl}\delta^{msj}(x,y)\big)\Big)y^j y^k \,\pi^*e_i =
\end{displaymath}
\begin{displaymath}
=\Big({\bf F}^i\,_{j}y^j\delta^k (x,y)y_{k}+{\bf F}^i\,
_m(x)\big(\delta^m(x,y)-y_{j}
y_{k}\delta^{mkj}(x,y)\big)\Big)\pi^*e_i,
\end{displaymath}
where $y_j =\eta_{jk}y^k$.

In the above subtractions the second contribution is of the same order in $\delta(y)$ as the first one. To show this, recall from the definitions
\begin{displaymath}
\delta^{msl}(x,y)y_s y_l =<\hat{y}^m (x)\hat{y}^s(x)\hat{y}^l (x)>\eta_{sj}\eta_{lk} y^j y^k\,- y^m.
\end{displaymath}
Then we can use the following relations:
\begin{displaymath}
\hat{y}^s=<\hat{y}^s>(x)-\delta^s(x,\hat{y}).
\end{displaymath}
One substitutes this relation and taking into account that $<\delta^k(x,\hat{y})>=0$, one gets
\begin{displaymath}
\delta^{msl}(x,y)y_sy_l= \Big(<\hat{y}^m>(x)<\hat{y}^s>(x)<\hat{y}^l>(x)
\,+<\hat{y}^m>(x)<\delta^s(x,\hat{y})\delta^l(x,\hat{y})>\,+
\end{displaymath}
\begin{displaymath}
+<\hat{y}^l>(x)\,<\delta^s(x,\hat{y})\delta^m(x,\hat{y})>
\,+<\hat{y}^s>(x)\,<\delta^m(x,\hat{y})\delta^l(x,\hat{y})>\,-
\end{displaymath}
\begin{displaymath}
-<\delta^m(x,\hat{y})\delta^s(x,\hat{y})\delta^l(x,\hat{y})>\Big)\eta_{sj}\eta_{lk}y^j y^k\, -y^m.
\end{displaymath}
Now we use a similar relation to go further in the calculation. Let us write
\begin{displaymath}
{y}^s=<\hat{y}^s>(x)-\delta^s(x,{y}).
\end{displaymath}
We introduce these expressions in the calculation of $\delta^{msl}y_sy_l$:
\begin{displaymath}
\delta^{msl}(x,y)y_sy_l = \Big(<\hat{y}^m>(x)\,(y^l+\delta^l(x,y))(y^s+\delta^s (x,y))+
<\hat{y}^m>(x)\,<\delta^s(x,\hat{y})\delta^l(x,\hat{y})>\,+
\end{displaymath}
\begin{displaymath}
+(y^l+\delta^l(x,y))<\delta^s(x,\hat{y})\delta^m(x,\hat{y})>\,+(y^s+\delta^s(x,y))
<\delta^m(x,\hat{y})\delta^l(x,\hat{y})>\,-
\end{displaymath}
\begin{displaymath}
-<\delta^m(x,\hat{y})\delta^s(x,\hat{y})\delta^l(x,\hat{y})>\Big)\eta_{sj}\eta_{lk}y^j y^k\, -y^m.
\end{displaymath}
Using again the fact that $y^l\eta_{lk}y^k=1$ we get (again using $<\hat{y}^m>=y^m+\delta^m(x,y)$
 to recombine the first and last term),
\begin{displaymath}
\delta^{msl}(x,y)y_sy_l =\delta^m(x,y)\,+2<\hat{y}^m>(x)\,\delta^s(y)\eta_{sj}y^j\,
+2\mathcal{O}^m_2(\delta^2)+2\mathcal{O}^m_3(\delta^3),
\end{displaymath}
the tensors $\mathcal{O}_2$ and $\mathcal{O}_3$ are given by the formulae
\begin{displaymath}
\mathcal{O}^i_2(\delta^2(y))(x,y)=\frac{1}{2}\,{\bf F}^i\,_{m}\Big( <\hat{y}^m>(x)\delta^s(x,y)
\delta^l(x,y)+<\hat{y}^m><\delta^s(x,\hat{y})\delta^l(x,\hat{y})>\,
\end{displaymath}
\begin{displaymath}
+2<\hat{y}^l>(x)<\delta^m (x,\hat{y})\delta^s(x,\hat{y})>\Big)\eta_{sj}\eta_{lk}y^jy^k,
\end{displaymath}
\begin{displaymath}
\mathcal{O}^i_3(\delta^3(y))(x,y)\,= \frac{1}{2}\,{\bf F}^i\,_{m}<\delta^m(x,\hat{y})\delta^s(x,\hat{y})\delta^l(x,\hat{y})>
\eta_{sj}\eta_{lk}y^jy^k.
\end{displaymath}

The {\it transversal contribution} to the difference between the connections is given by:
\begin{displaymath}
\frac{1}{2}{\bf F}^i\,
_m(x)\big(\delta^m(x,y)\eta_{jk}-\eta_{js}
\eta_{kl}\delta^{msj}(x,y)\,y^j y^k\big) =\frac{1}{2} {\bf
F}^i\,_{m}(x)\big(\delta^m(x,y)\,-\delta^m(x,y)\,
\end{displaymath}
\begin{displaymath}
-2<\hat{y}^m>(x)\delta^s(x,y)\eta_{sj}y^j\big)\,-(\mathcal{O}^m_2(\delta^2)-\mathcal{O}^m_3(\delta^3))\,{\bf F}^i\,_m=
\end{displaymath}
\begin{displaymath}
=-{\bf
F}^i\,_{m}(x)\Big(<\hat{y}^m>(x)\delta^s(x,y)\eta_{sj}y^j\,-\frac{1}{2}\mathcal{O}^m_2(\delta^2)-\,\frac{1}{2}\mathcal{O}^m_3(\delta^3)\Big)=
\end{displaymath}
\begin{displaymath}
=-{\bf
F}^i\,_{m}(x)\Big(<\hat{y}^m>(x)\delta^s(x,y)y_s\,-\frac{1}{2}\mathcal{O}^m_2(\delta^2)-\,\frac{1}{2}\mathcal{O}^m_3(\delta^3)\Big)=
\end{displaymath}
The {\it longitudinal contribution} is
\begin{displaymath}
\frac{1}{2}({\bf
F}^i\,_{j}(x)(<\hat{y}^m>(x)\,-y^m)\eta_{mk}+ {\bf
F}^i\,_{k}(x)(<\hat{y}^m>(x)\,-y^m)\eta_{mj})\,y^jy^k={\bf F}^i\,_j(x)\, y^j\delta^k(x,y)y_k.
\end{displaymath}
Adding together the {\it longitudinal} and {\it transversal} contributions
 and taking into account the formula $\delta^m(x,y)=<\hat{y}^m>(x)\,-y^{m}$
  we get the following expression:
\begin{displaymath}
({\bf F}^i\,_j y^j)(\delta^k(x,y)y_k) -\Big({\bf
F}^i\,_{m}(x)\big(<\hat{y}^m>(x)\,\delta^s(x,y)\eta_{sj}y^j\,
+\frac{1}{2}\mathcal{O}_2(\delta^2)\,+\frac{1}{2}\mathcal{O}_3(\delta^3)\Big)=
\end{displaymath}
\begin{displaymath}
=-\big({\bf
F}^i\,_{m}\delta^m(x,y)\big)\,\big(\delta^k(x,y) y_{k}\big)-\,\big(\frac{1}{2}\mathcal{O}^m_2(\delta^2)+\,\frac{1}{2}\mathcal{O}^m_3(\delta^3)\big)\,{\bf F}^i\,_m.
\end{displaymath}\hfill$\Box$

Let us consider a frame $\{e_i,\,i=0,...,n-1\}$ such that $\bar{\eta}_Z$ is diagonal at the point $x\in {\bf M}$ in this frame.
After calculating the distance
 of the connections using the formula $(4.6.3)$, the leading term in
  $\delta$ is quadratic. Also recall that in {\it section 1.3} we have fixed our units of energy and
  momentum in such a way that they are given by dimensionless numbers.

{\bf Remark}. Given a norm $\bar{\eta}_Z$ on ${\bf T}_x{\bf M}$, one can
 define an induced distance on $\pi^{-1}_1(u),\,u\in \pi^{-1}(x)$. This norm is just defined as $d_{\bar{\eta}_Z}(\xi,\zeta):=d_{\bar{\eta}_Z}(\pi_1(\xi),\pi_1(\zeta)),\,\,\zeta,\xi\in \pi^{-1}_1(u)$.
\begin{proposicion}
Let $({\bf M},\eta, [A])$ and $^L\nabla$ be as before and assume that
$f_x$ has compact and connected support for each fixed $x\in{\bf M}$
and that $\alpha:=\,sup \{\alpha_x,\,\,x\in {\bf M}\}<<1$. Then the following holds,
\begin{equation}
d_{\bar{\eta}_Z}(\,\pi^*<^L\nabla>,\,^L\nabla)(x)\leq \,\|{\bf F}\|_{\bar{\eta}_Z}(x)C(x){\alpha}^2+
\, 2C^2_2 (x){\alpha}^2(1+{\alpha})+\, C^3_3(x){\alpha}^3(1+{\alpha}),
\end{equation}
with $C(x), C_2(x), C_3(x)$ being functions depending only on $x$ with value of the order of unity.
\end{proposicion}
{\bf Proof}: Let $\{\pi^* e_0,...,\pi^* e_{n-1}\}$ be a local orthonormal frame for the induced
fiber metric on the pull-back bundle from the Riemannian metric $\bar{\eta}$. From equation $(4.6.3)$ one obtains
\begin{displaymath}
\|\, ^L\nabla_y y-\, \pi^*<\, ^L\nabla>_y y\|_{\bar{\eta}_Z}=\| {\bf F}^i\,_j(x)\delta^j(x,y)\delta^k (x,y) y_k\,\pi^*e_i\,
+\mathcal{O}_2(\delta^2)(x,y)\,+\mathcal{O}_3(\delta^3)(x,y)\|_{\bar{\eta}_Z}\leq
\end{displaymath}
\begin{displaymath}
\leq \| {\bf F}^i\,_j\delta^j(x,y)\delta^k(x,y) y_k \,e_i\|_{\bar{\eta}_Z}\,
+\|\mathcal{O}_2(\delta^2)(x,y)\|_{\bar{\eta}_Z}\,+\|\mathcal{O}_3(\delta^3)(x,y)\|_{\bar{\eta}_Z}.
\end{displaymath}
Each of these three terms can be bounded.

Recall that we are using a local frame such that $\|e_i\|_{\bar{\eta}_Z}=1$. Then we can bound the first term in the following way:
\begin{displaymath}
\| {\bf F}^i\,j \delta^j(x,y)\delta^k(x,y) y_k\, e_i\|_{\bar{\eta}_Z}\leq\,
\|{\bf F}\|_{\bar{\eta}_Z}\cdot \|\delta(x,y)\|_{\bar{\eta}_Z}\cdot |\delta^k (x,y) y_k|.
\end{displaymath}

For a fixed $x$ the support of the distribution function $f(x,y)$
 is compact and connected, thus one can write the decomposition $<\hat{y}>(x)=\epsilon (x)+z(x)$
 with the property $z(x)\in supp(f_x)$. In the case that $\eta$ is the Minkowski metric
 one can check that $\|\epsilon(x)\|_{\bar{\eta}_Z} \leq \,\alpha(x)\leq \alpha$ by geometric inspection.
 This bound of $\epsilon(x)$ follows from the shape of
 the unit hyperboloid and is proved in the following way. First, note that the domain
$\widehat{{\bf \Sigma}}_x:=\{y\in {\bf T}_x{\bf M}\,|\,\eta(y,y)\geq 1\,\,y^0>0\,\}$
is a convex set with respect to $\bar{\eta}$. Indeed, we note that
$\partial\big(\widehat{{\bf \Sigma}}_x\big)={\bf \Sigma}_x$ and
that $<\hat{y}>(x)\in \widehat{{\bf \Sigma}}_x$. Secondly, each
 $(\widehat{{\bf \Sigma}}_x, \bar{\eta}_{Z_x})$ is a Riemannian manifold,
  with $\bar{\eta}_{Z_x}:=(\bar{\eta}_Z)_{ij}(x)dy^i\otimes dy^j$.
Therefore, we can use the standard definition of center of mass [48],
 in this case with a measure given by $f(x,y,s)\, dvol(x,y)$. The function $f(x,y,s)$ is such that
\begin{displaymath}
\int^1_0 ds\,f(x,y,s)=f(x,y),
\end{displaymath}
$s$ is the parameter of the line
 connecting $y\in supp(f_x)$ with $<\hat{y}>$.
Let us denote by $\widehat{supp}(f_x)$ the convex hull of $supp(f_x)$.
By construction $<\hat{y}>\in \widehat{supp}(f_x)$.
One can check that $<\hat{y}>$ is the center of mass of the convex set $\widehat{supp}(f_x)$ [48].

We have the following bound
\begin{displaymath}
\|\delta(x,y)\|_{\bar{\eta}_Z}\,\leq\|<\hat{y}>(x)\,-y\|_{\bar{\eta}_Z}\,\leq \|\epsilon(x)+{z}(x)-y\|_{\bar{\eta}_Z}
\,\leq
\end{displaymath}
\begin{displaymath}
\leq \|\epsilon(x)\|_{\bar{\eta}_Z}+\,\|z(x)-y\|_{\bar{\eta}_Z}\,\leq \,\alpha\,+\alpha=2 \alpha.
\end{displaymath}

For the third factor, one has the following bound ($\delta(x,{y})=<{y}>-{y}$)
\begin{displaymath}
|\delta^k(x,y) y_k|=|<\hat{y}^k >(x)y_k -1|=|<\hat{y}^k>(x)\,(y_k\,-<\hat{y}^k>(x)\,+<\hat{y}^k>(x))\,-1|\leq
\end{displaymath}
\begin{displaymath}
\leq |<\hat{y}^k >(x)(y_k\,-<\hat{y}_k>(x))|\,+|<\hat{y^k}>(x)<\hat{y}_k>(x)-1|.
\end{displaymath}
Using Cauchy-Schwarz inequality for $\bar{\eta}_Z$ we obtain
\begin{displaymath}
|\delta^k(x,y) y_k|\leq \, \|<\hat{y}^k >(x)\|_{\bar{\eta}_Z}\,\|(y_k\,-<\hat{y}_k>(x))
\|_{\bar{\eta}_Z}+|<\hat{y}^k>(x)<\hat{y}_k>(x)-1|\,\leq
\end{displaymath}
\begin{displaymath}
\leq \|<\hat{y}_k>(x)\|_{\bar{\eta}_Z}\,{\alpha}\,+|<\hat{y}^k>(x)<\hat{y}_k>(x)\,-1|\,
\leq \sqrt{1+\,\|\epsilon\|_{\bar{\eta}_Z}}\,\alpha\,+(\sqrt{1+\,\|\epsilon(x)\|_{\bar{\eta}_Z}}-1)\,\leq
\end{displaymath}
\begin{displaymath}
\leq \sqrt{1+\,{\alpha}}\,\alpha\,+(\sqrt{1+\,{\alpha}}-1)\,
\leq ({1+\,{\alpha}})\alpha\,+({1+\,{\alpha}}-1)=\,2\alpha\,+{\alpha^2}.
\end{displaymath}
Therefore, one obtains
\begin{displaymath}
\| ({\bf F}^i\,_j \delta^j(x,y))(\delta^k (x,y)y_k)\,e_i\|_{\bar{\eta}_Z}\leq\,\|{\bf F}\|_{\bar{\eta}_Z}(x) C(x){\alpha}^2\,+\mathcal{O}(\alpha^4).
\end{displaymath}
The function $C(x)$ in equation $(4.6.6)$ is bounded by the constant $4$ in
 the coordinate frame determined by the vector field $Z$. This bound is universal,
  independent of the Lorentzian metric $\eta$, the vector field $Z$ and it has a geometric origin.

Using homogeneity properties on the variable $y$, one can see that the following relations hold:
\begin{equation}
\|\mathcal{O}_2(\delta^2)\|_{\bar{\eta}_Z}\leq\, C^2_2(x){\alpha}^2(1+B_2(x)\,{\alpha}),
\end{equation}
and
\begin{equation}
\|\mathcal{O}_3(\delta^3)\|_{\bar{\eta}_Z}\leq\, C^3_3(x)\,{\alpha}^3(1+B_3(x)\,{\alpha}).
\end{equation}
The functions $C_i(x)$ depend on the particular shape of the support of the distribution
 function $f$ and on the curvature of the metric $\eta$.
Using geometric arguments (and in particular compactness and connectedness of the $supp(f(x))$)in {\bf K},
one can bound these functions in terms of $\alpha$ in a similar way as we did for $C(x)$. The constants
 are of order 1 because this was the case for $C(x)$ and there are
 no new {\it divergence} factors in the functions $B_i(x)$ and $C_i(x)$.
\hfill$\Box$

\begin{corolario} Let $({\bf M},\eta, [A])$ be a (semi)-Randers space.
Let us consider a compact domain ${\bf K}\subset\bigsqcup_{x\in {\bf M}} supp(f_x)$
compact, with ${^L\nabla}$ and $<\,^L\nabla>$ as before. Then there is a global bound:
\begin{displaymath}
d_{\bar{\eta}_Z}(\, ^L\nabla,\pi^*<\,^L\nabla>)(x)\leq\, {C}\|{\bf F}\|_{\bar{\eta}_Z}{\alpha}^2\,
+2C^2_2{\alpha}^2(1+B_2{\alpha})\, +C^3_3{\alpha}^3(1+B_3{\alpha}), \quad \forall x\in {\bf M}.
\end{displaymath}
where the constants $C, C_2, C_3, B_2, B_3$ are bounded by a constant of order $1$.
\end{corolario}
{\bf Proof}: It follows from {\it proposition} $(4.6.4)$ and compactness of the domain ${\bf K}\subset
\bigsqcup_{x\in {\bf M}} supp(f_x)$ that we are considering.\hfill$\Box$

\subsection{Comparison between the geodesics of $^L\nabla$ and $\pi^*<\,^L\nabla>$}

There are several ways of defining the energy of a bunch of particles. We have chosen one which will
be useful for our comparison results.
We define the energy function $E$ of a distribution $f$ to be the real function
\begin{displaymath}
E:{\bf M}\longrightarrow {\bf R}
\end{displaymath}
\begin{equation}
x\mapsto E(x):=inf\{ y^0,\, y\in supp(f_x)\},
\end{equation}
where $y^0$ is the $0$-component of a tangent vector of a possible trajectory of a charged point particle,
 measured in the laboratory coordinate frame. The name {\it energy} for this function
 is deserved because of the choice of the units that we have adopted, even if energy is a function on the co-tangent bundle ${\bf T}^*{\bf M}$ instead of the bundle ${\bf TM}$, where the velocities are defined.
\paragraph{}
Let us restrict our attention to the case that the Lorentzian metric is the Minkowski metric in dimension $n$.
We can define $\theta ^2(t)=\vec{y}^2(t)-\,<\vec{\hat{y}}>^2(t)$
 and $\bar{\theta}^2(t)=<\vec{\hat{y}}>^2(t)-\vec{\tilde{y}}^2(t)$. Here $\vec{y}(t)$
   is the spatial component of the velocity tangent vector field along a solution of the Lorentz force equation
    and $\vec{\tilde{y}}(t)$ is the spatial component of the tangent vector field along a solution
     of the averaged Lorentz force equation, with both solutions having the same initial conditions. The spatial components are defined respect the observer $Z=\frac{\partial}{\partial t}.$ We will call this observer the laboratory frame. Since we are fixing this frame, the corresponding Riemannian metric $\|\cdot\|_{\bar{\eta}_Z}$ is denoted simply by $\|\cdot\|_{\bar{\eta}}$, simplifying the notation.
      The maximal values of these quantities on the compact domain {\bf K} of the space-time manifold {\bf M} are denoted
       by $\theta^2$ and $\bar{\theta}^2$.

\begin{teorema}
Let $({\bf M},\eta, [A])$ be a semi-Randers space and $\eta$ the Minkowski metric. Let us assume that
\begin{enumerate}

\item The auto-parallel curves of unit velocity of the connections $^L\nabla$ and
 $<\,^L\nabla>$ are defined for time $t$, the time coordinate measured in the laboratory frame $Z=\frac{\partial}{\partial t}.$

\item The ultra-relativistic limit holds: $E(x)>>1$ for all $x\in {\bf K}$.

\item The distribution function is narrow in the sense that ${\alpha}<<1$ for all $x\in {\bf K}$ in the laboratory frame.

\item The following inequality holds,
\begin{displaymath}
|\theta^2(t)\,-\bar{\theta}^2(t)|\ll 1,
\end{displaymath}
\item The support of the distribution function $f$ is invariant under the flow of the Lorentz force equation.

\item The change in the energy function is adiabatic: $\frac{d}{dt} log E<<1$.
\end{enumerate}
Then for the same arbitrary initial condition $(x(0),\dot{x}(0))$, the solutions of the equations
\begin{displaymath}
^L\nabla_{\dot{x}} \dot{x}=0,\, \quad <\,^L\nabla>_{\dot{\tilde{x}}} \dot{\tilde{x}}=0
\end{displaymath}
 are such that
\begin{equation}
\|\tilde{x}(t)-\, x(t)\|_{\bar{\eta}_Z}\leq\, 2\big(C(x(t))\|{\bf F}\|_{\bar{\eta}_Z}(x(t))\,+C^2_2(x(t))(1+B_2(x(t)){\alpha})\big)
{\alpha}^2\,E^{-2}(x)\,t^2\,+\mathcal{O}(\alpha^4),
\end{equation}
where the functions $C(x(t))$, $C_i(x(t))$ and $B_i(x(t))$ are bounded by constants of order $1$.
\end{teorema}
{\bf Proof}: At the instant $t$, we calculate the distance measured in the laboratory frame
between $x(t)$ and $\tilde{x}(t)$, solutions of the geodesic equations of the
 connections $^L \nabla$ and $<\,^L\nabla>$ respectively. Both geodesics have the same initial
 conditions $(x(0),\dot{x}(0))$. Let us start writing  the general expression of the solution of the Lorentz force equation for those initial conditions:
\begin{equation}
x^i (t)=x^i (0)+\int^t _0 ds \Big(\dot{x}^i(0) + \int^s _0
dl\ddot{x}^i(l)\Big).
\end{equation}
Since the initial conditions for both geodesics are the same, the equivalent
relation for the geodesics of the averaged connection is
\begin{equation}
\tilde{x}^i (t)=x^i (0)+\int^t _0 ds \Big(\dot{x}^i(0) + \int^s _0
dl\ddot{\tilde{x}}^i(l)\Big).
\end{equation}
We estimate the {\it distance} between both solutions at the instant $t$.
 Since we know
the distance between the connections $^L\nabla$ and $\pi^*<\,^L\nabla>$,
 it is possible to give a natural bound for the distance between the solutions. The main tool that we use
 is the smoothness theorem on the dependence of solutions of differential
 equations on the external parameters (for instance [21, {\it Appendix 1}] or [40, {\it chapter 1}] or in the {\it appendix} of this thesis).

Let us consider the family of connections depending on the distance between the two connections
\begin{displaymath}
\xi_{max}=d_{\bar{\eta}}(\,^L\nabla, <\,^L\nabla>)
\end{displaymath}
 given by the convex sum:
\begin{equation}
^{\xi}\nabla:=\frac{1}{\xi_{max}}(\xi_{max}-\xi)\,^L\nabla\,+
\frac{1}{\xi_{max}}\xi\pi^*<\,^L\nabla>, \quad \xi\in [0, \xi_{max}].
\end{equation}
For $\xi=0$ one has $^{\xi}\nabla=\,^L\nabla,$ while for $\xi=\xi_{max}$ one has
 the averaged connection. Using the result of the smoothness of the solutions
 of the differential equations, one can expand the solution $^{\xi}x^i$ of the
 geodesic equation for the connection with parameter $\xi$. The second derivative with respect to the coordinate time $t$ reads
\begin{equation}
^{\xi}\ddot{x}^i\,=\,^0\ddot{x}^i+\, (\partial_{\xi} \,^{\xi}\ddot{x}^i)|_{\xi =0}\cdot \xi +\mathcal{O}(\xi^2).
\end{equation}
We need to bound the derivative $(\partial_{\xi} \,^{\xi}\ddot{x}^i)|_{\xi =0}$. The first think is that if $\dot{x}$ (the tangent velocity vector to a Lorentz geodesic) is not on the support of the distribution $f$, this derivative can be done arbitrary large.
From the formula $(4.6.13)$ and $(4.6.14)$, one obtains:
\begin{displaymath}
(\partial_{\xi} \,^{\xi}\ddot{x}^i)|_{\xi =0}=\frac{1}{\xi_{max}}\cdot
 (\, ^L\nabla_{\dot{x}} \dot{x} -\, \pi^*<\,^L\nabla >_{\dot{x}}\dot{x})^i,
\end{displaymath}
where $\dot{x}(t)$ is the solution of $^L\nabla_{\dot{x}} \dot{x}=0$ with the given
 initial conditions. This is because the support of the distribution function $f$
  is invariant under the flow of the Lorentz force equation.
We are dividing by the distance $\xi_{\max}$, thus the derivative
is such that its norm is bounded by $1$,
\begin{displaymath}
\|(\partial_{\xi} \,^{\xi}\ddot{x})|_{\xi =0}\|_{\bar{\eta}}\,=
\frac{1}{\xi_{max}}\cdot \|(\, ^L\nabla_{\dot{x}} \dot{x} -\, \pi^*<\,^L\nabla >_{\dot{x}}\dot{x})\|_{\bar{\eta}}\leq 1,
\end{displaymath}
because of the definition of $\xi_{max}$ and the formula $(4.6.14)$.
 Note that for writing this condition, it is essential that the support
  of the distribution $f$ must be invariant under the flow defined by the Lorentz force equation: if this is not the case, the parameter $\xi_{max}$ can not be defined and the difference on the covariant derivatives cannot be bounded. On the other hand
the relations between proper times and
coordinate time in the laboratory frame are
\begin{displaymath}
d\tau=\gamma^{-1}dt, \quad d\tilde{\tau}=\tilde{\gamma}^{-1}dt.
\end{displaymath}
This implies the following relation between derivatives,
\begin{displaymath}
\frac{d}{dt}=\gamma^{-1}\frac{d}{d\tau},\quad
\frac{d}{dt}=\tilde{\gamma}^{-1}\frac{d}{d\tilde{\tau}}.
\end{displaymath}

Using the hypotheses $|\theta^2(t)\,-\bar{\theta}^2(t)|<<1$ and $\frac{d}{dt} log E<<1$, one obtains the following relation,
\begin{displaymath}
\|\tilde{x}(t)-{x}(t)\|_{\bar{\eta}}\leq \,2 t\int^t _0 dl E^{-2}\,\|
\frac{d^2\tilde{x}^i(l)}{dl^2}-\frac{d^2{x}^i(l)}{dl^2}\|_{\bar{\eta}},
\end{displaymath}
where by the adiabatic hypothesis, the time derivatives of the energy function
have been dropped out; the factor $2$ comes from the bound of the term
 which contains the derivative of the energy.\hfill$\Box$
\begin{corolario}
Let $({\bf M},\bar{\eta})$ be as before and such that there is a global bound for
$\|{\bf F}\|_{\bar{\eta}}(x)\leq \|{\bf F}\|_{\bar{\eta}}<\infty$, the energy
 function is bounded from below by a constant $E$ and the curves $x(t)$
  and $\tilde{x}(t)$ are compact. Then there are some constants $C_i$
  such that $C_i(x)<C_i$ and the following relation holds
\begin{equation}
\|\tilde{x}(t)-\, x(t)\|_{\bar{\eta}}\leq\,2 \big(C\|{\bf F}\|_{\bar{\eta}}\,+C^2_2(1+{\alpha})\big){\alpha}^2\,E^{-2}\,t^2\,+\mathcal{O}(\alpha^4),
\end{equation}
\end{corolario}
with ${\bf F}$ the maximal value of ${\bf F}(x)$ attached along the compact curves $x(t)$ and $\tilde{x}(t)$.

{\bf Remarks}
\begin{enumerate}

\item
 In the above result the $2$-form {\bf F} is physically interpreted as the Faraday form.

\item Global bounds occurs in two possible scenarios:

\begin{enumerate}

\item When manifold {\bf M} is compact. In this case, one has to consider spacelike boundaries, since it is well-known that if the space-time is compact without boundaries, there exists closed time-like curves ([23, pg 58]) and this violates causality.

\item The trajectories that we consider are compact in space and bounded in time between an initial time $t=0$ and a final time $t=T$. In this case, one can define
 an {\it effective compact space-time manifold} with a spurious boundary
  and apply the first case, with the exclusion of closed time-like curves, which do not occur in physical situations.
\end{enumerate}

\item We have assumed in our calculations that the external field {\bf F} does not depend
on the energy of the beam of particles. However, this is not necessarily
 the case in some situations (as in betatron accelerator machines [9,10]).

\item There are effects which could reduce the beam size (adiabatic damping
  and Landau Damping [9]). If this happens, there is a strong reduction of the size of the dispersion in energy
  and momenta of the beam. This implies that one can describe this as an effective
  exponent $E^{-2+\beta}$ with $\beta<0$. Then our results are safe under these kind of effects.

\item That the curves $x(t)$ and $\tilde{x}(t)$ have compact image has a physical
 interpretation: all the trajectories start in a source region and finish in a target region.

\item Although we have chosen $Z=\frac{\partial}{\partial t}$ to be the observer
 corresponding to the laboratory frame, the same kind of calculations can be performed in any frame.
\end{enumerate}

\begin{teorema} Under the same hypothesis as in {\it theorem
4.6.6}, the difference between the tangent vectors is given by
\begin{equation}
\|\dot{\tilde{x}}(t)-\dot{x}(t)\|_{\bar{\eta}}\leq \big(K(x)\|{\bf F}\|_{\bar{\eta}}(x)\,+K^2_2(1+D_2(x)
{\alpha})\big){\alpha}^2\,\big)\,E^{-1}\,t\,+\,\mathcal{O}(\alpha^4).
\end{equation}
with ${K}(x)$, $K_2(x)$ and $D_2(x)$ functions bounded by constants bounded by respective constants of order $1$.
\end{teorema}
{\bf Proof}: The proof of this theorem is similar to the proof of {\it Theorem} $(4.6.6)$,
 although based on the following formula for the tangent velocity field along a curve:
\begin{equation}
\dot{x}(t)=\dot{x}(0)+\int^t_0\ddot{x}(l)dl.
\end{equation}\hfill$\Box$
\begin{corolario}
Under the same hypothesis than in {\it Corollary 4.6.7}, there are some constants of order $1$, $K$, $K_2$ and $D_2$, such that $K(x)\leq K,$ $K(x)_2\leq K_2$, $D_2(x)\leq D_2$ and the following relation holds,
 \begin{equation}
\|\dot{\tilde{x}}(t)-\dot{x}(t)\|_{\bar{\eta}}\leq \big(K\|{\bf F}\|_{\bar{\eta}}(x)\,+K^2_2(1+D_2(x)
{\alpha})\big){\alpha}^2\,\big)\,E^{-1}\,t+\,\mathcal{O}(\alpha^4).
\end{equation}
\end{corolario}

\section{Discussion}
\subsection{Structural stability of the approximation \\
$^L\nabla\longrightarrow \pi^*<\,^L\nabla>$}

In the proof of proposition $(4.6.3)$ there is a cancelation of the leading orders
of transversal and longitudinal contributions when we calculate
 the difference between $<\,^L\nabla>$ and $^L\nabla$. The contribution coming from the {\it transversal} terms was:
\begin{displaymath}
\frac{1}{2}{\bf F}^i\,
_m(x)\big(\delta^m(x,y)\eta_{jk}-\eta_{js}
\eta_{kl}\delta^{msj}(x,y)\,y^j y^k\big) =\frac{1}{2} {\bf
F}^i\,_{m}(x)\big(\delta^m(x,y)\,-\delta^m(x,y)\,
\end{displaymath}
\begin{displaymath}
-2<\hat{y}^m>(x)\delta^s(x,y)\eta_{sj}y^j\big)\,-\mathcal{O}^m_2(\delta^2)-\mathcal{O}^m_3(\delta^3)\,{\bf F}^i\,_m(x)=
\end{displaymath}
\begin{displaymath}
=-{\bf
F}^i\,_{m}(x)\big(<\hat{y}^m>(x)\delta^s(x,y)\eta_{sj}y^j\big)\,-\mathcal{O}^m_2(\delta^2)-\mathcal{O}^m_3(\delta^3)\,{\bf F}^i\,_m(x).
\end{displaymath}
The contribution coming from the {\it longitudinal} terms was
\begin{displaymath}
\frac{1}{2}({\bf
F}^i\,_{j}(x)(<\hat{y}^m>(x)\,-y^m)\eta_{mk}+ {\bf
F}^i\,_{k}(x)(<\hat{y}^m>(x)\,-y^m)\eta_{mj})\,y^jy^k={\bf F}^i\,_j(x)\, y^j\delta^k(x,y)y_k.
\end{displaymath}
The reason for this cancelation of the first order term in $\delta$ is based on the formal structure of the
 connection $^L\nabla$. This structure has a two-fold origin:
\begin{enumerate}

\item The definition we have adopted for a general non-linear connection and

\item The structure of
  the Lorentz force equation.
\end{enumerate}
The cancelation is independent of the details of the distribution, even if  $\alpha$ is not small.
It is independent of the value of {\bf F}.

The cancelation of the linear terms can be written in the following way:
\begin{displaymath}
^L\nabla_y y\,-\pi^*<\,^L\nabla>_y y=\,\mathcal{O}(\alpha^2(x)).
\end{displaymath}

Let us consider the connection
\begin{displaymath}
\tilde{\nabla}=\,^L\nabla-T,\quad T^i\,_{jk}={\bf F}^i\,_m (x)
\frac{y^m}{2\sqrt{\eta(y,y)}}\Big(\eta_{jk}-\frac{1}{\eta(y,y)}\eta_{js}
\eta_{kl}y^s y^l\Big).
\end{displaymath}
The connection $\tilde{\nabla}$ is such that its auto-parallel curves coincide
with the solutions of the Lorentz force equation. It is also gauge invariant. Therefore, it is also a {\it good}
 candidate for a geometrization of the Lorentz force equation. However,
  if we calculate the analogous difference $ \tilde{\nabla}_y y\,-\pi^*<\tilde{\nabla}>_y y$
   we get that in general it is linear in $\delta$, which implies
 \begin{displaymath}
 \tilde{\nabla}_y y\,-\pi^*<\tilde{\nabla}>_y y=\,\mathcal{O}(\alpha(x)).
 \end{displaymath}
Finally, if we consider the covariant derivative $^{\chi}\nabla$ associated
 with a spray vector field $\chi \in \Gamma{\bf TN} $, we obtain
\begin{displaymath}
^{\chi}\nabla_y y\,-\pi^*<\,^{\chi}\nabla>=\,\mathcal{O}(\alpha(x)).
\end{displaymath}
We can write this relations as
\begin{equation}
\lim_{\alpha\rightarrow 0} \frac{ ^L\nabla_y y\,-\pi^*<\,^L\nabla>_y y}{\alpha}=0,
\end{equation}
\begin{equation}
\lim_{\alpha\rightarrow 0} \frac{ \tilde{\nabla}_y y\,-\pi^*<\tilde{\nabla}>_y y}{\alpha}\neq 0,
\end{equation}
\begin{equation}
\lim_{\alpha\rightarrow 0} \frac{ \,^{\chi}{\nabla}_y y\,-\pi^*<\,^{\chi}{\nabla}>_y y}{\alpha}\neq 0.
\end{equation}
This fact can be stated in the following way. Let us denote a
$1$-parameter family of linear connections by $\nabla(\alpha)$ such that both $\nabla(\alpha)$ and
$\nabla(\alpha+h)$ are defined. Then one can define the derivative operator
\begin{displaymath}
\lim_{h\rightarrow 0} \frac{\nabla(\alpha +h)_y\, y-\nabla(\alpha)_y \,y}{h},
\end{displaymath}
which is formally the differential (in the way we define, it is the Gateaux differential
 [52, {\it chapter 5}]) of the operator
\begin{displaymath}
\nabla(\alpha,y):{\bf I}\longrightarrow {\bf T}_x{\bf M}
\end{displaymath}
\begin{displaymath}
\alpha\mapsto ^L\nabla(\alpha)_y y.
\end{displaymath}
for a fixed $y\in  supp(f_x)$ and such that $\alpha, \alpha+h\in {\bf I}$.
 The notation makes sense such that for each $y$
 there is a given operator $\nabla_y(\alpha) y$. Then equations
 $(4.7.1)$, $(4.7.2)$ and $(4.7.3)$ can be stated in terms of derivatives.
  In particular, since $\nabla(\alpha=0,y)=\,^L\nabla_y y$, one has that equation
   $(4.7.1)$ is equivalent to the statement that $^L\nabla_y y$ is a critical value of $\nabla(\alpha,y)$. Therefore one can write
 \begin{displaymath}
\pi^*<\,^L\nabla_y y>=\,\pi^*<\nabla(0,y)>\,+\frac{\alpha^2}{2}\,\frac{d^2}{d\alpha^2}\big|_{\alpha=0}\nabla(\alpha, y)\,+\mathcal{O}(\alpha^3).
 \end{displaymath}
 When the support of the distribution is invariant under the flow of the Lorentz equation, for $\alpha=0$ one has that $^L\nabla=\pi^*<\,^L\nabla>$. Therefore,
  \begin{equation}
 <\,^L\nabla_y y>=\,^L\nabla_y y+\,\frac{\alpha^2}{2}\,\frac{d^2}{d\alpha^2}\big|_{\alpha=0}\nabla(\alpha, y)\,+\mathcal{O}(\alpha^3).
 \end{equation}
 for any $y\in {\bf \Sigma}$.
 For arbitrary distribution functions $\frac{\alpha^2}{2}\,\frac{d^2}{d\alpha}\big|_{\alpha=0}\nabla(\alpha, y)\neq 0$.

 On the other hand, equation $(4.7.2)$ can be rewritten, following the same steps
 \begin{equation}
  \pi^*<\tilde{\nabla}_y y>=\tilde{\nabla}_y y+\,{\alpha}\,\frac{d}{d\alpha}\big|_{\alpha=0}\tilde{\nabla}(\alpha, y)\,+\mathcal{O}(\alpha^2)
 \end{equation}
 for any $y\in {\bf \Sigma}$.
 For arbitrary distribution functions $\frac{d}{d\alpha}\big|_{\alpha=0}
 \tilde{\nabla}(\alpha, y)\neq 0$. Also note that $0=\,^L\nabla_y y
 =\,\tilde{\nabla}_y y$ for any $y\in {\bf \Sigma}$,
 which coincides with the Lorentz force equation.

 In the case of a connection $^{\chi}\nabla$ obtained from an arbitrary spray $\chi$, the relation is
  \begin{equation}
  \pi^*<\,^{\chi}{\nabla}_y y>=\,^{\chi}{\nabla}_y y+\,{\alpha}\,\frac{d}{d\alpha}\big|_{\alpha=0}\,^{\chi}{\nabla}(\alpha, y)\,+\mathcal{O}(\alpha^2).
 \end{equation}

 Apart from the relevance for calculational purposes, it is interesting to know
 if there is some reason why $(4.7.4)$ holds for the Lorentz connection
 $^L\nabla$ obtained from the Berwald connection associated to the spray
 $^L\chi$. If one changes the way we obtain the connection from the spray,
 or changes the spray (for instance using the connection $\tilde{\nabla}$
 instead of $^L\nabla$), one obtains conditions of the type $(4.7.5)$. If
 one performs a similar calculation for a general connection, a differential
 equation like $(4.7.6)$ is obtained. This suggests that there are two
 factors in obtaining the relation $(4.7.4)$:
 \begin{enumerate}

 \item The choice of the non-linear connection as the Berwald-type connection
 applied to the Lorentz spray $^L\chi$.

 \item The particular structure of the Lorentz equation. This is apparent
 in the calculations in the proof of {\it proposition} $(4.6.3)$.

 \end{enumerate}
 The conclusion is that the non-linear Berwald connection obtained from the
  Lorentz force equation is structurally stable with respect to the parameter $\alpha$.
   As compared with other connections, this property happens (maybe only) for the Berwald-type connection.
 This notion of stability is similar to the one presented in [61, {\it chapter 3}].

\subsection{On the hypotheses on which the approximation \\
$^L\nabla\longrightarrow \pi^*<\,^L\nabla>$ is based}

We would like to discuss some reflections and interpretations of the hypotheses
 of {\it theorem} $(4.6.6)$ and subsequent results. Some of the hypotheses are not
 essential to perform the approximation, but are very useful in the calculations
 and in writing the asymptotic expressions. Therefore, we can differentiate between fundamental hypotheses, which are
\begin{enumerate}

\item The auto-parallel curves of unit velocity of the connections $^L\nabla$ and
 $<\,^L\nabla>$ are defined for the time $t$, which is the coordinate time measured
  in the laboratory frame.
This is an hypothesis on existence, since nothing can be done if the curves are not
 defined for the parameter that we are speaking about.

 \item The support of the distribution function $f$ is invariant under the flow of the Lorentz force equation.
If this hypothesis does not hold, the relation between the averaged connection and
 the Lorentz connection is arbitrary. Therefore, this is a fundamental hypothesis.
  Note that one does not need to assume that $f$ is a solution of the Vlasov
   equation in the sense that the condition is weaker. This implies that our result
    holds for alternative kinetic models for the distribution function $f$.

 \item The dynamics occurs in the ultra-relativistic limit, $E(x)>>1$ for all $x\in {\bf M}$.
This is a hypothesis which in principle is not fundamental for the calculation but
 it is fundamental for the approximation $^L\nabla\longrightarrow \pi^*<\,^L\nabla>$
  to be good. If this hypothesis does not hold, the difference between the solutions
   of the two differential equations will not be as small as in the ultra-relativistic
    limit and the expressions will be more involved.

\item The distribution function is narrow, ${\alpha}<<1$ for all $x\in {\bf M}$.
 This is useful to interpret the formulas as asymptotic series in $\alpha$. Note
  that for current accelerators, this condition holds.
\end{enumerate}
Apart from discussed above, there are other hypotheses which are not fundamental
 to the results, although they are helpful in the calculations
\begin{enumerate}

\item The following inequality holds
\begin{displaymath}
|\theta^2\,-\bar{\theta}^2|\ll 1.
\end{displaymath}

This hypothesis is only used in the proof to simplify some expressions.
 Therefore, it is not fundamental. The approximation $^L\nabla\longrightarrow \pi^*<\,^L\nabla>$ can be good even if $|\theta^2\,-\bar{\theta}^2|\ll 1$ is non true, but the asymptotic expressions
  will be more involved.

\item The adiabatic hypothesis $\frac{d log E}{dt}<<1$ is not necessary for the approximation
 $^L\nabla\longrightarrow \pi^*<\,^L\nabla>$ to be good. However, this hypothesis simplifies
  the calculations and the final expression. Note that this is a condition
   which is satisfied in actual accelerator machines.

\end{enumerate}
There are limitations on the validity of the results that we have obtained in this chapter.
\begin{enumerate}
\item The $2$-form ${\bf F}$ is interpreted physically as the Faraday form of
 an external electromagnetic field.
We have not commented on the dependence on the strength of the electromagnetic
 field. It is clear that for any finite $\alpha$, if the electromagnetic field
  is too strong, the approximation will not be good. However, for delta
  distribution functions with invariant support by the flow of the Lorentz
   vector field $^L\chi$, the approximation is always valid.

\item The main results of this chapter ({\it theorems} $(4.6.6)$ and $(4.6.8)$
 and {\it corollaries} $(4.6.7)$ and $(4.6.9)$) are not Lorentz covariant.
  However, as we have said before, given an arbitrary observer defined by a time-like
   vector field $Z$, it is possible to obtain similar results.
\end{enumerate}
\subsection{Range of applicability of the approximation}

Let us discuss the limits of applicability of the averaged dynamics and in particular
of the formula $(4.6.10)$. We will make the assumption that during the time of evolution the
$\gamma$ factors are increasing.
Therefore, equation $(4.6.10)$ takes the form
\begin{displaymath}
\|{\tilde{x}}(t)-{x}(t)\|_{\bar{\eta}}\leq\,  C\,{\alpha}^{2}\,E^{-2}(t_0) \|{\bf F}\|_{\bar{\eta}}\,t^2.
\end{displaymath}
Note that we are assuming $E(t)\geq E(t_0)$.

Let us assume a natural maximal spatial distance $L_{max}$ between points on $supp(f)$.
For instance, in an accelerator machine, $L_{max}$ can be related with the diameter of the pipe: $L_{max}$ must be smaller than it. Assume that the averaged model is a good approximation if the distance $\|{\tilde{x}}^i(t)-{x}^i(t)\|_{\bar{\eta}}$ is less than $L_{max}$.
 This can start to happen after a time evolution such that the difference $\|{\tilde{x}}^i(\,^1t_{max})-{x}^i(\,^1t_{max})\|_{\bar{\eta}}=L_{max}$. The characteristic time where the averaged model loses validity is
\begin{equation}
^1t_{max}\sim \, \big(\frac{L_{max}}{C_1}\big)^{\frac{1}{2}}\cdot \frac{E(t_0)}{\alpha}\,\cdot\Big(\frac{1}{\|{\bf
F}\|}\Big)^{\frac{1}{2}}.
\end{equation}
There is a second constraint coming from the spread in velocities. Let us consider the relations $(4.6.16)$ and $(4.6.17)$. For any distribution function, one can expect that the approximation $^L\nabla\longrightarrow <\,^L\nabla>$ to be valid until $\|\dot{\tilde{x}}(t)-\dot{x}(t)\|_{\bar{\eta}}$ is of order $\alpha$. This is because the unit hyperboloid is strictly convex subset of the tangent space. One writes the condition
\begin{displaymath}
\alpha=\big(K\|{\bf F}\|_{\bar{\eta}_Z}(x)\,+K^2_2(1+D_2
{\alpha})\big){\alpha}^2\,\big)\,E^{-1}(t_0)\,^0t_{max}\,=\,K\|{\bf F}\|_{\bar{\eta}_Z}{\alpha}^2\,E^{-1}(t_0)\,^2t_{max},
\end{displaymath}
The maximal time calculated in this way is
\begin{equation}
^2t_{max}=\,K\cdot\frac{E(t_0)}{\alpha}\cdot \frac{1}{\|{\bf F}\|_{\bar{\eta}}}.
\end{equation}
The asymptotic behavior of the time where the approximation is broken is universal. We have to point out that the approximation can be broken before $t_{max}$. However,
\begin{proposicion}
The following consequences are true
\begin{enumerate}
\item $\lim_{E\rightarrow{\infty}} t_{max}= \infty,$ if all the other parameters are finite,

\item $\lim_{\alpha \rightarrow 0} t_{max}= \infty,$ if all the other parameters are finite,

\item $\lim_{\|{\bf F}\|_{\bar{\eta}} \rightarrow 0} t_{max}= \infty,$ if all the other parameters are finite.
\end{enumerate}
\end{proposicion}
The existence of two maximal times up to where the approximation is valid provides a definition of maximal length $L_{max}$: it is the length for which
\begin{displaymath}
\big(\frac{L_{max}}{C}\big)^{\frac{1}{2}}\cdot \frac{E(t_0)}{\alpha}\,\cdot\Big(\frac{1}{\|{\bf
F}\|_{\bar{\eta}}}\Big)^{\frac{1}{2}}=\,K\cdot\frac{E(t_0)}{\alpha}\cdot \frac{1}{\|{\bf F}\|_{\bar{\eta}}}.
\end{displaymath}
All the quantities appearing are known, therefore one has
\begin{equation}
L_{max}=A\cdot \frac{1}{\|{\bf F}\|_{\bar{\eta}}},
\end{equation}
with $A$ a constant which does not depend on $x$, ${\bf F}$, $\alpha$ or $E$.

If we compare this definition with $\bar{L}_{max}:=diam(\pi(supp(f)))$, we have a criterion for a definition of {\it weak} $2$-form:
\begin{definicion}
The $2$-form {\bf F} is said to be weak iff $
L_{max}\leq \bar{L}_{\max}.$
\end{definicion}
\begin{proposicion}
If the $2$-form ${\bf F}$ is not weak, then the approximation $^L\nabla\longrightarrow <\,^L\nabla>$ is not valid.
\end{proposicion}
{\bf Proof:} If $\bar{L}_{max}\geq L_{max}$ the system cannot reach the boundary $\partial (\pi(supp(f))$
before the approximation breaks down because the $2$-form {\bf F} produces undesirable effects on the
dynamics of the system.\hfill $\Box$

Finally, we have the following {\it corollary}:
\begin{corolario}
Under the same hypothesis of {\it theorem 4.6.6}, the following
hypothesis hold:
\begin{enumerate}
\item In the limit $\alpha\longrightarrow 0$ the Lorentz force
equation and the averaged Lorentz force equation coincide.

\item In the limit $E\longrightarrow \infty$ the Lorentz force
equation and the averaged Lorentz force equation coincide.
\end{enumerate}
\end{corolario}

\newpage

\chapter{Charged cold fluid model from the Vlasov model}
\section{Introduction}

Despite limitations concerning the mathematical description of the discrete nature of the particles comprising a plasma,
modeling the dynamics of relativistic non-neutral plasmas and charged particle beams by
fluid models is common place. The relative simplicity of these models, compared with the
 corresponding kinetic models makes them appealing.

 We propose in this {\it chapter} another justification for the use of fluid models
 in beam dynamics. We will concentrate on the charged cold fluid model. However,
 we should notice that the same philosophy is also applicable to more sophisticated models.

In high intensity beam accelerator machines, each bunch of a
 beam contains a large number of identical particles contained in a small
  phase-space region. In such conditions,
 a number of the order $10^9-10^{11}$ charged particles move {\it together} under
   the action of both external and internal electromagnetic fields. Often in modern applications,
   such bunches of particles move ultra-relativistically.

One is interested in modeling these physical systems in such a way that:
\begin{enumerate}
\item  The model for a bunch of particles must be {\it simple}, in order to be useful in numerical
simulations of beam dynamics and for analytical treatment,

\item  It allows for stability analysis and a qualitative understanding of the dynamical behavior of the
 system. Three dimensional numerical simulations can be also desirable.

\end{enumerate}

 The standard approach has been to use fluid models as an approximation to a kinetic model.
 These derivations of fluid models from kinetic models can be found for instance in [8, 37-39] and references therein.
They are based on some assumptions, usually in the form of equations of
  state for fluids or assumptions on the higher moments of the distribution function $f$.
  These constraints are necessary in order to close the hierarchy of moments of
   the distribution function and to have a sufficient number of differential relations
    to determine the remaining moments. This is a general feature of all the derivations of
    fluid models from kinetic theory: a truncation scheme is required for the fluid model to be predictive.

We present in this {\it chapter} a new {\it justification} of the charged cold fluid model from the framework
 of kinetic theory. The novelty of the new approach is that it uses {\it natural hypotheses} suitable for particle
  accelerator machines and exploits only the mathematical structure of the
   classical electrodynamics of charged point particles interacting with external electromagnetic fields. We estimate the covariant derivative of
   the mean velocity calculated with the one-particle distribution function. This is given as an
    asymptotic formula in terms of the time of the evolution, diameter of the distribution and the energy of the beam.
The charged cold fluid model is described by only one {\it dynamical} variable, the
  normalized mean velocity field. The variance and the heat flow tensor are not necessarily zero, but
  are finite and given externally. In our treatment both the fluid energy tensor and the flux tensor are assumed to be given.
  Our aim is not to give an equation for the mean velocity field, but to evaluate
   how much certain differential expressions (formally equivalent to the charged cold fluid model equations)
   differ from zero. Then one can stipulate the validity of the model from the estimates of the
    corresponding differential expressions. On the other hand, in the models presented for instance in
     [8, 37-39], the variance and the covariant heat flow are
     dynamical variables and a system of partial differential equations is used to
     determine the dynamics of these fields. We think the analysis performed
      in this {\it chapter}, can be extended to other fluids models in future work.

 The method used in this thesis to obtain these results is the following:
  \begin{enumerate}

 \item One considers the results from {\it chapter 4}, which compares the
  solutions of the Lorentz connection with the solutions of the averaged
   Lorentz force equation. In this sense, we are under the same hypotheses
    as in {\it theorems 4.6.6} and {\it 4.6.8}. We will use the bounds of
     the differences of the corresponding geodesics.

 \item  It happens that under the same assumptions as used for the particle dynamics, the
 corresponding solutions of the Vlasov equation $f$ and the averaged Vlasov
  equation $\tilde{f}$ are similar. This result is based on the comparison results of the point particle dynamics.

 \item Each of the distributions $f$ and $\tilde{f}$ determine a different
  mean velocity field. One can prove under the same hypotheses that these
   mean velocity vector fields are similar. This means that the difference
    between them is controlled by powers of small parameters.

 \item Finally, we show that the auto-parallel condition of the velocity field
  of the averaged Vlasov equation associated
  with the averaged dynamics is controlled by the diameter of the distribution $f$.
    Together with the above point, this result allows us to provide estimates for
     the auto-parallel condition of the mean velocity field of the solution of the Vlasov equation.
 \end{enumerate}

 Therefore, the methods presented here and the usual derivations of the fluid models contained in [8, 37-39]
 are different. The standard approaches assume an asymptotic expansion of the
  differential equations for the moments, in terms of a perturbation parameter
   which is similar to the diameter $\alpha$ of the distribution function.
    These asymptotic expansions are particularized at low orders as a {\it truncation scheme}
     in the hierarchy of moments. Those references discuss systems of partial
      differential equations which are self-contained and consistent with physical
       constraints and with the asymptotic expansions.
  On the other hand, our approach is based on the structure of the Lorentz force equation of a charged point particle, which lies at the basis of the kinetic models.
   Written in a geometric way, the Lorentz force equation is replaced by the averaged Lorentz force
    equation. The key point is that the averaged Lorentz connection admits normal coordinates, which simplifies in a fundamental way our calculations.
    Then under some regularity assumptions on the distribution function, we can place bounds on the
     differential expression of interest.

We have assumed that the distribution functions are smooth (at least of class $\mathcal{C}^1$) in the coordinates $x^i$.
 Although we do not currently have a proof that we can extend our results to bigger
functional spaces for the distribution functions, since the main results are written in terms of Sobolev norms, it is conjectured that they can be extended to Sobolev spaces. Indeed our proofs suggest that we require smoothness on the $x$ coordinates and the existence of weak derivatives on the $y$ coordinates.

\section{Comparison of the solutions of the Vlasov and averaged
Vlasov equations}

In this {\it section} we estimate the difference
between the solutions of the Liouville equations associated with the averaged Lorentz connection and
the original Lorentz connection. The Liouville equation associated to the Lorentz force equation is called Vlasov's equation in kinetic theory. In a similar way, we call to the Liouville equation associated with the averaged Lorentz force equation the averaged Vlasov equation.

\subsection{Examples of Liouville equations}

Given a non-linear connection characterized by the {\it second order}
 vector field $\chi\in {\bf TTN}$, the associated Liouville equation is $\chi (f)=0$.
In the following two examples presented here, which are related with our purposes.
\begin{enumerate}
\item From the coefficients of the Lorentz connection $^L\Gamma^i\,_{jk}(x,y)$ one can recover the spray coefficients $^LG^i(x,y)$, using the homogeneous properties on $y$ of $^LG^i(x,y)$ and Euler's theorem of positive homogeneous functions. In particular, the spray coefficients are
\begin{displaymath}
^LG^i(x,y)=\,^L\Gamma^i\,_{jk}(x,y)\,y^j y^k=\, \,\Big(\,^{\eta}\Gamma^i\,_{jk} +
\frac{1}{2{\sqrt{\eta(y,y)}}}({\bf F}^i\,_{j}(x)y^m\eta_{mk}+
\end{displaymath}
\begin{displaymath}
+{\bf
F}^i\,_{k}(x)y^m\eta_{mj})+{\bf F}^i\,_m (x)
\frac{y^m}{2\sqrt{\eta(y,y)}}(\eta_{jk}-\frac{1}{\eta(y,y)}\eta_{js}
\eta_{kl}y^s y^l)\Big)y^j y^k=
\end{displaymath}
\begin{displaymath}
=\, \Big(\,^{\eta}\Gamma^i\,_{jk} +
\frac{1}{2{\sqrt{\eta(y,y)}}}\,({\bf F}^i\,_{j}(x)y^m\eta_{mk}+ {\bf
F}^i\,_{k}(x)y^m\eta_{mj})\Big) y^j y^k=
\end{displaymath}
\begin{displaymath}
=\,^{\eta}\Gamma^i\,_{jk}\,y^j y^k +
{{\sqrt{\eta(y,y)}}}\,{\bf F}^i\,_{j}(x)y^j.
\end{displaymath}

Then one can define the vector field $^L\chi$:
\begin{displaymath}
^L\chi(x,y)=\, y^i\frac{\partial}{\partial x^i}- \,\big( \,^{\eta}\Gamma^i\,_{jk}(x)\,y^j y^k +
{{\sqrt{\eta(y,y)}}}{\bf F}^i\,_{j}(x)y^j\big)\,\frac{\partial}{\partial y^i}.
\end{displaymath}

\item A similar procedure applies to the averaged Lorentz Vlasov vector field. In this case, however, we do not have the simplifications given above. Therefore, the spray coefficients are
  \begin{displaymath}
<\,^LG^i>(x,y)=\,^L\Gamma^i\,_{jk}(x,y)\,y^j y^k=\, \,\Big(\,^{\eta}\Gamma^i\,_{jk}(x) +
<\frac{1}{2{\sqrt{\eta(y,y)}}}({\bf F}^i\,_{j}(x)y^m\eta_{mk}+
\end{displaymath}
\begin{displaymath}
 + {\bf
F}^i\,_{k}(x)y^m\eta_{mj})+{\bf F}^i\,_m (x)
\frac{y^m}{2\sqrt{\eta(y,y)}}(\eta_{jk}-\frac{1}{\eta(y,y)}\eta_{js}
\eta_{kl}y^s y^l)>\Big)y^j y^k.
\end{displaymath}
If $y\in {\bf \Sigma}$, semi-spray coefficients can be simplified to
 \begin{displaymath}
 <\,^LG^i>(x,y)|_{\bf \Sigma}=\,\,\Big(\,^{\eta}\Gamma^i\,_{jk}(x) +
<\frac{1}{2}({\bf F}^i\,_{j}(x)y^m\eta_{mk}+ {\bf
F}^i\,_{k}(x)y^m\eta_{mj})+
\end{displaymath}
\begin{displaymath} +{\bf F}^i\,_m (x)
\frac{y^m}{2}(\eta_{jk}-\eta_{js}
\eta_{kl}y^s y^l)>\Big)y^j y^k=
 \end{displaymath}
\begin{displaymath}
 =\,\,\Big(\,^{\eta}\Gamma^i\,_{jk}(x) +
\frac{1}{2}({\bf F}^i\,_{j}(x)<y^m>\eta_{mk}+ {\bf
F}^i\,_{k}(x)<y^m>\eta_{mj})+
\end{displaymath}
\begin{displaymath} +{\bf F}^i\,_m (x)
(\eta_{jk}<\frac{y^m}{2}>-\eta_{js}
\eta_{kl}\,<\frac{y^m}{2}y^s y^l>)\Big)\,y^j y^k.
 \end{displaymath}
\end{enumerate}
The averaged Vlasov vector field can be written in a similar way as before,
\begin{displaymath}
<\,^L\chi>|_{{\bf \Sigma}}=\, y^i\frac{\partial}{\partial x^i}- \,\Big( \,^{\eta}\Gamma^i\,_{jk}\,y^j y^k +\,
\frac{1}{2}({\bf F}^i\,_{j}(x)<y^m>\eta_{mk}+ {\bf
F}^i\,_{k}(x)<y^m>\eta_{mj})+
\end{displaymath}
\begin{displaymath} +{\bf F}^i\,_m (x)
(\eta_{jk}<\frac{y^m}{2}>-\eta_{js}
\eta_{kl}\,<\frac{y^m}{2}y^s y^l>)\Big)\,y^j y^k\, \frac{\partial}{\partial y^i}.
\end{displaymath}
\subsection{Comparison between the solutions of the Vlasov equation and the averaged Vlasov equation}

In the following $({\bf M},\eta)$ is Minkowski space, since we will use {\it theorems (4.6.6)}
 and {\it theorem (4.6.8)}. The Riemaniann metric $\bar{\eta}_Z$ is determined by the vector
  $Z=\frac{\partial}{\partial t}$. Since we will use the Euclidean metric associated with $Z=\frac{\partial}{\partial t}$,
   we simplify the notation and employ $\bar{\eta}$ in place of $\bar{\eta}_Z$. $Z=\frac{\partial}{\partial t}$
    will be the observer that we call laboratory frame. We will restrict our attention to a compact domain ${\bf K}\subset{\bf M}$.
\begin{proposicion}
Let $f$ and $\tilde{f}$ be solutions of the Vlasov equation $^L\chi (f) =0$ and the {\it averaged Vlasov equation} $<\,^L\chi
> (\tilde{f}) =0$, where $^L\chi$ and $<\,^L\chi
>$ are the spray vector fields obtained from the non-linear connections $^L\nabla$ and $<\, ^L\nabla>$.
Let us assume the same hypotheses as those in {\it theorem 4.6.6}.
Then for the solutions of the Vlasov and {\it averaged Vlasov's equation} with the same initial conditions, one has the relation
\begin{displaymath}
|f(t,x(t),\dot{x}(t))-\tilde{f}(t,{x}(t),\dot{{x}}(t))|<\,\big(\tilde{C}(x)\|{\bf F}\|_{\bar{\eta}}
C^2_2(x)(1+B_2(x){\alpha})\big){\alpha}^2\,E^{-2}\,t^2\,+
\end{displaymath}
\begin{equation}
+\big(\tilde{K}(x)\|{\bf F}\|_{\bar{\eta}}(x)\,K^2_2(1+D_2(x){\alpha})\big){\alpha}^2\,E^{-1}\,t
\end{equation}
for some functions $\tilde{C}(x(t))$ $\tilde{K}(x(t))$ along the geodesic of the Lorentz connection.
\end{proposicion}
{\bf Proof}: $f$ and $\tilde{f}$ are solutions of the
corresponding Vlasov and averaged Vlasov equations respectively. Therefore,  $f$ and
$\tilde{f}$ are constant along the corresponding auto-parallel curves;
$x(t)$ and $\tilde{x}(t)$ are the projections on the space-time manifold {\bf M} of the integral
curves of the vector fields $^L\chi$ and $<\,^L\chi>$. Here $t$ is the time-parameter in the laboratory frame
determined by the vector field $\frac{d}{dt}$. Then the Vlasov and averaged Vlasov equation can be written as
\begin{displaymath}
^L\chi_{} f^{}=\frac{d}{dt}f(x(t),\dot{x}(t))=0, \quad
<\,^L\chi_{}>
\tilde{f}^{}=\frac{d}{dt}\tilde{f}(\tilde{x}(t),\dot{\tilde{x}}(t))=0.
\end{displaymath}
For the same initial conditions, the geodesic curves corresponding
to the connections $^L\nabla$ and $<\,^L\nabla>$ are nearby curves
at the instant $t$ in the way described by {\it theorem 4.6.6}.

Let us introduce the family of {\it interpolating connections},
\begin{displaymath}
^L\nabla_{\epsilon}:=(1-\epsilon)\,^L\nabla +\,\epsilon <\,^L\nabla>,\,\quad \epsilon \in [0,1].
\end{displaymath}
Each of them has an associated spray vector field
$^L\chi_{\epsilon}$. Therefore, let us consider $f_{\epsilon}(x,y)$ to be
the solution of the following Liouville equation $^L\chi_{\epsilon} f_{\epsilon}=0$ for some given initial conditions.
Since the dependence on $(\epsilon,x,y)$ of the vector field
$^L\chi_{\epsilon} $ is $\mathcal{C}^1$, the solutions of
the Liouville equation are Lipschitz with respect to the parameter $\epsilon$. We can see this fact in
the following way. The Liouville equation can be written as
\begin{displaymath}
^L\chi_{\epsilon} f_{\epsilon}=0\,\Leftrightarrow \frac{d}{dt}
f(x_{\epsilon}(t), y_{\epsilon}(t))=0,
\end{displaymath}
where $(x_{\epsilon}(t), y_{\epsilon}(t))$ is an integral curve of the vector
field $^L\chi_{\epsilon}$ restricted to the unit hyperboloid
bundle and such that it is  parameterized by the coordinate time $t$.
Then one can use standard results from the theory of ordinary differential
equations to study the smoothness properties of the solutions of the above
equation. In particular, the connection coefficients for the
interpolating connection are,
\begin{displaymath}
(\,^L\Gamma_{\epsilon})^i\,_{jk}=(1-\epsilon)\,^L\Gamma^i\,_{jk}\,+\epsilon
<\,^L\Gamma^i\,_{jk}>.
\end{displaymath}
From the formula $(4.5.1)$ for the coefficients $^L\Gamma^i\,_{jk}(x,y)$
one can check that $(\,^L\Gamma_{\epsilon})^i\,_{jk}$ are smooth functions in an open set of
time-like vectors $y$ and the parameter $\epsilon$. From here it follows the Lipschitz condition for $f_{\epsilon}$ in $\epsilon$.

We will give an upper bound for the difference
$|f(t,x(t),\dot{x}(t))-\tilde{f}(t,{x}(t),\dot{{x}}(t))|$. Note that
 in this expression the point where both $f$ and $\tilde{f}$ are evaluated are $(t,x(t),\tilde{x}(t))$.
In order to achieve this,  standard results on
the smoothness of the solution of differential equations are used (see {\it chapter 1} of  [40]).
In particular we use that for each $ (\bar{\epsilon}, \bar{x}(s), \bar{y}(s))$, there is
an open neighborhood ${\bf U}_{\bar{\epsilon}}$ of $[0,1]\times { supp(f)}_{\bf TK}$ containing $ (\bar{\epsilon}, \bar{x}(t), \bar{y}(t))$
such that the solutions of the differential equations are Lipschitz  in ${\bf U}_{\bar{\epsilon}}$. Therefore, using the Lipschitz condition, one obtains the bound
\begin{displaymath}
|f^{\epsilon}(t,x_{\epsilon}(t),\dot{x}_{\epsilon}(t))-{f}^{\tilde{\epsilon}}(t,{x}_{\tilde{\epsilon}}(t),
\dot{{x}}_{\tilde{\epsilon}}(t))|\leq \,c_1{(\bar{\epsilon}, \bar{x}(t),\dot{\bar{x}}(t)) }\delta((\bar{\epsilon}, \bar{x}(t),\dot{\bar{x}}(t)))\,
+
\end{displaymath}
\begin{displaymath}
+c_2{(\bar{\epsilon}, \bar{x}(t),\dot{\bar{x}}(t)) }\, \|x_{\epsilon}(t)-{x}_{\tilde{\epsilon}}(t)\|_{\bar{\eta}}
+c_3{(\bar{\epsilon}, \bar{x}(t),\dot{\bar{x}}(t)) }\, \|\dot{x}_{\epsilon}(t)-\dot{x}_{\tilde{\epsilon}}(t)\|_{\bar{\eta}}.
\end{displaymath}
$c_i{(\bar{\epsilon}, \bar{x}(t),\dot{\bar{x}}(t)) }$ are constants which depend on the open neighborhood ${\bf U}_{\bar{\epsilon}}$;
$\delta((\bar{\epsilon}, \bar{x}(t),\dot{\bar{x}}(t)))$ is the diameter on the $\epsilon$ component
 where we are applying the Lipschitz condition.

One can always choose a refinement of an open cover of $[0,1]\times supp(f)|_{{\bf TK}}$ such that both the
 Lipschitz condition, {\it theorem} $(4.6.6)$ and {\it theorem} $(4.6.8)$ can be applied simultaneously.
Since $[0,1]$ is compact, we can consider a finite open covering of $[0,1]$ for each instant $t$.
Then using the above local bound in each of the open sets ${\bf U}_{\epsilon}$, one obtains the  global bound
\begin{displaymath}
|f(t,x(t),\dot{x}(t))-\tilde{f}(t,{x}(t),\dot{{x}}(t))|<\,c_1
+c_2\, \|x(t)-\tilde{x}(t)\|_{\bar{\eta}}+c_3\, \|\dot{x}(t)-\dot{\tilde{x}}(t)\|_{\bar{\eta}}.
\end{displaymath}
The constants $c_i$ are
finite (by definition of Liptschitz and by compactness of the
interval $[0,1]$). The functions $f$ and $\tilde{f}$ are constant
along the respective geodesics. Therefore,
\begin{displaymath}
|f(t,x(t),\dot{x}(t))-\tilde{f}(t,\tilde{x}(t),\dot{\tilde{x}}(t))|=\,
 |f(0,x(0),\dot{x}(0))-\tilde{f}(0,\tilde{x}(0),\dot{\tilde{x}}(0))|.
\end{displaymath}
Let us assume the same initial conditions $x(0)=\tilde{x}(0)$ and $\dot{x}(0)=\dot{\tilde{x}}(0)$ for the geodesics of the Lorentz connection.
Since the difference $|f(t,x(t),\dot{x}(t))-\tilde{f}(t,{x}(t),\dot{{x}}(t))|$ is a smooth function of
 $\|x(t)-\tilde{x}(t)\|_{\bar{\eta}}$ and $\|\dot{x}(t)-\dot{\tilde{x}}(t)\|_{\bar{\eta}}$, one obtains
\begin{displaymath}
0\leq \,c_1\leq \,\bar{K}_1\|x(t)-\tilde{x}(t)\|_{\bar{\eta}}+\,  \bar{K}_1\|\dot{x}(t)-\dot{\tilde{x}}(t)\|_{\bar{\eta}}
\end{displaymath}
for some constants $K_i$.
Then we have,
\begin{displaymath}
|f(t,x(t),\dot{x}(t))-\tilde{f}(t,{x}(t),\dot{{x}}(t))|\leq
\end{displaymath}
\begin{displaymath}
\leq \,|f(t,x(t),\dot{x}(t))-\tilde{f}(t,\tilde{x}(t),\dot{\tilde{x}}(t))|+\,
|\tilde{f}(t,x(t),\dot{x}(t))-\tilde{f}(t,\tilde{x}(t),\dot{\tilde{x}}(t))|.
\end{displaymath}
The first term is bounded by $c_1$, which is bounded by $\bar{K}_1\|x(t)-\tilde{x}(t)\|_{\bar{\eta}}+\,
 \bar{K}_1\|\dot{x}(t)-\dot{\tilde{x}}(t)\|_{\bar{\eta}}$. The second term can be developed in Taylor series in the differences $\|x(t)-\tilde{x}(t)\|_{\bar{\eta}}$
 and $\|\dot{x}(t)-\dot{\tilde{x}}(t)\|_{\bar{\eta}}$, since $\tilde{f}$ is smooth. Therefore,
\begin{displaymath}
|f(t,x(t),\dot{x}(t))-\tilde{f}(t,{x}(t),\dot{{x}}(t))|\leq\,\big(\tilde{C}(x(t))\|{\bf F}\|_{\bar{\eta}}(x(t))\,
{C}^2_2(1+B_2(x(t)){\alpha})\big){\alpha}^2\,E^{-2}\, t^2\,+
\end{displaymath}
\begin{displaymath}
+\big(\tilde{K}(x(t))\|{\bf F}\|_{\bar{\eta}}(x(t))\,{K}^2_2(1+{D}_2(x(t)){\alpha})\big){\alpha}^2\,E^{-1}\,t.
\end{displaymath}\hfill$\Box$

\section{The charged cold fluid model from the averaged Vlasov model}
In the following results $({\bf M},\eta)$ is Minkowski space, since we will use {\it theorem (4.6.6)} and {\it theorem (4.6.8)}. The Riemannian metric $\bar{\eta}_Z$ is determined by the vector $Z=\frac{d}{dt}$.

There is another local observer related with the vector field $<y>$. Since
the norm is not continuous on {\bf M}, one needs a local smoothing procedure.
 Given a subset of the paracompact manifold, ${\bf \Sigma}_x$,
we can take the induced bump  function from the bump functions defined on
 ${\bf \Sigma}_x$. Using these bump functions, we can smooth vector fields [51, pg 25].

\subsection{Comparison of the Vlasov model with the averaged Vlasov model}

\begin{definicion} Given a semi-Randers space $({\bf M}, \eta, [A])$,
the averaged Vlasov model is defined by the dynamical
variables $\tilde{f}$ determined by
\begin{equation}
 <\, ^L \chi> \tilde{f}=0,
\end{equation}
where $<\,^L\chi>$ is the Liouville vector field of the averaged
Lorentz dynamics associated with the external electromagnetic
field ${\bf F}$.
The dynamical variable $f(x,y)$ defines the following
\begin{equation}
\tilde{V}:=\int_{{\bf \Sigma}_x}\, y\tilde{f}(x,y)\,dvol(x,y)  ,\quad vol({\bf \Sigma}_x):=\int_{{
\bf \Sigma}_x} \,dvol(x,y)\tilde{f}(x,y).
\end{equation}
\end{definicion}

Since we will use the results of {\it chapter 4}, the distribution
 function $\tilde{f}$ is at least of type $\mathcal{C}^1$ in the $x$-coordinates and Lipschitz on the $y$-coordinates. Since the support $f\in f_x$ is compact, several
 Sobolev norms are defined [41, {\it chapter 3}]. We will write our results in terms of those norms.
\begin{proposicion}
Let $<\,^L\chi>\tilde{f}(x,y)=0$ and $^L\chi f(x,y)=0$ be such that the domain of definition of the vector field $<\, ^L\chi>$ is an open sub-manifold of ${\bf \Sigma}$.
Then one can reduce $supp(\tilde{f}_x)\longrightarrow supp(f_x)$ for all $x\in {\bf M}$.
\end{proposicion}
{\bf Proof}:
Let us consider the product of the functions $\tilde{f}(x,y)g(x,y)$, where $<\,^L\chi>\tilde{f}=0$,
 the function $g(x,y)$ is a {\it bump} function adapted to the support in {\bf K} of the vector field $<\,^L\chi>$.
 Since both the support of $<\,^L\chi>$ and $supp(f_x)$ are sub-sets of the paracompact manifold ${\bf \Sigma}$, this function exists [32]. Therefore, we select the function $g(x,y)$ such that
\begin{displaymath}
g_x(y)=0, \quad (x,y)\in supp(\tilde{f}_x)\setminus U(f_x),\quad g_x(y)=0,\quad (x,y)\in \partial{supp({f}_x)}
\end{displaymath}
where $U(f_x)\supset supp(f_x)$ and all the derivatives are zero on $\partial supp({f}_x)$. Then one can perform the following calculation:
\begin{displaymath}
<\,^L\chi>(\tilde{f}g)=\,g<\,^L\chi>\tilde{f}\,+\tilde{f}<\,^L\nabla>g=0.
\end{displaymath}
We can always restrict the solutions of $<\,^L\chi>\tilde{f}=0$ in such a way that formally $supp(\tilde{f}_x)=supp(f_x)$ and points $2$ and $3$ are proved.\hfill$\Box$

 Using equation $(5.2.1)$, it follows that
 the error induced by the substitution $\tilde{f}\longrightarrow f$ is of order $\alpha^2$.
 Hence, in the following calculations, when it is useful, we can use $dvol(x,y)$ as a measure
  and substitute $supp(\tilde{f}_x)$ by $supp(f_x)$ and $\tilde{f}$ by $f$.

Let ${\bf F}$ be a closed differential $2$-form defining the Liouville vector field $^L\chi$.
 Let us consider the Sobolev spaces $(\mathcal{W}^{1,1}({\bf \Sigma}_x),\|\cdot\|_{1,1})$
  and $(\mathcal{W}^{0,2}({\bf \Sigma}_x),\,\|\cdot\|_{0,2})$ [41].
   Recall that the space of smooth functions is denoted by $\mathcal{F}({\bf \Sigma}_x)$
   (an introduction to the notions of Sobolev spaces can be found in {\it appendix 5} or [40, {\it chapter 3}]).

Recall that we have denoted $\delta(x,y)=<y>(x)-y$. For the next results we will restrict
to the Minkowski space-time $({\bf M},\eta)$. Let us denote $\Upsilon(supp(\tilde{f}_x))$
 the characteristic function of $supp(\tilde{f}_x)$.
\begin{teorema}
Let ${\bf M}$ be an $n$-dimensional space-time manifold, ${\bf K}\subset {\bf M}$ a compact domain and $<\,^L\chi>$ the vector
 field associated with the averaged Lorentz force equation. Assume that:
\begin{enumerate}
\item The distribution function is such that $\tilde{f}_x,\, \partial_j \tilde{f}(x,\cdot)\in
 \mathcal{F}({\bf \Sigma}_x)\, \subset\mathcal{W}^{1,1}({\bf \Sigma}_x)$,

\item The function  $\delta(x,\cdot), (\partial_j \delta)(x,\cdot)\in \mathcal{W}^{0,2}({\bf \Sigma}_x)$.

\end{enumerate}
Then
\begin{equation}
\|<\, ^L\nabla>_{\tilde{V}} \tilde{V}(x)\|_{\bar{\eta}}\leq \,\frac{vol^{\frac{1}{2}}_E(supp(\tilde{f}_x))}{vol(supp(\tilde{f}_x))}
\,(\sum_{k}\,\|\partial_0\, log(\delta^k_x)\|_{0,2})\,\,
\cdot \|\tilde{f}_x\|_{1,1}\cdot {\alpha}^2\,+ O(\alpha^3),
\end{equation}
where $\delta_x(\cdot):=\delta(x,\cdot)$ and
\begin{displaymath}
\tilde{V}^i(x)=<\hat{y}^i>_{\tilde{f}}(x)\,:=\frac{1}{\int_{{\bf \Sigma}_x} f(x,y)
 dvol(x,{y})}\,\int_{{\bf \Sigma}_x}\,dvol(x,y)\,f(x,y) y^i.
\end{displaymath}
The volumes are
\begin{displaymath}
vol(supp(\tilde{f}_x)):=vol({\bf \Sigma}_x);\quad vol_E({\bf \Sigma}_x):=\int_{{\bf \Sigma}_x}\, \Upsilon(supp(\tilde{f}_x))\cdot dvol(x,\tilde{y});
\end{displaymath}
the derivative in equation $(5.3.3)$ refers to the local frame such that the vector $U:=\frac{<y>}{\sqrt{\eta(<y>,<y>)}}=(U_0,\vec{0})$.
\end{teorema}
{\bf Proof:} Because the averaged Lorentz connection is an affine
connection on {\bf M}, given a point $x\in {\bf M}$, there is a coordinate system where the
connection coefficients are zero at that point,  $<\, ^L\Gamma>^i\, _{jk}(x)=0$.
Therefore, for any given point $x\in {\bf M}$ one can choose a {\it normal coordinate system}
such that the {\it averaged Vlasov condition} holds,
\begin{equation}
y^j \partial_j \tilde{f}(x,y)|_{x}=0.
\end{equation}
Using this normal coordinate system, one can get a simplified expression
for the covariant derivative of $\tilde{V}$ along the integral curve of $\tilde{V}$:
\begin{equation}
<\, ^L\nabla>_{\tilde{V}} \tilde{V}=(\tilde{V}^j\partial_j \tilde{V}^k )\frac{\partial}{\partial
x^k},
\end{equation}
using a coordinate frame $\{\frac{\partial}{\partial
x^k},\,\,k=0,...,n-1\}$. Note that this expression is not a partial differential equation
because it only holds at the point $x$.

From the relation $(5.3.5)$  we obtain that
\begin{displaymath}
<\, ^L\nabla>_{\tilde{V}} \tilde{V}(x)=\tilde{V}^i\partial_i \tilde{V}^k=\,\frac{1}{vol({\bf \Sigma}_x)}\int_{{\bf \Sigma}_x}\,dvol(x,y)\,
y^j \tilde{f}(x,{y})\cdot
\end{displaymath}
\begin{displaymath}
\cdot \partial_j \big(\frac{1}{vol({\bf \Sigma}_x)}\,\int_{{\bf \Sigma}_x}dvol(x,\hat{y}) \hat{y}^k
\tilde{f}(x,\hat{y})\big).
\end{displaymath}
It is the right hand of this equation that we shall estimate,
\begin{displaymath}
\frac{1}{vol(supp(\tilde{f}_x))}\,\Big(\int_{{\bf \Sigma}_x}\,dvol(x,y)\, y^j \tilde{f}(x,{y})\cdot
\end{displaymath}
\begin{displaymath}
\cdot \partial_j \big(\frac{1}{vol({\bf \Sigma}_x)}\,\int_{{\bf \Sigma}_x}\,dvol(x,\hat{y}) \hat{y}^k
\tilde{f}(x,\hat{y})\big)\,\Big)=
\end{displaymath}
\begin{displaymath}
=\frac{1}{vol({\bf \Sigma}_x)}\Big( \,\int_{{\bf \Sigma}_x}\,dvol(x,y)\, y^j
\tilde{f}(x,y)\,\big(-\,\frac{\int_{{\bf \Sigma}_x}\, dvol(x,\hat{y})\,
\tilde{f}(x,\hat{y})\,\hat{y}^k}{vol^2( {\bf \Sigma}_x)}\cdot
\end{displaymath}
\begin{displaymath}
\cdot\partial_j\,(\int_{{\bf \Sigma}_x}\, dvol(x,\tilde{y})\, f(x,\tilde{y}))\big)\Big)
+\frac{1}{vol({\bf \Sigma}_x)}\,\Big(\, \int_{{\bf \Sigma}_x} \,dvol(x,y)\,y^j\,
\tilde{f}(x,y)\,\cdot
\end{displaymath}
\begin{displaymath}
\cdot\partial_j\,\big(\, \int_{{\bf \Sigma}_x}\,dvol(x,\hat{y})\,
\hat{y}^k\,\tilde{f}(x,\hat{y})\,\big)\Big)=
\end{displaymath}
\begin{displaymath}
=-\frac{1}{vol^2({\bf \Sigma}_x)}\Big(\,\int_{{\bf \Sigma}_x}\,dvol(x,y)\, y^j \tilde{f}(x,y)\,
\partial_j\,\big(\,\int_{{\bf \Sigma}_x}\, dvol(x,\hat{y})\,\tilde{f}(x,\hat{y})\,\,\big)\,<y^k>\,\Big)\,+
\end{displaymath}
\begin{displaymath}
+\frac{1}{vol^2({\bf \Sigma}_x)}\,\Big(\,\int_{{\bf \Sigma}_x}\, dvol(x,y)\,y^j\,
\tilde{f}(x,y)\,\cdot
\end{displaymath}
\begin{displaymath}
\cdot\partial_j\,(\int_{{\bf \Sigma}_x}\, dvol(x,\hat{y})\,\hat{y}^k\,
\tilde{f}(x,\hat{y})\,\big)\,\Big).
\end{displaymath}
Shifting the variable of integration $-\hat{y}^k+<\hat{y}^k>=-\delta^k (x,\hat{y})$, one obtains the following for the above expression
\begin{displaymath}
<\, ^L\nabla>_{\tilde{V}} \tilde{V}(x)=-\frac{1}{vol^2({\bf \Sigma}_x)}\Big(\,\int_{{\bf \Sigma}_x}\,dvol(x,y)\, y^j
\tilde{f}(x,y)\cdot
\end{displaymath}
\begin{displaymath}
\cdot\partial_j\,\big(\,\int_{{\bf \Sigma}_x}\, dvol\hat{y}\,
\tilde{f}(x,\hat{y})\,\,\big)\,<y^k>\,\Big)\,+
\end{displaymath}
\begin{displaymath}
+\frac{1}{vol^2({\bf \Sigma}_x)}\,\Big(\,\int_{{\bf \Sigma}_x}\, dvol(x,y)\,y^j\,
\tilde{f}(x,y)\,\cdot
\end{displaymath}
\begin{displaymath}
\cdot \partial_j\,(\int_{{\bf \Sigma}_x}\, dvol(x,\hat{y})\,(<y^k>+\delta^k(x,\hat{y}))\,
\tilde{f}(x,\hat{y})\,\big)\,\Big)\partial_k=
\end{displaymath}
\begin{displaymath}
=\frac{1}{vol^2({\bf \Sigma}_x)}\,\Big(\,\int_{{\bf \Sigma}_x}\, dvol(x,y)\,y^j\,
\tilde{f}(x,y)\,\partial_j\,(\int_{{\bf \Sigma}_x}\, dvol(x,\hat{y})\,\delta^k(x,\hat{y})\,
\tilde{f}(x,\hat{y})\,\big)\,\Big)\partial_k.
\end{displaymath}
Since $y^i\partial_i f(x,y)=0$ at $x$ and since
$\tilde{f}_x$ is a smooth function of $y$, we can Taylor expand the integrand, obtaining:
\begin{displaymath}
\frac{1}{vol^2({\bf \Sigma}_x)}\,\Big(\,\int_{{\bf \Sigma}_x}\, dvol(x,y)\,y^j\,
\tilde{f}(x,y)\,\cdot
\end{displaymath}
\begin{displaymath}
\cdot\partial_j\,(\int_{{\bf \Sigma}_x}\, dvol(x,\hat{y})\,\delta^k(x,\hat{y})\,
\big(\tilde{f}(x,{y})+\frac{\partial f}{\partial \hat{y}^l}\, (\hat{y}^l-y^l)\,\big)\,\Big).
\end{displaymath}
There is a coordinate system such that $ <\,^L\Gamma>^i\,_{jk}(x) =0$ at the point $x$.
 This is reflected in the averaged Vlasov equation, which has the form $y^j\,\partial _j
\tilde{f}(x,y)=0$ at one given point $x\in {\bf M}$. Then we get the following for the above expression
\begin{displaymath}
\Big(<\, ^LD>_{\tilde{V}} \tilde{V}(x)\Big)^k=\frac{1}{vol^2({\bf \Sigma}_x)}\,\Big(\,\int_{{\bf \Sigma}_x}\, dvol(x,y)\,y^j\,
\tilde{f}(x,y)\,\cdot
\end{displaymath}
\begin{displaymath}
\cdot \partial_j\,(\int_{{\bf \Sigma}_x}\, dvol(x,\hat{y})\,\delta^k(x,\hat{y})\,
\frac{\partial f}{\partial \hat{y}^l}\, (\hat{y}^l-y^l)\,\big)\,\Big).
\end{displaymath}
$(\hat{y}^l-y^l)$ and ${\delta}^k(x,y)$ are bounded by the
diameter ${\alpha}(x)$ (remember that in taking the moments we can substitute the pair
$(\tilde{f}_x,\,  supp(\tilde{f}_x))$ by $(f_x,\, supp(f_x))$ if we desire, since by {\it proposition (5.2.1)} the difference between the two distributions functions is small and because by {\it proposition (5.3.2)} we can replace the supports as well). Therefore,
\begin{displaymath}
\Big\|<\, ^L\nabla>_{\tilde{V}}
\tilde{V}\Big\|_{\bar{\eta}}\,
=\frac{1}{vol^2({\bf \Sigma}_x)}\,\Big\|\Big(\,\int_{{\bf \Sigma}_x}\,
dvol(x,y)\,y^j\,\tilde{f}(x,y)\,
\cdot
\end{displaymath}
\begin{displaymath}
\cdot \partial_j\,(\int_{{\bf \Sigma}_x}\,
dvol(x,\hat{y})\,\delta^k(x,\hat{y})\,\frac{\partial f}{\partial
\hat{y}^l}\, (\hat{y}^l-y^l)\partial_k\,\big)\,\Big)\Big\|_{\bar{\eta}}\,\leq
\end{displaymath}
\begin{displaymath}
\leq\frac{1}{vol^2({\bf \Sigma}_x)}\,\Big|\Big(\,\int_{{\bf \Sigma}_x}\,
dvol(x,y)\,y^j\,\tilde{f}(x,y)\,\partial_j\,\frac{\partial f}{\partial
\tilde{y}^l}\,\Big)\Big|\,\cdot
\end{displaymath}
\begin{displaymath}
\cdot\Big\|\Big(\int_{{\bf \Sigma}_x}\,
dvol(x,\hat{y})\,\delta^k(x,\hat{y})\,\,
(\hat{y}^l-y^l)\partial_k\,\Big)\,\Big\|_{\bar{\eta}}+
\end{displaymath}
\begin{displaymath}
+\frac{1}{vol^2({\bf \Sigma}_x)}\,\Big|\Big(\,\int_{{\bf \Sigma}_x}\,
dvol(x,y)\,y^j\,\tilde{f}(x,y)\,\frac{\partial f}{\partial
\tilde{y}^l}\,\Big)\Big|\,\cdot
\end{displaymath}
\begin{displaymath}
\cdot \Big\|\Big(\int_{{\bf \Sigma}_x}\,
dvol(x,\hat{y})\,\,\partial_j\delta^k(x,\hat{y})\,\,
(\hat{y}^l-y^l)\partial_k\,\Big)\,\Big\|_{\bar{\eta}}\,\leq
\end{displaymath}
\begin{displaymath}
\leq \frac{1}{vol^2({\bf \Sigma}_x)}\,\Big|\Big(\,\int_{{\bf \Sigma}_x}\, dvol(x,y)\,y^j\,\tilde{f}(x,y)\,\partial_j\,\frac{\partial
f}{\partial \tilde{y}^l}\,\Big)\Big|\,\cdot
\end{displaymath}
\begin{displaymath}
\cdot\Big(\int_{{\bf \Sigma}_x}\,
dvol(x,\hat{y})\,\big\|\delta^k(x,\hat{y})\,\,
(\hat{y}^l-y^l)\partial_k\Big\|_{\bar{\eta}}\,\Big)+
\end{displaymath}
\begin{displaymath}
+\frac{1}{vol^2({\bf \Sigma}_x)}\,\Big|\Big(\,\int_{{\bf \Sigma}_x}\, dvol(x,y)\,y^j\,\tilde{f}(x,y)\,\frac{\partial
f}{\partial \tilde{y}^l}\,\Big)\Big|\,\cdot
\end{displaymath}
\begin{displaymath}
\cdot\Big(\int_{{\bf \Sigma}_x}\,
dvol(x,\hat{y})\,\partial_j\big\|\delta^k(x,\hat{y})\,\,
(\hat{y}^l-y^l)\partial_k\big\|_{\bar{\eta}}\,\Big).
\end{displaymath}
One can find a bound for each of these integrals.
For instance, using the Hoelder inequality for integrals in an arbitrary space {\bf X} [13]
\begin{displaymath}
\Big|\int_{{\bf X}} \lambda \phi\, \,d\mu\,\Big|\leq\,\Big(\int_{{\bf X}}
\big|\lambda\, d\mu\big|^{p}\,\Big)^{1/p} \Big(\int_{{\bf X}}  \big|\phi\,d\mu \big|^{q}\,\Big)^{1/q},\,\,\,
\frac{1}{p}+\frac{1}{q}=1,\,\,1\leq p,q\leq\infty.
\end{displaymath}
We will use this inequality several times for the case $p=q=2$, obtaining
\begin{displaymath}
\Big\|\Big(\int_{{\bf \Sigma}_x}\, dvol(x,\tilde{y})\,\,
(\tilde{y}^l-y^l)\,\delta^k(x,y)\partial_k\,\Big\|_{\bar{\eta}}\,\Big)\leq
\end{displaymath}
\begin{displaymath}
\leq\,\Big(\int_{{\bf \Sigma}_x}\, dvol(x,\hat{y})\,
\big|(\hat{y}^l-y^l)\big|\,\big\|\,
\,\delta^k(x,\hat{y})\partial_k\,\big\|_{\bar{\eta}}\,\Big)\leq
\end{displaymath}
\begin{displaymath}
\leq \,\Big(\int_{{\bf \Sigma}_x}\, dvol(x,\hat{y})\big|\,
(\hat{y}^l-y^l)\big|^2\,\Big)^{\frac{1}{2}}
\cdot\, \Big(\int_{{\bf \Sigma}_x}\, dvol(x,\hat{y})\,
\big\|\,\delta^k(x,\hat{y})\partial_k\,\big\|^2_{\bar{\eta}}\,\Big)^{\frac{1}{2}}.
\end{displaymath}
Note that the index $l$ is contracted with a factor $\frac{\partial f}{\partial \tilde{y}^l}$. Therefore $y^l\,\frac{\partial f}{\partial \tilde{y}^l}$ is Lorentz invariant and it can be computed in any inertial system, in particular in the laboratory frame. If we do this computation on this frame, we can continue with the above bound in the following way:
\begin{displaymath}
\Big(\int_{{\bf \Sigma}_x}\, dvol(x,\hat{y})\big|\,
(\hat{y}^l-y^l)\big|^2\,\Big)^{\frac{1}{2}}
\cdot\, \Big(\int_{{\bf \Sigma}_x}\, dvol(x,\hat{y})\,
\big\|\,\delta^k(x,\hat{y})\partial_k\,\big\|^2_{\bar{\eta}}\,\Big)^{\frac{1}{2}}.
\end{displaymath}
\begin{displaymath}
 \leq \, vol^{\frac{1}{2}}_E({\bf \Sigma}_x)\,\alpha \cdot
  \Big(\int_{{\bf \Sigma}_x}\, dvol(x,\hat{y})\,\big\|\,\delta^k(x,\hat{y})\partial_k\,\big\|^2_{\bar{\eta}}\Big)^{\frac{1}{2}}.
\end{displaymath}
In order to bound the second factor we use the following argument (that we used already in {\it section 4.6}),
\begin{displaymath}
\|\delta(x,y)\|_{\bar{\eta}}\,\leq\|<\hat{y}>(x)\,-y\|_{\bar{\eta}}\,\leq \|\epsilon+\hat{y}-y\|_{\bar{\eta}} \,
\leq \|\epsilon\|_{\bar{\eta}}+\,\|\hat{y}-y\|_{\bar{\eta}}\,\leq \frac{1}{2}\,\alpha\,+\alpha=\frac{3}{2} \alpha.
\end{displaymath}
 $\hat{y}$
 is in the support of the distribution $f$. Therefore, a bound on the integral is
\begin{displaymath}
\Big\|\Big(\int_{{\bf \Sigma}_x}\, dvol(x,\tilde{y})\,
(\tilde{y}^l-y^l)\,\delta^k(x,\tilde{y})\partial_k\,\Big)\Big\|_{\bar{\eta}}\,\,\leq \frac{3}{2}\cdot vol_E({\bf \Sigma}_x)\cdot \alpha^2.
\end{displaymath}

Similarly one obtains the following bound:
\begin{displaymath}
\Big\|\Big(\int_{{\bf \Sigma}_x}\, dvol(x,\hat{y})\,\partial_j (\delta^k(x,y))\,
(\hat{y}^l-y^l)\partial_k\,\Big)\Big\|_{\bar{\eta}}\,\leq \, \Big(\int_{{\bf \Sigma}_x}\, dvol(x,\hat{y})\,\big|
(\hat{y}^l-y^l)\big|^2\,\Big)^{\frac{1}{2}}\cdot \,
\end{displaymath}
\begin{displaymath}
\cdot\,\Big(\int_{{\bf \Sigma}_x}\, dvol(x,{y})\,
\big\|\,\partial_j (\delta^k(x,{y}))\partial_k\big\|^2_{\bar{\eta}}\,\Big)^{\frac{1}{2}}\,\leq
\end{displaymath}
\begin{displaymath}
\leq \, vol_E^{\frac{1}{2}}({\bf \Sigma}_x)\,\alpha
\cdot \, \Big(\int_{{\bf \Sigma}_x}\, dvol(x,{y})\,\big\|\,
\partial_j (\delta^k(x,y))\partial_k\big\|^2 _{\bar{\eta}}\,\Big)^{\frac{1}{2}}.
\end{displaymath}
Then because of the definition of the corresponding Sobolev norm $\|\cdot \|_{0,2}$:
\begin{displaymath}
 \Big(\int_{{\bf \Sigma}_x}\, dvol(x,\hat{y})\,\big\|\partial_j (\delta^k(x,\hat{y}))\,
(\hat{y}^l-y^l)\partial_k\big\|_{\bar{\eta}}\,\Big)\leq \,vol^{\frac{1}{2}}_E(supp(\tilde{f}_x))
\,{\alpha}\cdot\|\partial_j \delta^k_x\|_{0,2}\,=
\end{displaymath}
\begin{displaymath}
=\,vol^{\frac{1}{2}}_E({\bf \Sigma}_x)
\,{\alpha}\cdot\sum_k\,\|\delta^k_x\partial_j log(\delta^k_x)\|_{0,2}
\leq \,vol^{\frac{1}{2}}_E({\bf \Sigma}_x)
\,{\alpha}^2\cdot\sum_k\,\|\partial_j log(\delta^k_x)\|_{0,2}.
\end{displaymath}
Similarly,
\begin{displaymath}
\Big|\Big(\,\int_{{\bf \Sigma}_x}\, dvol(x,y)\,y^j\,\tilde{f}(x,y)\,\partial_j\,\frac{\partial
f}{\partial \tilde{y}^l}\,\Big)\Big|\,\leq \,\sum^{n-1}_{j=0}\,\Big(\,\int_{{\bf \Sigma}_x)}\, dvol(x,y)\,
\big|y^j\,\tilde{f}(x,y)\big|^2\,\Big)^{\frac{1}{2}}\cdot
\end{displaymath}
\begin{displaymath}
\cdot\Big(\,\int_{{\bf \Sigma}_x}\, dvol(x,y)\,\big|\partial_j\,\frac{\partial \tilde{f}(x,y)}{\partial y^k}\big|^2\,\Big)^{\frac{1}{2}}.
\end{displaymath}
The second factor is equal to the Sobolev norm $\|\partial_j \tilde{f}_x\|_{1,1}$. The first factor is bounded in the following way:
\begin{displaymath}
\Big(\,\int_{{\bf \Sigma}_x}\, dvol(x,y)\,\big|y^j\,\tilde{f}(x,y)\big|^2\,\Big)^{\frac{1}{2}}\, \leq
 \Big(\,\int_{{\bf \Sigma}_x}\, dvol(x,y)\,|y^j|^2\,\tilde{f}(x,y)\,\Big)^{\frac{1}{2}}=
 \end{displaymath}
 \begin{displaymath}
 =vol({\bf \Sigma}_x)\,\cdot(<|y^j|^2>)^{\frac{1}{2}}.
\end{displaymath}
Therefore, we get the bound:
\begin{displaymath}
\Big|\Big(\,\int_{{\bf \Sigma}_x}\, dvol(x,y)\,y^j\,\tilde{f}(x,y)\,\partial_j\,\frac{\partial
f}{\partial \tilde{y}^l}\,\Big)\Big|\,\leq vol({\bf \Sigma}_x)\,\Big(\sum^{n-1}_{j=0}(<|y^j|^2>)^{\frac{1}{2}}\,\Big)\, \|\partial_j  \tilde{f}_x\|_{1,1}.
\end{displaymath}
In a local frame where the vector field  $U=(U^0,\vec{0})$, this contraction can be re-written as
\begin{displaymath}
(<(y^0)^2>)^{\frac{1}{2}}\cdot\, \|\partial_0 \tilde{f}_x\|_{1,1}=
\|(<|y^0|^2>)^{\frac{1}{2}}\cdot\, \partial_0 \tilde{f}_x\|_{1,1}=\|(<(y^0)^2>)^{\frac{1}{2}}\cdot\, \partial_0 \tilde{f}_x\|_{1,1}=
\end{displaymath}
\begin{displaymath}
=\|(<(y^0)^2\cdot\, (\partial_0 \tilde{f}_x)^2>)^{\frac{1}{2}}\|_{1,1}=
\|(<(y^j\cdot\, \partial_j \tilde{f}_x)^2>)^{\frac{1}{2}}\|_{1,1}.
\end{displaymath}
The last expression is covariant. Using normal coordinates associated with the affine connection $<\,^L\nabla>$ we obtain $\|(<(y^j\cdot\, \partial_j \tilde{f}_x)^2>)^{\frac{1}{2}}\|_{1,1}=0$.

Finally, we can bound the following integral
\begin{displaymath}
\Big|\Big(\,\int_{{\bf \Sigma}_x}\, dvol(x,y)\,y^j\,\tilde{f}(x,y)\,\frac{\partial
f}{\partial {y}^l}\,\Big)\Big|\,\leq \Big(\,\int_{{\bf \Sigma}_x}\, dvol(x,y)|\,y^j\,\tilde{f}(x,y)\,\frac{\partial
f}{\partial {y}^l}|\,\Big)\,\leq
\end{displaymath}
\begin{displaymath}
\leq \,\sum^{n-1}_{j=0}\,\Big(\,\int_{{\bf \Sigma}_x}\, dvol(x,y)\,|y^j\,\tilde{f}(x,y))|^2\,\Big)^{\frac{1}{2}}\
\cdot(\,\int_{{\bf \Sigma}_x}\, dvol(x,y)\,|\frac{\partial \tilde{f}(x,y)}{\partial y^k}|^{\frac{1}{2}}\Big)^{\frac{1}{2}}.
\end{displaymath}
As in the previous integral, we get
\begin{displaymath}
|\Big(\,\int_{{\bf \Sigma}_x}\, dvol(x,y)\,y^j\,\tilde{f}(x,y)\,\frac{\partial
f}{\partial {y}^l}\,\Big)|\leq vol({\bf \Sigma}_x)\,\cdot\, (<|y^j|^2>)^{\frac{1}{2}}\cdot\, \|\tilde{f}_x\|_{1,1}.
\end{displaymath}

Using these bounds, we obtain the following relation:
\begin{displaymath}
\|<\, ^LD>_{\tilde{V}} \tilde{V}(x)\|_{\bar{\eta}}\leq \frac{1}{vol^2({\bf \Sigma}_x)}\,\cdot
\end{displaymath}
\begin{displaymath}
\Big|\Big(\,\int_{{\bf \Sigma}_x}\,
dvol(x,y)\,y^j\,\tilde{f}(x,y)\,\partial_j\,\frac{\partial f}{\partial
{y}^l}\,\Big)\Big|\,\cdot\,\frac{3}{2}\,vol_E(supp(\tilde{f}_x))\,{\alpha}^2+
\end{displaymath}
\begin{displaymath}
+\,\frac{1}{vol^2({\bf \Sigma}_x)}\,\sum^{n-1}_{j=0}\Big|\Big(\,\int_{{\bf \Sigma}_x}\,
dvol(x,y)\,y^j\,\tilde{f}(x,y)\,\frac{\partial f}{\partial
\tilde{y}^l}\,\Big)\Big|\cdot
 \end{displaymath}
 \begin{displaymath}
\cdot \,vol^{\frac{1}{2}}_E({\bf \Sigma}_x)\cdot {\alpha}^2\cdot(\sum_k\,\|\partial_j\, log(\delta^k_x)\|_{0,2})\,\leq
\end{displaymath}
\begin{displaymath}
\leq\frac{1}{vol^2({\bf \Sigma}_x)}\, \cdot vol({\bf \Sigma}_x)\,\cdot
\end{displaymath}
\begin{displaymath}
\Big(\sum^{n-1}_{j=0} \, (<|y^j|^2>)^{\frac{1}{2}}\,
\cdot \|\tilde{f}_x\|_{1,1}\cdot\,\,vol^{\frac{1}{2}}_E({\bf \Sigma}_x)\cdot {\alpha}^2\cdot(\sum_k\,\|\partial_j\, log(\delta^k_x)\|_{0,2})\Big)\,=
\end{displaymath}
\begin{displaymath}
=\frac{vol^{\frac{1}{2}}_E({\bf \Sigma}_x)}{vol({\bf \Sigma}_x)}\,
\cdot\Big(\,\sum^{n-1}_{j=0}(<|y^j|^2>)^{\frac{1}{2}}
\,(\sum_k\,\|\partial_j\, log(\delta^k_x)\|_{0,2})\big)\,
\cdot \|\tilde{f}_x\|_{1,1}\cdot {\alpha}^2.
\end{displaymath}
In a local frame where the vector field $U=<Y>$ has components $(U^0,\vec{0})$, the following relation holds:
\begin{displaymath}
\sum^{n-1}_{j=0} (<|y^j|^2>)^{\frac{1}{2}}
\,(\sum_k\,\|\partial_j\, log(\delta^k_x)\,\|_{0,2}=(<|y^0|^2>)^{\frac{1}{2}}
\,(\sum_k\,\|\partial_0\, log(\delta^k_x)\,\|_{0,2}=
\end{displaymath}
\begin{displaymath}
=\,\sum_k\,\|<(y^0)^2>\, \partial_0\, log(\delta^k_x)\,\|_{0,2}.
\end{displaymath}
Note that the normal coordinate system (that we are using) coincides with the adapted coordinate system,
 associated with the vector field $U=(U^0,\vec{0})$. In this coordinate system, there is a bound $<y^0>\leq 1+\tilde{\alpha}$, where $\tilde{\alpha}$ is the diameter measured in the co-moving frame. It is of order $1$ or smaller than $1$. Therefore,
 \begin{displaymath}
 \sum^{n-1}_{j=0}\,(<|y^j|^2>)^{\frac{1}{2}}
\,(\sum_k\,\|\partial_j\, log(\delta^k_x)\,\|_{0,2}\leq (<|y^0|^2>)^{\frac{1}{2}}\cdot \,(\sum_k\,\|\partial_0\, log(\delta^k_x)\,\|_{0,2}\,\cdot \,(1+\tilde{\alpha}).
 \end{displaymath}

 Then we have the following result:
\begin{displaymath}
 \|<\, ^L\nabla>_{\tilde{V}} \tilde{V}(x)\|_{\bar{\eta}}\leq \,\frac{vol^{\frac{1}{2}}_E({\bf \Sigma}_x)}{vol({\bf \Sigma}_x)}
\,(\sum_{k}\,\|\partial_0\, log(\delta^k_x)\|_{0,2})\,\,
\cdot \|\tilde{f}_x\|_{1,1}\cdot {\alpha}^2\,+\mathcal{ O}(\alpha^3).
 \end{displaymath}\hfill$\Box$
 \begin{corolario}
 For compact domains ${\bf K}\subset {\bf M}$ and under the same hypotheses as in {\it theorem 4.3}, the following relation holds:
 \begin{displaymath}
 \|<\, ^L\nabla>_{\tilde{V}} \tilde{V}(x)\|_{\bar{\eta}}\leq n\,\cdot \tilde{C}({\bf K})\cdot \alpha^2\,  +\mathcal{O}(\alpha^3),
 \end{displaymath}
 for some constant $\tilde{C}({\bf K})$.
 \end{corolario}
 {\bf Proof}: Take the constant $\tilde{C}({\bf K})$ to be
 \begin{displaymath}
 \tilde{C}({\bf K})=max_{x\in {\bf K}} \Big\{\,\frac{vol^{\frac{1}{2}}_E({\bf \Sigma}_x)}{vol({\bf \Sigma}_x)}
\,(\sum_{k}\,\|\partial_0\, log(\delta^k_x)\|_{0,2})\,\,
\cdot \|\tilde{f}_x\|_{1,1}\,\Big\}.
 \end{displaymath}
 \hfill$\Box$

These expressions are asymptotic formulas if  $1>> \alpha$.

{\bf Remarks}
\begin{enumerate}
\item In the preceding results the ultra-relativistic limit ($E>>1$) was not essential. However, the series in power of the energy has asymptotic meaning if $E>>1$.

\item There are several notions of normal coordinates, since we have several affine connections: $^{\eta}\nabla$, $^{\bar{\eta}}\nabla$ and $<\,^L\nabla>$. However, we have only used the normal coordinates associated with $<\,^L\nabla>$.

\end{enumerate}

\subsection{Bound on the auto-parallel condition of the unitary mean vector field of the averaged Vlasov model}

Let us consider the normalized mean velocity vector field:
\begin{displaymath}
\tilde{u}=\frac{\tilde{V}}{\eta(\tilde{V},\tilde{V})^{1/2}}.
\end{displaymath}
Since $<\,^L\nabla>$ does not preserve the Minkowski metric $\eta$,
the covariant derivative of
$\tilde{u}$ in the
direction of $\tilde{u}$ using the Lorentz connection $^LD$ is
\begin{equation}
<\, ^L\nabla>_{\tilde{u}} \tilde{u} = \frac{1}{\eta(\tilde{V},\tilde{V})}\,
<\, ^L\nabla>_{\tilde{V}} \tilde{V} \,
+\frac{1}{2}\Big(\tilde{V}\cdot\big(log(\eta(\tilde{V},\tilde{V}))\big) \Big)\tilde{V}.
\end{equation}
The first term is bounded by {\it theorem 5.3.3}, since $\eta(\tilde{V},\tilde{V})>1$.
The total derivative of $\eta(\tilde{V},\tilde{V})$ along a trajectory of $\tilde{V}$
is
\begin{displaymath}
\mathcal{L}_{\tilde{V}}
\big(\eta(\tilde{V},\tilde{V})\big)=\tilde{V}\cdot\big(\eta(\tilde{V},\tilde{V})\big)=
\end{displaymath}
\begin{displaymath}
=2\eta\Big(<\, ^L\nabla>_{\tilde{V}} \tilde{V} ,\tilde{V}\Big)+\Big(<\, ^L\nabla>_{\tilde{V}} \eta
\Big) (\tilde{V},\tilde{V}).
\end{displaymath}
We have proved that the first term is of order $\alpha^2$.
Using normal coordinates for $<\, ^L\nabla>$, one can compute the
second term:
\begin{displaymath}
\big(<\, ^L\nabla>_{\tilde{V}} \eta\,\big) (\tilde{V},\tilde{V}) =
\eta(\tilde{V},\tilde{V}) {\bf F}_{j m} <\delta^m(x,y)\,
\delta^s(x,y)\, \delta^l(x,y)> \tilde{V}^{j}\tilde{V}_s\tilde{V}_l.
\end{displaymath}
We can estimate these contributions
\begin{proposicion}
Under the same assumptions as in {\it theorem 5.3.3}, the following relation holds:
\begin{equation}
<\,^L\nabla>_{\tilde{u}} \tilde{u}\leq \,\frac{vol^{\frac{1}{2}}_E({\bf \Sigma}_x)}{vol({\bf \Sigma}_x)}
\,(\sum_{k}\,\|\partial_0\, log(\delta^k_x)\|_{0,2})\,\,
\cdot \|\tilde{f}_x\|_{1,1}\cdot {\alpha}^2\,+ \mathcal{O}(\alpha^3).
\end{equation}
\end{proposicion}
{\bf Proof}: The first term of the right hand of the equation ${\it 5.3.6}$ is bounded by {\it theorem 5.3.3}.
 The second term is bounded using Hoelder's inequality for integrals [42, pg 62]
 \begin{displaymath}
 \Big|\int_{\bf X} dvol(z)\, f_1(z)\cdot\cdot\cdot\,f_m(z)\Big|\,\leq \prod^{m}_{k=1}\,
 \Big(\int_{\bf X} dvol(z)\, \big|f_k(z)\big|^{p_k}\Big)^{\frac{1}{p_k}},\,\,\,\, \sum_{k} p_k =1,\,\, 1\leq p_k\leq \infty.
 \end{displaymath}
In particular one can apply this inequality to the third order
moment\\
$<\delta^m(x,y)\,\delta^s(x,y)\, \delta^l(x,y)>$:
\begin{displaymath}
\Big|<\delta^m(x,y)\,
\delta^s(x,y)\, \delta^l(x,y)>\Big|\,=\,\frac{1}{vol({\bf \Sigma}_x)}\,\cdot
\end{displaymath}
\begin{displaymath}
\cdot \Big|\int_{{\bf \Sigma}_x} dvol(x,{y})\,f(x,y)\,\delta^m(x,y)\,
\delta^s(x,y)\, \delta^l(x,y)\Big|\,
\end{displaymath}
\begin{displaymath}
\leq \frac{1}{vol({\bf \Sigma}_x))}\,
\Big(\int_{{\bf \Sigma}_x} dvol(x,y)\,|\tilde{f}(x,y)\,\delta^m(x,y)|^3\,\Big)^{\frac{1}{3}}\cdot
\end{displaymath}
\begin{displaymath}
\cdot\Big(\int_{{\bf \Sigma}_x} dvol(x,y)\,|\tilde{f}(x,y)\delta^s(x,y)|^3\,\Big)^{\frac{1}{3}}\cdot
\end{displaymath}
\begin{displaymath}
\cdot\Big(\int_{{\bf \Sigma}_x} dvol(x,y)\,|\tilde{f}(x,y)\delta^l(x,y)|^3\,\Big)^{\frac{1}{3}}.
\end{displaymath}
The distribution function is positive on $supp(\tilde{f})$. Also one can choose a distribution function such that $\tilde{f}_x\leq 1$.
By proposition $(5.3.1)$, one can substitute in the integrations $\tilde{f}_x\longrightarrow f_x$,
which implies that
\begin{displaymath}
\Big|<\delta^m(x,y)\,
\delta^s(x,y)\, \delta^l(x,y)>\Big|\,=O(\alpha^3).
\end{displaymath}
 Since the norm $\bar{\eta}(<y>,<y>)\ge
1$, one gets a third degree monomial term
in ${\alpha}$ for the covariant derivative $<\,
^LD>_{\tilde{u}} \tilde{u} $.\hfill$\Box$

\begin{corolario}
Under the same assumptions as in {\it theorem 5.3.3} in a compact domain ${\bf K}\subset {\bf M}$,
 one obtains
\begin{equation}
<\,^L\nabla>_{\tilde{u}} \tilde{u}\leq \,\cdot \tilde{C}({\bf K})\cdot \alpha^2 \,+\mathcal{O}(\alpha^3),
\end{equation}
for a convenient constant $\tilde{C}({\bf K})$.
\end{corolario}
\subsection{Bound on the auto-parallel condition of the mean velocity field of the Vlasov model}

Let us consider a local {\it Lorentz congruence}, which is a set of
auto-parallel curves of the Lorentz connection $^L\nabla$, for a set
of initial conditions at each $(t_0,\vec{x})$, $\vec{x}\in{\bf
M}_{t_0}$ where ${\bf M}_{t_0}\hookrightarrow {\bf M}$ is a $3$-dimensional
spatial sub-manifold. One can consider in a similar way the congruence
associated with the averaged Lorentz connection for the same initial conditions.
Note that, while the Lorentz connection preserves the Lorentz norm
  $\eta(\dot{x},\dot{x})$ of the tangent vectors of the geodesics,
  this is not the case for the averaged Lorentz connection.
\begin{teorema}
Let {\bf F} be a closed $2$-form and $^L\nabla$
the associated non-linear Lorentz connection. Under the same assumptions as in {\it theorem 5.3.3} the solutions of
 the Lorentz force equation $^{\eta}\nabla_{\dot{x}} {\dot{x}}= (\iota _{\dot{x}} {\bf F})^\sharp$
can be approximated by the integral curves of the normalized mean velocity vector field $u=\frac{V(x)}{\sqrt{\eta(V(x), V(x))}}$ of the
distribution function $f(x,y)$, where $f(x,y)$ is a solution of the associated Vlasov equation
$^L\chi f=0$. The difference is controlled by polynomial functions at least of order $2$ in ${\alpha}$,
\begin{equation}
\|\,^L\nabla_{{u}} {u}\|(x)\,\leq \,{a}_2(x)\, {\alpha}^2\,+\mathcal{O}(\alpha^3)
\end{equation}
where the function $a_2(x)$ is a bounded function of $x$.
\end{teorema}
{\bf Proof:} We repeat an argument that we have used before.  By {\it proposition 5.3.1}, both distribution
functions $f$ and $\tilde{f}$, solutions of $^L\chi f=0$ and $<\,^L\chi>\tilde{f}$, are such that
\begin{displaymath}
|f(t,x(t),\dot{x}(t))-\tilde{f}(t,{x}(t),\dot{{x}}(t))|\leq\,\big(\tilde{C}(x)\|{\bf F}\|_{\bar{\eta}}
C^2_2(x)(1+B_2(x){\alpha})\big){\alpha}^2\,E^{-2}\,t^2\,+
\end{displaymath}
\begin{displaymath}
+\big(\tilde{K}(x)\|{\bf F}\|_{\bar{\eta}}(x)\,K^2_2(1+D_2(x){\alpha})\big){\alpha}^2\,E^{-1}\,t.
\end{displaymath}
Therefore, the corresponding mean velocity fields  are nearby as well,
because of the linearity of the averaging operation and because of the above relation.
Then their corresponding
integral curves and the associated local congruences are also similar. By {\it corollary 5.3.6},
 for narrow distributions,
 the normalized mean field $\tilde{u}$ associated with $\tilde{f}$ is such that
\begin{displaymath}
\|<\,^L\nabla>_{\tilde{u}} \tilde{u}(x)\|_{\bar{\eta}}\,\leq\, \tilde{a}_2\, {\alpha}^2\,+\,\mathcal{O}({\alpha}^3).
\end{displaymath}
for some function $\tilde{a}_2(x)$. Remember that one can interpolate smoothly between the connections $^L\nabla$ and $<\,^L\nabla>$.
 Therefore, locally, one can interpolate smoothly between their integral curves.
 Also, because of the smoothness of the solutions of the geodesic equations with respect to the parameter of interpolation, there is a function ${a}_2$
 in a small open neighborhood of {\bf M} such that
\begin{displaymath}
^L\nabla_{{u}} {u}\,\leq \,{a}_2(x)\, {\alpha}^2\,+O(\alpha^3).
\end{displaymath}\hfill$\Box$
\section{Discussion}

{\it Theorem 5.3.7} shows when the charged cold fluid
model is a good approximation to the Vlasov equation in the description of the dynamics of a collection of
particles interacting with an external electromagnetic field, in the ultra-relativistic regime.
It is interesting that we have obtained this result without using
additional hypotheses on the higher moments of the distribution function, except that the distribution is
 narrow and smooth enough for our calculations (we need some smoothness conditions in order
 to use Taylor expansion for the function $f(x,y)$ in the velocity coordinates. Indeed, it seems that one
 only requires weak differentiability in $y$).

 One of the important hypothesis on which the calculation is physically relevant is
 the requirement that the diameter of the distribution $\alpha$ must be small in the
 laboratory frame. Also, the ultra-relativistic regime $E>>1$ is useful in order to have good estimates and holds in current particle accelerators.

There are some technical issues that we would like to mention briefly:
\begin{enumerate}

\item We have assumed that the distribution functions $\tilde{f}$ and $f$ are at least $\mathcal{C}^1$ in x. However, let us consider the Dirac
 delta distribution with support invariant by the flow of the Lorentz force,
\begin{equation}
f(x,y)=\Psi(x)\,\delta(y-V(x)).
\end{equation}
Since the width of the distribution is zero, ${\alpha}=0$. One can use this distribution as a solution
 of the Vlasov equation in $2$-dimensional space-times, for a proper value of the function $\Psi$.  This example and the fact that the bounds found in {\it section 5.3}
  are formulated using Sobolev norms suggest the possibility of generalize the results to bigger function spaces. The results to use in this case are Sobolev embedding theorems [41, 57, 58]. However, we are not investigating this question in this thesis.

\item The same method can be applied to other fluid equations.
Depending on the specific bounds and parameters, one can decide
which model is better in each particular situation.

\subsection{On the validity of the truncation schemes in fluid
models}

Given a kinetic model, usually the Maxwell-Vlasov system of
differential equations, one defines a fluid model in terms of the
low moments (typically, first, second, third and fourth moments) of
the distribution function $f$ [8, 37-39]. All the higher moments are
set equal to zero.
Usually, the typical reasoning is that with the low moments one can
write down models that are consistent and explain a reasonable
number of phenomena in plasma physics.

We can argue that the reason why these models work is that in some
situations the underlying Vlasov model can be substituted by the
averaged Vlasov model. Then the Vlasov model depends only on the
first, second and third moments of the distribution function $f$. Therefore, as soon as the
the hypothesis of a given fluid model are compatible with the
hypothesis of the approximation {\it Maxwell-Vlasov model
$\longrightarrow$ averaged Maxwell-Vlasov model}, fluid models whose dynamical fields
 can be written in terms of the first, second and third
moment, are {\it equivalent} to the underlying averaged Vlasov model. The
equivalence must be understood in an approximated way, since there
is an approximation in this argument.

The variables that one considers in fluid models are the mean
velocity field (2.1.6), the covariant kinetic energy-momentum tensor
(2.1.7) and the covariant energy-momentum flux tensor (2.1.8). Therefore, one can propose the following
\begin{definicion}
Two kinetic models are equivalent if their corresponding mean
velocity field,  covariant kinetic energy-momentum tensor and
covariant energy-momentum flux tensor are the same.
\end{definicion}
\begin{definicion}
Two fluid models are the same if their corresponding mean velocity
field, covariant kinetic energy-momentum tensor and covariant
energy-momentum flux tensor are the same.
\end{definicion}
We propose the following conjecture in the form of a {\it theorem}
\begin{teorema}
If two kinetic models are equivalent, the corresponding fluid models
are the same. If two fluid models are the same, the underlying
kinetic models are equivalent up to the order of approximation of
the kinetic model by the averaged kinetic model.
\end{teorema}
The first implication is trivial. The second implication is true for the Vlasov model, as we have proved in
this {\it chapter}.

Therefore, when one works with a kinetic model, there is an
underlying equivalence class of fluid models. We can call this an
{\it universal class}. The elements of an universal class are, by
construction, fluid models of ultra-relativistic narrow
distributions. Then it is a useful idea to consider for each class the
simplest model possible. In practice, the simplest model will dismiss
higher order moments.

\end{enumerate}
\newpage

\chapter{The Jacobi equation of the averaged Lorentz connection and applications in beam dynamics}
\section{Introduction}

\subsection{The Jacobi equation of an affine connection on {\bf M}}

Let ${\bf M}$ be an $n$-dimensional manifold.
Given an affine connection $\nabla$ on the tangent bundle
$\pi:{\bf TM}\longrightarrow {\bf M}$, the auto-parallel curves of $\nabla$
are the  solutions $c:{\bf I}\longrightarrow {\bf M}$ of the system of
differential equations
\begin{displaymath}
\nabla _T T=0,\quad \quad T=\frac{d c}{dt},
\end{displaymath}
where $t$ is an affine parameter of $\nabla$. The curvature tensor $R$ of the
affine connection ${\bf \nabla}$ is the tensor field defined by the expression
\begin{displaymath}
R:\Gamma{\bf T}{\bf M}\times\Gamma{\bf T}{\bf M}\times\Gamma{\bf
T}{\bf M}\longrightarrow \Gamma{\bf T}{\bf M}
\end{displaymath}
\begin{equation}
(X,Y,Z)\mapsto R(X,Y,Z)=\nabla_X \nabla_Y Z
-\nabla_Y \nabla_X Z -\nabla_{[X,Y]} Z,\,
\quad \forall
X,Y,Z\in \Gamma{\bf TM}.
\end{equation}
The curvature tensor has an associated family of curvature
endomorphisms $\{R_x(X,Y),\quad X,\, Y \in \Gamma{\bf T M}, x\in {\bf M}\}$
defined by
\begin{displaymath}
R(X,Y):\Gamma{\bf TM}\longrightarrow \Gamma{\bf TM}
\end{displaymath}
\begin{displaymath}
Z\mapsto R_x(X,Y)Z=R(X,Y,Z)(x),
\,\quad Z\in\Gamma{\bf TM},\, \forall x\in{\bf M}.
\end{displaymath}

A vector field $J$ along the parameterized geodesic $c:{\bf I}\longrightarrow {\bf M}$ of an
affine, torsion-free connection is a Jacobi field if it satisfies
the Jacobi equation
\begin{equation}
\nabla_X \nabla_X J -R(X,J)X=0, \quad X=\frac{d c}{dt}.
\end{equation}
This equation can be re-written using a local frame $\{e_i,\,i=0,1,2,...,n-1\}$.
It corresponds to the system of the second order differential equations
\begin{equation}
\frac{D^2 J^{i}}{dt^2} - R^{i}\,_{jkm}(c(t))J^{k}\frac{d
X^{j}}{dt}\frac{dX^m}{dt}=0,
\end{equation}
where $\frac{D J}{dt}$ is the covariant derivative along the reference geodesic and
the curvature tensor is given by the expression
\begin{displaymath}
R^i\,_{jkm}=\partial_m \Gamma^i\,_{jk}- \partial_k \Gamma^i\,_{jm}+
\big(\Gamma^r\,_{jk}\Gamma^i\,_{rm}-\Gamma^r\,_{jm}\Gamma^i\,_{rk}\big).
\end{displaymath}

For the next fundamental results one can consult [35,
{\it section 10.1}; 54, $\S$ 14]. It is well known that for an affine connection
there are $2n$ linear independent Jacobi fields along any given
central geodesic $c$. This
fact is a consequence of the existence and uniqueness of the solutions to
second order differential equations and their smoothness properties on the
initial values. A Jacobi field is completely determined by
the value of $J(0)$ and $\frac{DJ}{dt}(0)$.

A smooth map
\begin{displaymath}
C: (-\lambda,\lambda)\times {\bf I}\longrightarrow {\bf M}\quad
t\longrightarrow c_s (t)\subset {\bf M}
\end{displaymath}
is a geodesic variation of $c(t)$ if each $c_s(t) :(-\lambda, \lambda)\longrightarrow {\bf M}$ is an affine parameterized geodesic curve for
all $s\in {\bf I}$. The variation vector field of a geodesic
variation is the vector field along the curve $c$ defined by $dC(\frac{\partial}{\partial s})$, with $dC:{\bf T}((-\lambda,\lambda)\times {\bf I})\longrightarrow {\bf TM}$ the differential of the smooth function $C$. The vector field $[dC(\frac{\partial}{\partial s}),\frac{dX}{dt}(c(t))]$ along the curve $c(t)$ vanish.
The variation vector field acts on an
arbitrary smooth function as a derivation
\begin{displaymath}
 dC(f)=\frac{\partial f(
C_t(s))}{\partial x^k}\frac{\partial C^k}{\partial t}.
\end{displaymath}
The geometric interpretation of a Jacobi field is obtained through
the following [35, 54]:
\begin{proposicion}
Let ${\bf M}$ be a manifold equipped with an affine torsion-free
connection $\nabla$. Then each Jacobi field $J$ is a variation vector field of
a geodesic variation $C$. Conversely, any
variation field of a geodesic variation $C$ defines a Jacobi
field.
\end{proposicion}
\subsection{Jacobi equation for linear connections defined on the pull-back
bundle $\pi^*{\bf TM}\longrightarrow {\bf N}$}

Let {\bf M} be a smooth $n$-dimensional manifold, ${\bf N}\hookrightarrow{\bf TM}$
a sub-bundle of the tangent bundle {\bf TM} and let us consider an Ehresmann connection defined on ${\bf TN}$.

In order to formulate a Jacobi equation for
linear connections on $\pi^*{\bf TM}\longrightarrow {\bf N}$, we mimic the standard derivation of the Jacobi equation for affine connections [35]. The {\it bending term} is determined by
the $hh$-curvature endomorphisms:
\begin{displaymath}
R(h(X),h(Y)):\pi^*{\bf TM}\longrightarrow \pi^*{\bf TM}
\end{displaymath}
\begin{displaymath}
\zeta\mapsto R(h(X),h(Y))\zeta\,=\,\big(\nabla_{h(X)}\nabla_{h(Y)}-\nabla_{h(Y)}\nabla_{h(X)}-\nabla_{h([X,Y])}\big)\zeta,
\end{displaymath}
\begin{displaymath}
\quad \forall \zeta\in \Gamma {\bf TM},
\end{displaymath}
where the horizontal lift $h$, $h(X^i\partial_i) =X^i(x)\frac{\delta}{\delta x^i}$ was
introduced in {\it section 3.3}.

Similar to the case of affine connections, one obtains the Jacobi
equation for linear connections on $\pi^*{\bf TM}\longrightarrow {\bf N}$.
A basic fact is that we require that the Jacobi vector field
commutes with the vector field $X(t)$ along the curve $c(t)$, which means that
\begin{displaymath}
[J,X]|_{c(t)}=0.
\end{displaymath}
Then the torsion-free condition along the curve $c$ is
\begin{displaymath}
\nabla_{h({J})}\pi^* X=\nabla_{h({X})}\pi^* J,
\end{displaymath}
and the curvature endomorphism along the curve $c$ is such that
\begin{displaymath}
R(h({X}),h({J}))\,\zeta=\big(\nabla_{h({X})}\nabla_{h({J})}
-\nabla_{h({J})}\nabla_{h({X})}\big)\,\zeta,\quad \zeta\in \Gamma\pi^*{\bf TM}.
\end{displaymath}
We can compute the second covariant derivatives along the curve $c$,
\begin{equation}
\nabla_{h(X)} \nabla_{h(X)} \pi^* J=\nabla_{h(X)} \nabla_{h(J)}\pi^* X=
\nabla_{h(J)} \nabla_{h(X)}\pi^* X - R(h({X}),h({J}))\,\pi^* X.
\end{equation}
Let us assume that $X\in \Gamma{\bf TM}$ is a auto-parallel respect to $\nabla$, which means $\nabla_{h(X)}\pi^* X =0$. One obtains the following second order differential equation for $J(t)$
\begin{equation}
\nabla_{h(X)} \nabla_{h(X)} \pi^* J\,- R(h({X}),h({J}))\,\pi^* X\,=0.
\end{equation}
\begin{definicion}
A field $J(t)$ along the curve $c:{\bf I}\longrightarrow {\bf M}$ satisfying equation $(6.1.5)$ is a Jacobi field of $\nabla$. The corresponding vector field $\pi_2 (J(t))$ along the curve is the associated vector field.
\end{definicion}
We can identify the vector field $\pi_2(J(t))$ with $J(t)$.
\begin{definicion}
Let $\nabla$ be a linear connection on the bundle $\pi^*{\bf TM}$. An auto-parallel variation of the auto-parallel curve $c:{\bf I}\longrightarrow {\bf M}$ is a map $C:(-\lambda,\lambda)\times {\bf I}\longrightarrow {\bf M}$ such that for each value of the parameter $s$, the curve $c_s(t):=C(s,t)$ is an auto-parallel curve, $\nabla_{h(\dot{c}_s(t))}\pi^*\dot{c}_s(t)=0$.
\end{definicion}
\begin{proposicion}
Let $\nabla$ a linear connection on $\pi^*{\bf TM}$.
The variation field along $c(t)$ of a variation $C(t,s)$ is a Jacobi field of $\nabla$.
\end{proposicion}
{\bf Proof}: It is clear from the deduction of the Jacobi equation for $\nabla$.\hfill$\Box$

{\bf Remark}. For covariant derivatives such that they are zero along the vertical direction,
the covariant derivatives does not depend on the particular lift of $X\in {\bf T}_x{\bf M}$ to ${\bf T}_u{\bf N}$.
 Therefore for those covariant derivatives the expression that one obtains in this
 case for the Jacobi equation is defined as before and is independent of the
vertical component of the lift that we are using.

\subsection{Physical Interpretation of the Jacobi equation}

For an affine connection, the Jacobi field represents
the deviation vector of a given trajectory from the reference geodesic.
However, if the reference trajectory is observable, the assumption that the
 reference trajectory coincides with the central geodesic provides physical
 meaning to the Jacobi field. Let us consider a geodesic variation $C(s,t)$. Each of the geodesics $c_s(t)=C(s,t)$ corresponds to a possible trajectory for a charged point particle. Then
 the Jacobi field corresponds to the deviation variable from a particular trajectory follow by a particle
  to the reference trajectory.

The above property holds at least for affine connections. Therefore, let us fix a central geodesic $c(t)$ and consider the set of all geodesic variations of the central geodesic $c(t)$. Hence there is a relation between the set of geodesic variations (which is equivalent to the set of Jacobi fields along $c(t)$ by {\it proposition} 6.1.1) and the set of all the trajectories allowable by the dynamics and the topology of the space-time manifold {\bf M}.

 If the Jacobi vector field is given by $J(t)$, one identifies the components of $J(t)$
 with the relative coordinates of a given geodesics respect the central affine geodesic $c(t)$, $J^k(t)=u^k(t)$.

Assuming this interpretation, the Jacobi Equation (for both affine and non-affine connections) is a second order Riccati equation
 \begin{displaymath}
 \frac{d^2 {u}}{dt^2}+R(t)u=0.
 \end{displaymath}
 This type of equation appears when one considers small deviations from a solution
  of another differential equation.
One example is Hill's equation in celestial mechanics [55].

However, in the case where the connection is affine, the form of the Riccati equation is the same,
but the endomorphism $R(t)$ along the curve $c(t)$ is simpler, since the curvature
endomorphism depends only on the point $c(t)$ and not on the derivative $\frac{dc(t)}{dt}$
as in the general case of a non-linear connection.

\section{Jacobi equation of the averaged lorentz connection}

In this {\it section} {\bf M} is a $n$-dimensional manifold.
The averaged Lorentz connection $<\,^L\nabla>$ is an affine connection on {\bf M}.

Using local coordinates, a Jacobi field along the central geodesic
can be written as $J=\xi^j(s)\partial_j $. The reference trajectory will be $X(\tau)$,
 which will be assumed to be a geodesic of the averaged Lorentz connection. Then the Jacobi equation for
an affine connection on the tangent bundle ${\bf TM}\longrightarrow
{\bf M}$ can be expressed as
\begin{displaymath}
\frac{d^2 \xi^{i}}{d\tau^2}+2\Gamma^{i}\,_{j
k}(X(\tau))\frac{d\xi^j}{d\tau}\frac{dX^k}{d\tau}+\xi^{l}\partial_{l}
\Gamma^{j}\,_{jk}(X)\frac{d
X^j}{d\tau}\frac{dX^k}{d\tau}=0.
\end{displaymath}
This equation is called the geodesic deviation equation.
The central geodesic is denoted by $X(\tau)$ and a neighborhood geodesic is
given by $x(\tau)=\xi(\tau)+X(\tau)$. The parameter $\tau$ is the proper-time of the central
geodesic measured with the metric $\eta$.

The averaged Lorentz connection $<\,^L\nabla>$ is an affine connection on
the tangent bundle {\bf TM}. Therefore, we can apply the standard
Jacobi equation to the averaged Lorentz connection. Given an arbitrary semi-Randers space
$({\bf M}, \eta, [A])$, in a local natural coordinate system,
 the averaged Lorentz connection has the connection coefficients
\begin{displaymath}
<\,^L\Gamma^i\,_{jk} >=\, ^{\eta}\Gamma^i\,_{jk}+
\frac{1}{2}({\bf F}^i\,_{j}<y^m>\eta_{mk}+ {\bf
F}^i\,_{k}<y^m>\eta_{mj})+
\end{displaymath}
\begin{displaymath}
+{\bf F}^i\,_m\big(
{<y^m>}\eta_{jk}-\eta_{js}
\eta_{kl}<y^m y^s y^l>\big),
\end{displaymath}
with ${\bf F}=dA,$ with $A$ being a representative of $[A]$, $\quad A\in[A]$.
The tangent vector $y$ are on the unit hyperboloid ${\bf \Sigma_x}$.

The Jacobi equation of the averaged Lorentz connection is
\begin{displaymath}
\frac{d^2 \xi^i}{d\tau^2}+
2\frac{d\xi^j}{d\tau}\frac{dX^k}{d\tau}\Big(\frac{1}{2}({\bf
F}^i\,_{j}<y^m>\eta_{mk}+ {\bf F}^i\,_{k}<y^m>\eta_{mj})+
\end{displaymath}
\begin{displaymath}
+{\bf F}^i\, _m\big( {<y^m>}\eta_{jk}-\eta_{js} \eta_{kl}<y^m y^s
y^l>\big)\Big) +2\xi^l \partial_l \Big(\frac{1}{2}({\bf
F}^i\,_{j}<y^m>\eta_{mk}+ {\bf F}^i\,_{k}<y^m>\eta_{mj})+
\end{displaymath}
\begin{displaymath}
+{\bf F}^i\, _m\big( {<y^m>}\eta_{jk}-\eta_{js} \eta_{kl}<y^m y^s
y^l>\big)\Big)\frac{dX^j}{d\tau}\frac{dX^k}{d\tau}+
\end{displaymath}
\begin{equation}
+\big(\,^{\eta}\Gamma^i\,_{jk}+\xi^l\partial_l
\,^{\eta}\Gamma^i_{jk}\big)
\big(\frac{dX^j}{d\tau}\frac{dX^k}{d\tau}+2\frac{dX^j}{ds}
\frac{d\xi^k}{d\tau}\big)=0.
\end{equation}
From the form of the system of differential equations $(6.2.1)$ we
conclude that:
\begin{enumerate}
\item There is a term representing the inertial acceleration:
\begin{equation}
\mathcal{A}_I:=\big(\,^{\eta}\Gamma^i\,_{jk}+\xi^l\partial_l
\,^{\eta}\Gamma^i\,_{jk}\big)
\big(\frac{dX^j}{d\tau}\frac{dX^k}{d\tau}+2\frac{dX^j}{d\tau}\frac{d\xi^k}{d\tau}\big).
\end{equation}
The inertial acceleration $\mathcal{A_I}$ is universal, in the sense that it is independent of the particle mass.

\item We can not say that $\mathcal{A_I}$ is independent of the electromagnetic field,
since $\frac{dX^j}{d\tau}$ can depend implicitly on the electromagnetic field
when defining the reference trajectory. The typical example is the reference orbit of a betatron [9, {\it chapter 3}].
\end{enumerate}
These properties of the inertial acceleration $\mathcal{A_I}$ will help us to
write it without doing explicit calculations, using establish formulae for elementary cases.

\subsection{The Jacobi equation of the Lorentz connection versus the Jacobi
equation of the averaged Lorentz connection}

We have shown that the deviation equation from a given reference trajectory defines a Jacobi equation.
 However the non-linearity of the Lorentz force equation creates difficulties
in view of the applicability of the above interpretation:
\begin{enumerate}
\item The evaluation of the covariant derivatives with respect to the
original Lorentz connection requires a {\it reference vector}. Due to the dependence of the
connection coefficients $\Gamma^i\,_{jk}(x,y)$ on the direction $y$, there
must be assigned a particular point in the tangent space $y_0\in {\bf
T}_x{\bf M}$ at which the connection coefficients should be evaluated. This implies a specification of the direction
where the connection coefficients are evaluated. Attaching a
physical significance to the choice of the vector $y_0$ implies the
selection of a particular model, which requires additional justification.
This difficulty is resolved using the averaged Lorentz connection,
which is an affine connection and whose connection coefficients do not depend on the {\it
reference vector} $y_0$.

\item It was proved in [22] (although in the category of
Finsler spaces and for connections which covariant derivative vanish along vertical directions) that the averaged curvature of the original
connection is the curvature of the averaged connection.
This result can be extended to arbitrary linear connections on $\pi^*{\bf TM}$ with vanishing covariant derivative in the vertical directions. Hence, we can apply this result to the Lorentz connection.
In the corresponding averaged Jacobi equation appears the averaged
curvature, which is the same as the curvature of the averaged
connection. Therefore we can work with the Jacobi equation for the
averaged connection, as an attempt to give an {\it averaged
description} of the dynamics of a beam of particles where the {bending term} is the {\it averaged force}.

\end{enumerate}

Motivated by the above reasons, in this chapter we replace the Jacobi equation of the Lorentz
connection by the Jacobi equation of the averaged Lorentz connection
as a description of beam dynamics of bunches of particles in
accelerators. We consider systems in the ultra-relativistic regime and we also assume
that the distribution function $f$ is narrow in velocity space. We
show how the Jacobi equation for the averaged connection
$<\,^L\nabla>$ provides a geometrical formulation of the {\it transversal} (in the case of dipole and
quadrupole fields)
and {\it longitudinal beam dynamics} (when the external fields are
linearizable in the relative coordinates).
 We then provide a method to introduce
corrections to the averaged Lorentz dynamics caused by the composed nature of the bunch of particles.
These corrections are expressed in terms of known or observable quantities.

\section{Transversal beam dynamics from the
Jacobi equation of the averaged connection}

Let us assume that $({\bf M},\eta)$ is Minkowski space with {\bf M} being $4$-dimensional.
There is a global coordinate system denoted by $(\tau,x^1,x^2,x^3)$.  $\tau$ is the proper
time considered from a given initial point of the reference
trajectory, the coordinate $x^2$
is given by the Euclidean length of the path of the reference
trajectory measured from the initial position in the reference frame defined by the vector field $\frac{d}{dt}$, which corresponds to the laboratory frame, $x^1$ is the horizontal coordinate and $x^3$ the vertical coordinate
 respect to the central geodesic. The longitudinal direction at each
instant $\tau$ is given by the vector $\frac{\partial}{\partial x^2}$.
By definition $(x^1, x^3)$ are the transverse coordinates, while
$x^2$ is the longitudinal coordinate.
\subsection{Relation between the transverse dynamics and the Jacobi equation}

We choose the laboratory reference frame for our calculations. Under the transverse
 dynamics, the difference $\frac{dx^2}{dt}-\frac{dX^2}{dt}$ will be constant. The external
electromagnetic fields are static magnetic fields in Minkowski space.
In the subsequent calculations we will only consider the lower order terms in the degree $a+b+c$ of the monomials $\xi^a(\frac{d\xi}{d\tau})^b\epsilon^c$ appearing in the expressions, with $\xi=x(\tau)-X(\tau),\quad \epsilon=<y>-\frac{dX}{d\tau}$.

Firstly, we linearize the equations with respect to the degree defined
by the vector fields along the central geodesic $\xi$ and the powers of the difference $\epsilon$ appearing on each term
Recall that
the {\it transverse component}, which are the terms in the geodesic
equation proportional to the tensor $
{T}^i\,_{jk}={\bf F}^i \,_m\big( {<y^m>}\eta_{jk}-\eta_{js}
\eta_{kl}<y^m y^s y^l>\big)$, are neglected systematically because they are of higher order in the
degree $(a+b+c)$ than the {\it longitudinal component}. The longitudinal component is proportional
to
\begin{displaymath}
{L}^i\,_{jk}=\frac{1}{2}({\bf F}^i\,_{j}<y^m>\eta_{mk}+ {\bf
F}^i\,_{k}<y^m>\eta_{mj})
\end{displaymath}
After linearizing, the system of differential equations are
\begin{displaymath}
\frac{d^2 \xi^i}{d\tau^2}+
2\frac{d\xi^j}{d\tau}\frac{dX^k}{d\tau}\Big(\frac{1}{2}({\bf
F}^i\,_{j}<y^m>\eta_{mk}+ {\bf F}^i\,_{k}<y^m>\eta_{mj})\Big) +
\end{displaymath}
\begin{displaymath}
 +2\xi^l
\partial_l \Big(\frac{1}{2}({\bf F}^i\,_{j}<y^m>\eta_{mk}+ {\bf
F}^i\,_{k}<y^m>\eta_{mj})\Big)\cdot\frac{dX^j}{d\tau}\frac{dX^k}{d\tau}
\end{displaymath}
\begin{displaymath}
+\big(\,^{\eta}\Gamma^i\,_{jk}+\xi^l\partial_l
\,^{\eta}\Gamma^i\,_{jk}\big)
\big(\frac{dX^j}{d\tau}\frac{dX^k}{d\tau}+2\frac{dX^j}{d\tau}
\frac{d\xi^k}{d\tau}\big)=0, \quad i,j,k,m=0,...,n-1.
\end{displaymath}
Since $\epsilon$ is small, we can replace
$<y>\longrightarrow \frac{dX}{ds}$, obtaining the following differential equation:
\begin{displaymath}
\frac{d^2 \xi^i}{d\tau^2}+
2\frac{d\xi^j}{d\tau}\frac{dX^k}{d\tau}\Big(\frac{1}{2}({\bf F}^i\,_{j}
\frac{dX^m}{d\tau}\eta_{mk}+{\bf F}^i\,_{k}\frac{dX^m}{d\tau}\eta_{mj})\Big)+
\end{displaymath}
\begin{displaymath}
 +2\xi^l
\partial_l \Big(\frac{1}{2}({\bf F}^i\,_{j}\frac{dX^m}{d\tau}\eta_{mk}+ {\bf
F}^i\,_{k}\frac{dX^m}{d\tau}\eta_{mj})\Big)\cdot\frac{dX^j}{d\tau}\frac{dX^k}{d\tau}+
\end{displaymath}
\begin{displaymath}
+\big(\,^{\eta}\Gamma^i\,_{jk}+\xi^l\partial_l
\,^{\eta}\Gamma^i\,_{jk}\big)
\big(\frac{dX^j}{d\tau}\frac{dX^k}{d\tau}+2\frac{dX^j}{d\tau}
\frac{d\xi^k}{d\tau}\big)=0.
\end{displaymath}
{\bf Remark}. This equation is the geodesic deviation equation of the
Lorentz force equation. Therefore, the difference between the averaged Jacobi
equation and the deviation equation associated with the Lorentz
force is of higher order in the degree $(a+b+c)$. At leading order both equations coincide.

The condition of transversal dynamics is
\begin{displaymath}
\frac{d \xi^j}{d\tau} \frac{dX_j}{d\tau} \simeq \mathcal{O}^2,
\end{displaymath}
where $\mathcal{O}^2$ indicates first order in the general degree $(a+b+c)$.
This is a more general condition than what is usually stated as the transverse dynamics in accelerator physics: the
magnetic field is perpendicular to the velocity field $\frac{dX}{d \tau}$ of the particles in the beam [9].

 Due to this condition, we suppress the respective term in the differential equations, getting the equations
\begin{displaymath}
\frac{d^2 \xi^i}{d\tau^2}+
2\frac{d\xi^j}{d\tau}\frac{dX^k}{d\tau}\Big(\frac{1}{2}{\bf F}^i\,_{j}
\frac{dX^m}{d\tau}\eta_{mk}\Big)+ 2\xi^l
\partial_l \Big(\frac{1}{2}{\bf
F}^i\,_{j}\frac{dX^m}{d\tau}\eta_{mk}\Big)\frac{dX^j}{d\tau}\frac{dX^k}{d\tau}+
\end{displaymath}
\begin{displaymath}
+\big(\,^{\eta}\Gamma^i\,_{jk}+\xi^l\partial_l
\,^{\eta}\Gamma^i\,_{jk}\big)
\big(\frac{dX^j}{d\tau}\frac{dX^k}{d\tau}+2\frac{dX^j}{d\tau}\frac{d\xi^k}{d\tau}\big)=0.
\end{displaymath}
Therefore, at first order, the differential equations are
\begin{displaymath}
\frac{d^2 \xi^i}{d\tau^2}+ \frac{d\xi^j}{d\tau}{\bf F}^i\,_{j} + \frac{d
X^j}{d\tau}\xi^l
\partial_l {\bf F}^i\,_{j}
+\,^{\eta}\Gamma^i\,_{jk}\frac{dX^j}{d\tau}\frac{dX^k}{d\tau}+\xi^l\partial_l
\,^{\eta}\Gamma^i\,_{jk}\frac{dX^j}{d\tau}\frac{dX^k}{d\tau}+
2\,^{\eta}\Gamma^i\,_{jk}\frac{dX^j}{d\tau}\frac{d\xi^k}{d\tau}=0.
\end{displaymath}
In the transverse dynamics, one assumes by construction that $\frac{d\xi^j}{d \tau}{\bf F}^i\,_j=0$, if there is not {\it dispersion}. Hence, the differential equations in this regime are
\begin{equation}
\frac{d^2 \xi^i}{d\tau^2}+ \frac{d\xi^j}{d\tau}{\bf F}^i\,_{j}
+\,^{\eta}\Gamma^i\,_{jk}\frac{dX^j}{d\tau}\frac{dX^k}{d\tau}+\xi^l\partial_l
\,^{\eta}\Gamma^i\,_{jk}\frac{dX^j}{d\tau}\frac{dX^k}{d\tau}+
2\,^{\eta}\Gamma^i\,_{jk}\frac{dX^j}{d\tau}\frac{d\xi^k}{d\tau}=0.
\end{equation}
As we mentioned before, the last term of this equation corresponds
to the inertial term. In a similar way as in the circular motion, the inertial
terms is already well known [9,10]; the components of the {\it inertial acceleration} will be assumed to be
\begin{displaymath}
\big(\,^{\eta}\Gamma^0\,_{jk}+\xi^l\partial_l
\,^{\eta}\Gamma^0\,_{jk}\big)
\big(\frac{dX^j}{d\tau}\frac{dX^k}{d\tau}+2\frac{dX^j}{d\tau}\frac{d\xi^k}{d\tau}\big)=0,
\end{displaymath}
\begin{displaymath}
\big(\,^{\eta}\Gamma^3\,_{jk}+\xi^l\partial_l
\,^{\eta}\Gamma^3\,_{jk}\big)
\big(\frac{dX^j}{d\tau}\frac{dX^k}{d\tau}+2\frac{dX^j}{d\tau}
\frac{d\xi^k}{d\tau}\big)=0,
\end{displaymath}
\begin{displaymath}
\big(\,^{\eta}\Gamma^2\,_{jk}+\xi^l\partial_l
\,^{\eta}\Gamma^2\,_{jk}\big)
\big(\frac{dX^j}{d\tau}\frac{dX^k}{d\tau}+2\frac{dX^j}{d\tau}\frac{d\xi^k}{d\tau}\big)=0,
\end{displaymath}
\begin{displaymath}
\big(\,^{\eta}\Gamma^1\,_{jk}+\xi^l\partial_l
\,^{\eta}\Gamma^1_{jk}\big)
\big(\frac{dX^j}{d\tau}\frac{dX^k}{d\tau}+2\frac{dX^j}{d\tau}\frac{d\xi^k}{d\tau}\big)=
\big(\frac{d\vec{X}}{d\tau}\big)^2\,\frac{1}{\xi+\rho}\,-\big(\frac{d\vec{X}}{d\tau}\big)^2\,\frac{1}{\rho}=
\big(\frac{d\vec{X}}{d\tau}\big)^2\,\frac{1}{\rho}(-\frac{\xi^1}{\rho}).
\end{displaymath}
In the last expression we  consider $\xi$ to be small in relation to
 the curvature radius of the central geodesic $\rho$. Note that $\rho$ is not necessarily
  constant. However, we are assuming a planar trajectory.
In the usual formalism, the last term corresponds to the {\it relative centripetal force}
 between two particles following close trajectories.
\subsection{Examples of transverse linear dynamics}

We study some examples of transverse dynamics using the {\it linearized version}
 of the averaged Jacobi equation.
\begin{enumerate}
\item {\bf Motion in a normal magnetic dipole}

As we said before, the reference frame is the laboratory frame. The reference trajectory is
a solution of the averaged Lorentz force equation.
In this case the electromagnetic field is given by the expression
\begin{displaymath}
{\bf F}=\left(%
\begin{array}{cccc}
  0 & 0 & 0 & 0 \\
  0 & 0 & b_0 & 0 \\
  0 & -b_0 & 0 & 0 \\
  0 & 0 & 0 & 0 \\
\end{array}%
\right)
\end{displaymath}
where $b_0$ is the dipole strength. Since the magnetic field is constant, $\xi^l\partial_l {\bf F}^i\,_j =0$. Therefore, the equations
of motion for the transverse degrees of freedom $(\xi^1,\xi^3)$ are
\begin{displaymath}
\frac{d^2 \xi^1}{d\tau^2}
+\big(\frac{d\vec{X}}{d\tau}\big)^2\,\frac{1}{\rho}(-\frac{\xi^1}{\rho})=0,\quad
\frac{d^2 \xi^3}{d\tau^2}=0.
\end{displaymath}
Changing the parameter of the curve from $\tau\longrightarrow x^1$, one has  that:
\begin{displaymath}
\frac{d^2 \xi^1}{dl^2}-\frac{\xi^1}{\rho^2}=0,\quad \frac{d^2
\xi^3}{dl^2}=0.
\end{displaymath}
These are the standard equations for a normal dipole.

\item {\bf Motion in a skew magnetic dipole}

The electromagnetic field is given by the expression
\begin{displaymath}
{\bf F}=\left(%
\begin{array}{cccc}
  0 & 0 & 0 & 0 \\
  0 & 0 & -b_0 & 0\\
  0 & b_0 & 0 & 0 \\
  0 & 0 & 0 & 0 \\
\end{array}%
\right)
\end{displaymath}
In this case, the Jacobi equations are
\begin{displaymath}
\frac{d^2 \xi^1}{d\tau^2}
-\big(\frac{d\vec{X}}{d\tau}\big)^2\,\frac{1}{\rho}(-\frac{\xi^1}{\rho})=0,\quad
\frac{d^2 \xi^3}{d\tau^2}=0.
\end{displaymath}
Following the same procedure as before we end with the
equations for the deviation equation in a skew magnetic field:
\begin{displaymath}
\frac{d^2 \xi^1}{dl^2}-\frac{\xi^1}{\rho^2}=0,\quad \frac{d^2
\xi^3}{dl^2}=0.
\end{displaymath}

\item {\bf Motion in a normal quadrupole field combined with a dipole}

In this case the electromagnetic field has the form
\begin{displaymath}
{\bf F}(x)=\left(%
\begin{array}{cccc}
  0 & 0 & 0 & 0 \\
  0 & 0 & b_0-b_1 \xi^1 & 0\\
  0 & -b_0+b_1 \xi^1 & 0 & b_1 \xi^3 \\
  0 & 0 & -b_1 \xi^3 & 0\\
\end{array}%
\right)
\end{displaymath}
The Jacobi equation reduces to t
\begin{displaymath}
\frac{d^2 \xi^1}{d\tau^2}-
\frac{d X^j}{d\tau}\xi^l
\partial_{l} {\bf F}^1\,_{j}
+\big(\frac{d\vec{X}}{d\tau}\big)^2\,\frac{1}{\rho}(-\frac{\xi^1}{\rho})=0,
\end{displaymath}
\begin{displaymath}
\frac{d^2 \xi^3}{d\tau^2} +
\frac{dX^j}{d\tau}\xi^l
\partial_l {\bf F}^3\,_{j}=0.
\end{displaymath}
Let us consider the respective contributions $ \frac{d\xi^j}{d\tau}
\xi^l
\partial_l {\bf F}^1\,_{j}$ and $ \frac{d\xi^j}{ds} \xi^l
\partial_l {\bf F}^3\,_{j}$. Using Euler's theorem on homogenous
functions one gets the relations:
\begin{displaymath}
\frac{d X^j}{d\tau}\xi^l
\partial_l {\bf F}^3\,_{j}=\frac{d X^j}{d\tau}{\bf F}^3_{j}|_{\xi=0}.
\end{displaymath}
Then the differential equations are
\begin{displaymath}
\frac{d^2 \xi^1}{d\tau^2}-\xi^1 b_1
+\big(\frac{d\vec{X}}{d\tau}\big)^2\,\frac{1}{\rho}(-\frac{\xi^1}{\rho})=0,\quad
\frac{d^2 \xi^3}{d\tau^2}+\xi^3 b_1=0.
\end{displaymath}
Using the Euclidean length as a parameter of the
curve, one obtains
\begin{displaymath}
\frac{d^2\xi^1}{dl^2}-\xi^1\frac{\partial B^3}{\partial \xi^1}
+\big(\frac{d\vec{X}}{dl}\big)^2\,\frac{\xi^1}{\rho^2}=0,\quad
\frac{d^2 \xi^3}{dl^2}+
\frac{d\xi^2}{dl}{b}^1_{0}+\xi^3\frac{\partial B^1}{\partial
\xi^3}=0.
\end{displaymath}
In the second differential equation, the second term is zero, since
$({\bf b}^1_{0},{\bf b}^2_{0}, {\bf b}^3_{0})=(0,0,{\bf
b}^3_{0})$. Then we obtain the following differential equations
for the transverse motion:
\begin{displaymath}
\frac{d^2\xi^1}{d \tau^2}-\xi^1\,b_1
+\big(\frac{d\vec{X}}{d\tau}\big)^2\,\frac{\xi^1}{\rho^2}=0,\quad
\frac{d^2 \xi^3}{d\tau^2}+\xi^3\,b_1=0.
\end{displaymath}
These are the equations of the linear transverse dynamics in
quadrupoles combined with magnetic dipole fields using the proper
time parameter $\tau$. If we use the Euclidean length $l$, the equations
are
\begin{equation}
\frac{d^2\xi^1}{dl^2}-\xi^1\,b_1 +\frac{\xi^1}{\rho^2}=0, \quad
\frac{d^2 \xi^3}{dl^2}+\xi^3\,b_0=0,
\end{equation}
which are the standard equations in transverse dynamics in accelerator physics [9, 10].

\item {\bf Motion in a normal dipole combined with a 45 degrees quadrupole}

In this case the electromagnetic field is
\begin{displaymath}
{\bf F}=\left(%
\begin{array}{cccc}
  0 & 0 & 0 & 0 \\
  0 & 0 & b_0+b_1 \xi^3 & 0\\
  0 & -b_0-b_1 \xi^3 & 0 & b_1 \xi^1 \\
  0 & 0 & -b_1 \xi^1 & 0\\
\end{array}%
\right)
\end{displaymath}
Following the same procedure as before, we get the Jacobi equations
\begin{equation}
\frac{d^2\xi^1}{dl^2}+\xi^1\,b_1 +\frac{\xi^1}{\rho^2}=0, \quad
\frac{d^2 \xi^3}{dl^2}-\xi^3\,b_0=0.
\end{equation}
\end{enumerate}
The above examples show how the linear transverse dynamics can be
obtained from the Jacobi equation of the averaged Lorentz connection.

It is not possible to use this formulation for higher
multipole magnetic fields because the linear approximation breaks down. One possibility to incorporate higher modes is to consider the generalized Jacobi equation [59], which is a non-linear geodesic deviation equation.

We remark
again that with the approximation $<y>\longrightarrow
\frac{dX}{d\tau}$ the deviation equation of the averaged connection
and the Lorentz connection coincide. If one considers higher order effects, one can obtain  differences between the Jacobi equations of the Lorentz connections and averaged connection.

\section{Calculation of the averaged off-set effect between the reference trajectory and the central geodesic of the averaged Lorentz connection}

In this section we calculate the averaged difference between the reference trajectory and the solutions of the averaged connection.

\subsection{Calculation of the dispersion function in beam dynamics}

We follow the formalism developed in [9] for the treatment
of linear perturbations and dispersion. However we will
maintain the proper time $\tau$ as the parameter of the curves, in contrast
with the usual treatment, which uses the Euclidean length along the reference trajectory. In the following,
primes indicate derivatives with respect to the proper time.

It follows from {\it section 6.3} that the transverse dynamics is determined by equations of the form
\begin{equation}
u'' \, +K(\tau)u=0
\end{equation}
The general solution is of the form
\begin{displaymath}
u(\tau)=C(\tau)u_0\,+S(\tau)u'_0,\quad u'(\tau)=C'(\tau)u_0+S'(\tau)u'_0,
\end{displaymath}
with initial conditions
\begin{displaymath}
C(0)=1,\quad C'(0)=0;\quad S(0)=0,\quad S'(0)=1,
\end{displaymath}
 for arbitrary initial values $u_0$ and
$u'_0$. The functions $C(\tau)$ and $S(\tau)$ satisfy
\begin{displaymath}
C''(\tau)+K(\tau)S(\tau)=0,\quad S''(\tau)+K(\tau)S(\tau)=0.
\end{displaymath}

However, small perturbations can change the dynamics. The perturbed equation has the form:
\begin{equation}
u''(\tau) \, +K(\tau)u(\tau)=p(\tau)
\end{equation}
A particular solution for $(6.4.2)$ is
\begin{equation}
P(\tau)=\int^{\tau}_0 p(\tilde{\tau})G(\tau,\tilde{\tau})d\tilde{\tau},
\end{equation}
where $G(\tau,\tilde{\tau})$ is the Green function associated to the differential equation $(6.4.1)$.
One can prove that in the absence of dissipative forces (that is, which do not depend on the velocity of the particle), the Green
function of the differential equation is given by the following combination:
\begin{equation}
G(\tau,\tilde{\tau})=S(\tau)C(\tilde{\tau})-C(\tau)S(\tilde{\tau}).
\end{equation}
Therefore, the general solution for the equation (6.4.2) is
\begin{equation}
u(\tau)=a\,C(\tau)+b\,S(\tau)\,+P(\tau).
\end{equation}
 This solution breaks down if there are synchrotron radiation or other dissipative effects.

We will use standard notation of beam dynamics. If all the particles in a bunch do not have the same energy, one obtains for the transverse degrees of
freedom the following differential equation [9, pg 109], [10]:
\begin{displaymath}
u'' \, +K(\tau)u\,=\frac{1}{\rho_0}(\tau)\Delta_u, \quad
\Delta=\frac{\delta p}{p_0},\quad \delta p=\sqrt{(\delta
p_1)^2\,+(\delta p_2)^2\, +(\delta p_3)^2}.
\end{displaymath}
We need to assign a value to $\delta p$. One natural value is the maximal
value of $\{\|\vec{\xi(x)}\|_{\bar{\eta}},\,x\in {\bf M}\}$. Since {\bf M} is non-compact, we restrict to a compact domain ${\bf K}\subset {\bf M}$.
This definition does not depend on the particular
trajectory of each particle. The general solution for $u$ is
linear in the perturbation, therefore
\begin{displaymath}
u(\tau)=a_u\,C(\tau)+b_u\,S(\tau)\,+\Delta\,D(\tau):=a_u\,C(\tau)+b_u\,S(\tau)\,
+Off_u(\tau),\quad P(\tau)=\Delta D(\tau).
\end{displaymath}
where $a$ and $b$ depend on the initial values.

\subsection{Calculation of the off-set due to the deviation $\epsilon^j$}

In this subsection $({\bf M},\eta)$ is the $4$-dimensional Minkowski space-time.
We have shown in the previous section that in first order of approximation with respect to the degree $(a+b)$ of the monomials $\xi^a\cdot (\epsilon^m)^b$ and its derivatives, when we take the approximation $<y>\longrightarrow
\frac{d X}{dt}$, the differential
equation for the transverse motion is the Jacobi equation of the
averaged connection. Therefore we can consider the terms on $\epsilon^k$
in the averaged Jacobi equation as a perturbation and apply the method of the Green function.

From the definition of the off-set function for the transverse degrees of
freedom, we obtain
\begin{displaymath}
Off^{1,3}_u(\tau)=u^{1,3}(\tau)-a_{1,3}\,C^{1,3}(\tau)-\,
b_{1,3}\,S^{1,3}(\tau),
\end{displaymath}
the super-index refers to the transverse components $x^2$ and $x^3$ in the laboratory frame defined previously.
Using the corresponding Green function we obtain
\begin{displaymath}
Off^{1,3}_{u}(\tau) = \int^{\tau} _0\, p^{1,3}(\tau)G(\tau,\tilde{\tau})d\tilde{\tau}.
\end{displaymath}
The perturbation $p(\tau)$ is in this case defined by all the terms of the averaged
Jacobi equation which are not contained in the linearized equation respect to the degree $(a+b+c)$. Therefore let us re-write the
Jacobi equation of the averaged connection. Using $\epsilon_k\,=<y_k>-\frac{dX_k}{d\tilde{\tau}}$ we get
\begin{displaymath}
Off^{1,3}_{u}(\tau)= \int^{\tau}_0\,d\tilde{\tau}\, 2\,\frac{d
\xi^j}{d\tilde{\tau}}(\tilde{\tau})\cdot
\frac{dX^k}{d\tilde{\tilde{\tau}}}\Big(\frac{1}{2}\big(\,{\bf
F}^{1,3}\,_j(\tilde{\tau})\,\epsilon_k(\tilde{\tau})\,+{\bf F}^{1,3}\,_k(\tilde{\tau})\, \epsilon_j(\tilde{\tau})\,\big)\, +
\end{displaymath}
\begin{displaymath}
+ \frac{dX^j}{d\tilde{\tau}}\frac{dX^k}{d\tilde{\tau}}\Big( \,{\bf F}^{1,3}\,_m(\tilde{\tau})\, \big(\,
<y^m>(\tilde{\tau})\eta_{jk}\,-<y^m y^a y^k>(\tilde{\tau})\eta_{ja}\eta_{lk}\big)+
\end{displaymath}
\begin{equation}
+\xi^l\partial_l \,\big(\,{\bf F}^{1,3}\,_m(\tilde{\tau})\, \big(\,
<y^m>(\tilde{\tau})\eta_{jk}\,-<y^m y^a y^k>(\tilde{\tau})\eta_{ja}\eta_{lk}\big)\Big)\Big).
\end{equation}
This is an integro-differential equation for $Off^{1,3}_{u}$ as we can show. In the integrand of equation $(6.4.6)$ we can make the substitution
\begin{displaymath}
\frac{d \xi^{1,3}}{d\tilde{\tau}}(\tilde{\tau})\longrightarrow
\big(a_{1,3}\,C'^{1,3}(\tilde{\tau})+b_{1,3}\,S'^{1,3}(\tilde{\tau})+
Off_u(\tilde{\tau})'\big).
\end{displaymath}
We can also consider the derivatives in the longitudinal and
temporal direction using this notation, with a convenient choice
of the coefficients $a_{0,2}$ and $b_{0,2}$. Then we can write
\begin{displaymath}
\frac{d \xi^{j}}{d\tilde{\tau}}(\tilde{\tau})\longrightarrow
\big(a_j\,C'^j(\tilde{\tau})+b_j\,S'^j(\tilde{\tau})+Off^j_u(\tilde{\tau})'\big),\quad
j=0,1,2,3.
\end{displaymath}
In this expression repeated indices are not summed!
For the transverse degrees of freedom, the unperturbed solutions
are the same as before [10],
\begin{displaymath}
u^{1,3}(\tau)=u_0\,C(\tau)+u'_0\,S(\tau)\quad C''(\tau)+K(\tau)C=0,\,\,
S''(\tau)+K(\tau)S=0.
\end{displaymath}
For the longitudinal $j=2$ and temporal $j=0$ degrees of freedom,
one gets the following relations by comparison with the Jacobi
equation,
\begin{displaymath}
\frac{d \xi^{2}}{d\tilde{\tau}}(\tilde{\tau})\longrightarrow
\big(a_2\,C'^2(\tilde{\tau})+b_2\,S'^2(\tilde{\tau})+Off^2_u(\tilde{\tau})'\big),
\end{displaymath}
\begin{displaymath}
\frac{d \xi^{0}}{d\tilde{\tau}}(\tilde{\tau})\longrightarrow
\big(a_0\,C'^0(\tilde{\tau})+b_0\,S'^0(\tilde{\tau})+Off^0_u(\tilde{\tau})'\big)
\end{displaymath}
Let us consider the regime where $Off^0_u=Off^2_u =0,\,\, \forall u$.
Then the off-set function is
\begin{displaymath}
Off^{1,3}_u (\tau) = \int^{\tau}_0\,d\tilde{\tau}\,\Big(\sum^{3}_{j=0}
2\big(a_j\,C'^{j}(\tilde{\tau})+b_j\,S'^{j}(\tilde{\tau})+
Off^{j}_u(\tilde{\tau})'\big)(\tilde{\tau})\cdot
\frac{dX^k}{d\tilde{\tau}}\Big(\frac{1}{2}\big(\,{\bf
F}^{1,3}\,_j(\tilde{\tau})\,\epsilon_k(\tilde{\tau})\, +
\end{displaymath}
\begin{displaymath}
+ {\bf F}^{1,3}\,_k(\tilde{\tau})\, \epsilon_j(\tilde{\tau})\,\big)\,+
\frac{dX^j}{d\tilde{\tau}}\frac{dX^k}{d\tilde{\tau}}\Big({\bf F}^{1,3}\,_m(\tilde{\tau})\, \big(\,
<y^m>(\tilde{\tau})\eta_{jk}\,-<y^m y^a y^k>(\tilde{\tau})\eta_{ja}\eta_{lk}\big)+
\end{displaymath}
\begin{displaymath}
+\big(a_l\,C^{l}(\tilde{\tau})+b_l\,S^{l}(\tilde{\tau})+
Off^{l}_u(\tilde{\tau})\big)\partial_l \,\big(\,{\bf F}^{1,3}\,_m\,
\big(\, <y^m>(\tilde{\tau})\eta_{jk}\,-<y^m y^a
y^l>(\tilde{\tau})\eta_{ja}\eta_{lk}\big)\Big)\Big).
\end{displaymath}
This is an integro-differential equation for $Off^{1,3}_u$ that we formally can
solve iteratively. In {\it the Born approximation} one puts $Off^{l}_u(\tilde{\tau})=0$ in the integrand:
\begin{displaymath}
Off^{1,3}_{\xi}(\tau) = \int^{\tau}_0\,d\tilde{\tau}\,\Big(
2\big(a_j\,C'^{j}(\tilde{s})+b_j\,S'^{j}(\tilde{\tau})\big)\cdot
\frac{dX^k}{d\tilde{\tau}}\Big(\frac{1}{2}\big(\,{\bf
F}^{1,3}\,_j(\tilde{\tau})\,\epsilon_k(\tilde{\tau})\, + {\bf F}^{1,3}\,_k(\tilde{\tau})\, \epsilon_j(\tilde{\tau})\,\big)\,+
\end{displaymath}
\begin{displaymath}
+
\frac{dX^j}{d\tilde{\tau}}\frac{dX^k}{d\tilde{\tau}}\Big( \,{\bf F}^{1,3}\,_m(\tilde{\tau})\, \big(\,
<y^m>(\tilde{\tau})\eta_{jk}\,-<y^m y^a y^k>(\tilde{\tau})\eta_{ja}\eta_{lk}\big)+
\end{displaymath}
\begin{displaymath}
+\big(a_l\,C^{l}(\tilde{\tau})+b_l\,S^{l}(\tilde{\tau})\big)\partial_l
\,\big(\,{\bf F}^{1,3}\,_m(\tilde{\tau})\, \big(\, <y^m>(\tilde{\tau})\eta_{jk}\,-<y^m y^a
y^l>(\tilde{\tau})\eta_{ja}\eta_{lk}\big)\Big)\Big).
\end{displaymath}
Therefore, we get the expression
\begin{displaymath}
Off^{1,3}_{\xi}(\tau) = \int^{\tau}_0\,d\tilde{\tau}\,\Big(
2\frac{d\xi^j}{d\tilde{\tau}}\cdot
\frac{dX^k}{d\tilde{\tau}}\Big(\frac{1}{2}\big(\,{\bf
F}^{1,3}\,_j(\tilde{\tau})\,\epsilon_k(\tilde{\tau})\, + {\bf F}^{1,3}\,_k(\tilde{\tau})\, \epsilon_j(\tilde{\tau})\,\big)\,
\end{displaymath}
\begin{displaymath}
+
\frac{dX^j}{d\tilde{\tau}}\frac{dX^k}{d\tilde{\tau}}\Big( \,{\bf F}^{1,3}\,_m\, \big(\,
<y^m>(\tilde{\tau})\eta_{jk}\,-<y^m y^a y^k>(\tilde{\tau})\eta_{ja}\eta_{lk}\big)+
\end{displaymath}
\begin{equation}
+\xi^l\partial_l \,\big(\,{\bf F}^{1,3}\,_m(\tilde{\tau})\, \big(\,
<y^m>(\tilde{\tau})\eta_{jk}\,-<y^m y^a y^l>(\tilde{\tau})\eta_{ja}\eta_{lk}\big)\Big)\Big).
\end{equation}
This expression depends on the particular solution $u$. A way to eliminate this dependence is to take the following {\it average}
\begin{displaymath}
<Off^{1,3}_{u}>(\tau) = \int^{\tau}_0\,d\tilde{\tau}\,\Big(
2\epsilon^j(\tilde{\tau})\cdot
\frac{dX^k}{d\tilde{\tau}}\Big(\frac{1}{2}\big(\,{\bf
F}^{1,3}\,_j(\tilde{\tau})\,\epsilon_k(\tilde{\tau}) +{\bf F}^{1,3}\,_k(\tilde{\tau})\, \epsilon_j(\tilde{\tau})\big)\,
\end{displaymath}
\begin{displaymath}
+ \frac{dX^j}{d\tilde{\tau}}\frac{dX^k}{d\tilde{\tau}}\Big( \,{\bf F}^{1,3}\,_m(\tilde{\tau})\, \big(\,
<y^m>(\tilde{\tau})\eta_{jk}\,-<y^m y^a y^k>(\tilde{\tau})\eta_{ja}\eta_{lk}\big)+
\end{displaymath}
\begin{equation}
+<\xi^l>(\tilde{\tau})\partial_l \,\big(\,{\bf F}^{1,3}\,_m(\tilde{\tau})\, \big(\,
<y^m>(\tilde{\tau})\eta_{jk}\,-<y^m y^a y^l>(\tilde{\tau})\eta_{ja}\eta_{lk}\big)\Big)\Big).
\end{equation}

The averaged off-set is therefore an observable quantity. It is determined by:
\begin{enumerate}
\item The reference trajectory $X(\tau)$, which is a geodesic of the averaged connection and as we have discussed before, it is known theoretically.

\item The tangent velocity field $\frac{dX}{d\tau}$ along the reference trajectory. This is known theoretically.

\item The external electromagnetic field ${\bf F}^{1,3}\,_m(x)$,

\item The value of the vector field $\epsilon^k(\tau)=\,<y^k(\tau)>\,-\frac{dX}{d\tau}$,

\item The first, second and third moments of the distribution function
$f(x(\tau),p(\tau))$ along the reference trajectory.
\end{enumerate}
Finally, in the case that the perturbation does not change
significatively along the trajectory, we obtain that the term
containing derivatives are neglected. Therefore,
\begin{displaymath}
<Off^{1,3}_{u}>(\tau) = \int^{\tau}_0\,d\tilde{\tau}\,<\Big(
2\epsilon^j(\tilde{\tau})\frac{dX^k}{d\tilde{\tau}}\cdot \big(\,{\bf
F}^{1,3}\,_j(\tilde{\tau})\,\epsilon_k (\tilde{\tau})+{\bf F}^{1,3}\,_k(\tilde{\tau})\, \epsilon_j(\tilde{\tau})\big)>\, +
\end{displaymath}
\begin{equation}
+ \frac{dX^j}{d\tilde{\tau}}\frac{dX^k}{d\tilde{\tau}}\Big( \,{\bf F}^{1,3}\,_m(\tilde{\tau})\, \big(\,
<y^m>(\tilde{\tau})\eta_{jk}\,-<y^m y^a y^k>(\tilde{\tau})\eta_{ja}\eta_{lk}\big)\Big)\Big).
\end{equation}

In the case of a delta function distribution we have$<Off^{1,3}_{u}>(\tau)=0$. This means that the averaged off-set effect is a collective effect.

\section{Longitudinal beam dynamics and corrections
from the Jacobi equation of the averaged connection}

Let $({\bf M},\eta)$ be the Minkowski space-time and consider an inertial coordinate
 system defined by the vector field $Z=\frac{\partial}{\partial t}$, that corresponds to the laboratory frame.
The interaction of an ultra-relativistic bunch of particles with an external longitudinal
 electric field is described by the Faraday tensor
\begin{displaymath}
{\bf F}=\left(
\begin{array}{cccc}
  0 & 0 & E_2(x) &  0 \\
  0 & 0 & 0 & 0 \\
  -E_2(x) & 0 & 0 & 0 \\
  0 & 0 & 0 & 0 \\
\end{array}
\right)
\end{displaymath}
For narrow distributions one obtains the following
condition,
\begin{displaymath}
\frac{dX^j}{d\tau}\frac{d\xi_j}{d\tau}=\mathcal{O}^1.
\end{displaymath}
This relation can be seen as follows.
For the linear dynamics $\xi=(\xi,0,-\xi,0)$ in the laboratory frame.
 Using the ultra-relativistic limit $\frac{dX^k}{ds}=(1+E,0,E,0)$,
with $E>>1$.

Then the averaged Jacobi equation for the
limit $\epsilon^j\longrightarrow 0$ in the ultra-relativistic regime are
\begin{displaymath}
\frac{d^2 \xi^i}{d\tau^2}+
2\frac{d\xi^j}{d\tau}\frac{dX^k}{d\tau}\Big(\frac{1}{2}({\bf F}^i\,_{j}
\frac{dX^m}{d\tau}\eta_{mk}+ {\bf
F}^i\,_{k}\frac{dX^m}{d\tau}\eta_{mj})\Big) +2\xi^l
\partial_l \Big(\frac{1}{2}({\bf F}^i\,_{j}\frac{dX^m}{d\tau}\eta_{mk}+ {\bf
F}^i\,_{k}\frac{dX^m}{d\tau}\eta_{mj})\Big)\cdot
\end{displaymath}
\begin{displaymath}
\cdot\frac{dX^j}{d\tau}\frac{dX^k}{d\tau}+
\big(\,^{\eta}\Gamma^i\,_{jk}+\xi^l\partial_l
\,^{\eta}\Gamma^i\,_{jk}\big)
\big(\frac{dX^j}{d\tau}\frac{dX^k}{d\tau}+2
\frac{dX^j}{d\tau}\frac{d\xi^k}{d\tau}\big)=0.
\end{displaymath}

For the above longitudinal electric field, the equations of motion are
\begin{displaymath}
\frac{d^2 \xi^0}{d\tau^2}+
\frac{d\xi^2}{d\tau}\frac{dX^k}{d\tau}E_2<y^m>\eta_{mk}+
\frac{d\xi^k}{d\tau}\frac{dX^2}{d\tau}E_2<y^m>\eta_{mk}+
\end{displaymath}
\begin{displaymath}
+\xi^l \partial_l
\Big(\frac{dX^k}{d\tau}\frac{dX^2}{d\tau}E_2<y^m>\eta_{mk}+
\frac{dX^2}{d\tau}\frac{dX^j}{d\tau}E_2<y^m>\eta_{mj}\Big)+
\end{displaymath}
\begin{displaymath}
+\big(\,^{\eta}\Gamma^0_{jk}+\xi^l\partial_l
\,^{\eta}\Gamma^0\,_{jk}\big)
\big(\frac{dX^j}{d\tau}\frac{dX^k}{d\tau}+2\frac{dX^j}{d\tau}
\frac{d\xi^k}{d\tau}\big)=0,
\end{displaymath}
\begin{displaymath}
\frac{d^2 \xi^2}{d\tau^2}+
\frac{d\xi^0}{d\tau}\frac{dX^k}{d\tau}E_2<y^m>\eta_{mk}-
\frac{d\xi^k}{d\tau}\frac{dX^0}{d\tau}E_2<y^m>\eta_{mk}-
\end{displaymath}
\begin{displaymath}
\xi^l \partial_l
\Big(\frac{dX^j}{d\tau}\frac{dX^0}{d\tau}E_2<y^m>\eta_{mj}+
\frac{dX^0}{d\tau}\frac{dX^j}{d\tau}E_2<y^m>\eta_{mj} \Big)+
\end{displaymath}
\begin{displaymath}
+\big(\,^{\eta}\Gamma^2\,_{jk}+\xi^l\partial_l
\,^{\eta}\Gamma^2\,_{jk}\big)
\big(\frac{dX^j}{d\tau}\frac{dX^k}{d\tau}+2\frac{dX^j}{d\tau}
\frac{d\xi^k}{d\tau}\big)=0,
\end{displaymath}
\begin{displaymath}
\frac{d^2 X^1}{d\tau^2}+\big(\,^{\eta}\Gamma^1\,_{jk}+\xi^l\partial_l
\,^{\eta}\Gamma^1\,_{jk}\big)
\big(\frac{dX^j}{d\tau}\frac{dX^k}{d\tau}+
2\frac{dX^j}{d\tau}\frac{d\xi^k}{d\tau}\big)=0,
\end{displaymath}
\begin{displaymath}
\frac{d^2 X^2}{d \tau^2}+\big(\,^{\eta}\Gamma^2\,_{jk}+\xi^l\partial_l
\,^{\eta}\Gamma^2_{jk}\big)
\big(\frac{dX^j}{d\tau}\frac{dX^k}{d\tau}+2\frac{dX^j}{d\tau}
\frac{d\xi^k}{d\tau}\big)=0.
\end{displaymath}
In an inertial coordinate system, the inertial terms are zero. Therefore, the
system of equations in the linear longitudinal dynamics in the
ultra-relativistic regime is
\begin{displaymath}
\frac{d^2 \xi^0}{d\tau^2}+
\frac{d\xi^2}{d\tau}\frac{dX^k}{d\tau}E_2<y^m>\eta_{mk}+
\frac{d\xi^k}{d\tau}\frac{dX^2}{d\tau}E_2<y^m>\eta_{mk} +
\end{displaymath}
\begin{displaymath}
+2\xi^l
\partial_l (\frac{dX^k}{d\tau}\frac{dX^2}{d\tau}E_2<y^m>\eta_{mk})=0
\end{displaymath}
\begin{displaymath}
\frac{d^2 \xi^2}{d\tau^2}-
\frac{d\xi^2}{d\tau}\frac{dX^k}{d\tau}E_2<y^m>\eta_{mk}-
\frac{d\xi^k}{d\tau}\frac{dX^0}{d\tau}E_2<y^m>\eta_{mk}-
\end{displaymath}
\begin{displaymath}-2\xi^l
\partial_l (\frac{dX^j}{d\tau}\frac{dX^0}{d\tau}E_2<y^m>\eta_{mj})=0
\end{displaymath}
\begin{displaymath}
\frac{d^2 X^1}{d\tau^2}=0,
\end{displaymath}
\begin{displaymath}
\frac{d^2 X^2}{d \tau^2}=0.
\end{displaymath}
If $\epsilon^k=<y^k>-\frac{dX^k}{d\tau}\approx 0$ and since
the distribution function has support on the unit hyperboloid,  $<y^k>\frac{dX_k}{d\tau}\approx 1+\alpha$.
Using also
the decoupling condition $\frac{dX^k}{d\tau}\frac{d\xi_k}{d\tau}\approx
0$, we have that
\begin{equation}
\frac{d^2 \xi^0}{d\tau^2}+ \frac{d\xi^2}{ds}E_2 +2\xi^l
\partial_l (\frac{dX^2}{d\tau}E_2)=0,
\end{equation}
\begin{equation}
\frac{d^2 \xi^2}{d\tau^2}- \frac{d\xi^0}{d\tau}E_2-2\xi^l
\partial_l (\frac{dX^0}{d\tau}E_2)=0,
\end{equation}
\begin{equation}
\frac{d^2 X^1}{d\tau^2}=0,
\end{equation}
\begin{equation}
\frac{d^2 X^2}{d \tau^2}=0.
\end{equation}
The only non-trivial equation has the form
\begin{displaymath}
\frac{d^2 \xi^2}{d\tau^2}+ \frac{d\xi^2}{d\tau}E_2-2\xi^l
\partial_l (\frac{dX^0}{d\tau}E_2)=0.
\end{displaymath}
In the ultra-relativistic limit the velocity field
$\frac{dX^0}{d\tau}=\gamma $ (in units where the speed of light is equal to $1$). Therefore, the equation above can be written as
\begin{equation}
\frac{d^2 \xi^2}{d\tau^2} +\frac{d\xi^2}{d\tau}E_2-2\gamma \xi^l
\partial_l
E_2({X}+\xi^2)=0.
\end{equation}
We perform the following approximation in equation $(6.5.5)$:
\begin{displaymath}
E_2({X}+\xi^2)=E_2(X)+\xi^k\frac{\partial}{\partial \xi^k}E_2.
\end{displaymath}
Due to the translational invariance of the partial derivatives,
$\partial_l \equiv\frac{\partial}{\partial \xi^l}$ in the above expressions, by the chain rule. Then we have
\begin{displaymath}
\frac{d^2 \xi^2}{d\tau^2} +\frac{d\xi^2}{d\tau}E_2-2\gamma
\xi^k\frac{\partial}{\partial \xi^k}(E_2(X+\xi)-E_2(X))=0.
\end{displaymath}
If $E_2(X+\xi)$ can be
approximated linearly on $\xi$, using Euler's theorem of
homogeneous functions one gets the following expression:
\begin{displaymath}
\frac{d^2 \xi^2}{d\tau^2} +\frac{d\xi^2}{d\tau}E_2(\tau)-2\gamma(\tau)
(E_2((X+\xi)-E_2(X))=0.
\end{displaymath}

\subsection{Examples}
\begin{enumerate}

\item {\bf Constant longitudinal electric field.}

In this case the
equation of motion is
\begin{displaymath}
\frac{d^2 \xi^2}{d\tau^2} +\frac{d\xi^2}{d\tau}E_2=0.
\end{displaymath}
A particular solution is
\begin{displaymath}
\xi^2=-\frac{\xi^2}{E_2}(e^{-E_2(\tau-\tau_0)}-1).
\end{displaymath}

\item {\bf Alternate longitudinal electric field}. In this case,
the electric field is
\begin{displaymath}
E_2(X^2+\xi^2)=E_2(0)sin(w_{rf}(X^2+\xi^2)).
\end{displaymath}
The differential equation is
\begin{displaymath}
\frac{d^2 \xi^2}{d\tau^2}
+\frac{d\xi^2}{d\tau}E_2(0)sin(w_{rf}(X^2+\xi^2))-2\gamma
E_2(0)(sin(w_{rf}(X^2+\xi^2))-sin(w_{rf}X^2))=0.
\end{displaymath}
We can expand this equation in $\xi$, since $\xi$ is small
\begin{displaymath}
\frac{d^2 \xi^2}{d\tau^2}
+\frac{d\xi^2}{ds}E_2(0)(sin(w_{rf}X^2)+cos(w_{rf}X^2)\xi^2)-2\gamma
 E_2(0)(cos(w_{rf}X^2)\xi^2)=0.
\end{displaymath}
At first order in $\xi^2$ we have the equivalent expression
\begin{displaymath}
\frac{d^2 \xi^2}{d\tau^2}
+\frac{d\xi^2}{d\tau}E_2(0)sin(w_{rf}X^2)-2\gamma
E_2(0)(cos(w_{rf}X^2)\xi^2)=0.
\end{displaymath}
We choose the initial phase such that $sin(w_{rf}X^2)\simeq 0$; therefore
$cos(w_{rf}X^2)\simeq 1$ and the equation is

\begin{equation}
\frac{d^2 \xi^2}{d\tau^2} -2\gamma  E_2(0)\xi^2=0.
\end{equation}

This equation is a linearized version of the ordinary linearized longitudinal
dynamics and has exactly the same structure [11].
\end{enumerate}

\section{Conclusion}

We have seen that the
linear beam dynamics in accelerator physics can be obtained from the Jacobi equation of the averaged Lorentz connection
for electromagnetic fields ${\bf F(\xi)}$ linear in $\xi$. In particular we have proved that in the case where the magnetic fields are linear on the deviation $\xi$, like in a dipole and quadrupole magnetic fields, the transverse dynamics can be interpreted as the dynamics of the Jacobi equation. A similar conclusion follows for the so-called longitudinal dynamics.

The advantages of this derivation are that it involves
 only observable quantities. Another advantage is that the averaged dynamics is linked through the distribution
function to the collective behavior of the system.

The $off$-$set$ effect calculated in {\it section 6.4} can be
of relevance in the control and diagnostics of the beam parameters, since it is a direct observable
quantity and because it is related with the collective description of the bunch of particles.

\chapter{Conclusions}
\section{General conclusions}

This thesis describes the foundations of the averaged Lorentz force equation
 and its applications in the mathematical modeling of ultra-relativistic bunches of charged particles.
Through the averaged Lorentz dynamics, a new theoretical justification of the use of fluid models
in beam dynamics has been obtained. We have seen that for relativistic dynamics and
for narrow probability distribution functions (in the sense that the diameter $\alpha$ of the distribution function obtained
using the Euclidean metric in the laboratory frame is very small compared with the mass of the
particles at rest), it is justified to substitute the original kinetic model based on the Vlasov equation
    by an averaged charged cold fluid model. One can control this approximation
    in terms of the {\it energy} of the bunch, the diameter of the distribution $\alpha$ and the time of the evolution of the bunch, all these variables measured in the laboratory frame.

     Our method does not provide a system of differential equations for the fluid models.
     Instead it provides estimates of the differential operations which occur in the
     definition of the charged cold fluid model. Given a particular physical situation one
     can decide whether the given model is satisfactory or is a bad approximation using those bounds.

      The averaged model has been an essential tool in obtaining those results. The reason
      is that the averaged Lorentz force equation is simpler than the Lorentz force equation.
      The existence of normal coordinate systems associated to the averaged
      connection $<\,^L\nabla>$ has been crucial for the calculations performed in {\it chapter 5}.

\paragraph{}
There are some advantages using
 the averaged Lorentz equation in the description of the dynamics of a bunch
  of particles instead of the Lorentz force equation:
\begin{enumerate}
\item At the classical level the electromagnetic field is
measured by the effect on charged point particles. In case of the electromagnetic
interaction of charged particles with an external electromagnetic field,
the Lorentz force equation is the geodesic equation of a complicated non-linear {\it almost-connection} (for the notion of {\it almost-connection} see the appendix).
The equation can be simplified by considering the associated averaged Lorentz connection. The averaged Lorentz
connection is simpler than the original one, since the averaged
connection is an affine connection on ${\bf M}$. This property allows
 us to have important technical tools (in particular normal coordinates).

\item Since the averaged Lorentz connection is simpler than the original Lorentz connection,
one can use it to perform numerical simulations of the dynamics of a
bunch of particles. The simplified model must allow a better
 numerical implementation in the simulation of the dynamics of a bunch
  containing a large number of charged particles.

\end{enumerate}

In a similar way, there are advantages using the averaged Vlasov equation instead of the original Vlasov equation:
\begin{enumerate}
\item The calculations using the averaged Vlasov equation can be simplified using normal coordinates
systems. This is basically because the underlying averaged Lorentz
connection is an affine and torsion free connection.

\item It involves only the low moments of the distribution
function $\tilde{f}$. This dependence on low moments, together with
the fact that $\tilde{f}$ is an approximation of $f$ in the regime
when the dynamics is ultra-relativistic and the distributions are
narrow, can explain why current fluid models need only to consider
differential equations for the first, second and third moments,
while the fourth moments are neglected. That is, the justification
for the truncation schemes in fluid models comes from the structure
of the averaged Lorentz connection.
\end{enumerate}

\paragraph{}
Since the metric approach to geometrization contains intrinsic problems,
we have considered an alternative geometric treatment of the Lorentz force.
The framework has been the theory of non-linear connections associated with second order
differential equations. We have applied this theory to the Lorentz force equation $(4.1.1)$.

However, starting a dynamical model from a differential equation can be insufficient
for some purposes. For instance, one could not recover a canonical Hamiltonian formalism. From
a geometric point of view, if one has only
a differential equation, one can not speak of variational problems.

The Lorentz force equation has been interpreted as the non-linear Berwald connection of
a spray vector field $^L\chi$. However, the above points imply that one has to look
for a variational formulation. A consistent formulation has been described in
{\it section 4.3}. This formulation is technically complicated and requires sheaves
and pre-sheaves theory. It is a non-metric approach to semi-Randers spaces.
\subsection{Brief discussion of the main problem}

The main problem formulated in {\it section 2.3} has been addressed: we have obtained
a recipe such that we can decide when the charged cold fluid model is a good approximation
or not to the underlying kinetic model, in the context of accelerator physics.
The conclusion is that for the actual accelerator machines, the charged cold fluid
model is a good approximation to the underlying Vlasov model and that it can be
used in the description of the beam dynamics.

As a byproduct we have obtained an averaged Lorentz force equation which is simpler
than the original Lorentz force equation. We have proved that in physical situations,
the Lorentz force equation can be substituted by the averaged Lorentz equation.

\section{Generalizations and open problems}

Some open problems that the present thesis leaves for
future investigation are the following:
\begin{enumerate}
\item The geometric averaged method can be applied to other dynamical systems.
As an example, let us consider the structure of the Lorentz
connection $^L\nabla$. If $\eta(y,y)=1$, the connection coefficients are polynomial in $y$ up to third
order. Therefore, it could be interesting to consider it in a
similar way as an effective connection that has semi-spray
coefficients of the form:
\begin{equation}
G^i (x,y)=a^i\,_{jk}(x)y^jy^k \,+a^i\,_{jkl}(x)y^jy^ky^l\,+a^i\,_{jklm}(x)y^jy^ky^ly^m\,+
a^i\,_{jklmn}(x)y^jy^ky^ly^my^n.
\end{equation}
This is the simplest generalization of the semi-spray coefficients of
the connection $^L\nabla$.

There are some restrictions on these semi-sprays coefficients
\begin{enumerate}

\item The resulting geodesic equations must be gauge invariant. This means that there is an
intrinsic gauge symmetry transformation and the tensors depend only of gauge invariant quantities.

\item The resulting equation must be Lorentz invariant.

\item The study of the basic dynamics of these connections compatible with
the laws of the Electrodynamics (in particular with the Larmor law
[3, pg 469]). A second order
dynamics of a charged point particle must be a particular class of
these dynamics, since it is well known and experimentally checked.

\item It is well known that if the back-reaction force is taken into account,
then the Lorentz-Dirac equation follows from a balance equation [1-4]. One
can investigate if a generalization of the type $(7.2.1)$ can accommodate a second
order equation which remains second order, after considering back-reaction.

\item It could be interesting to clarify whose semi-sprays of the
type $(7.2.1)$ or possible generalizations of the Lorentz
force coming from non-linear electrodynamics [53].

\end{enumerate}

\item The method used in this thesis to justify the use of the charged cold fluid model
 is applicable to other fluid models, like the warm fluid model. Indeed one
  can discriminate which model is better in some particular application,
depending on the diameter of the given distribution function and the energy of the beam.

\item One can use the averaged Lorentz equation as a model in numerical simulations.
Since the structure of the averaged equation is simpler than the original equation,
it could be convenient to use it in numerical simulations in beam dynamics.

\item We have assumed some technical hypotheses.
Although these assumptions are well defined and hold for the physical examples that
we have in mind, it could be interesting to reduce the number of assumptions, obtaining more general results.
\end{enumerate}

Apart from these points, directed to the core of this thesis, there are several points
which could deserve more attention:

\begin{enumerate}

\item The notion of {\it almost-connection}. As is explained in the appendix,
it is a natural generalization of the notion of connection. We think that it
is non-trivial, since the non-extensibility of the covariant derivatives and parallel transport is a difficult property to prove.

\item The notion of semi-Randers space [14, work in progress]. This a basic notion that we have need to discuss but which is of interest on its own. Also it is interesting to generalize the notion of semi-Rander space associated with non-abelian symmetries.

\item The notions of structural stability introduced in {\it section 6.4}. It indicates a topological structure behind current fluid models which deserves investigation.

\end{enumerate}

\newpage

\appendix
\chapter{Mathematical appendix}

\section{Proofs for {Chapter 3}} The following proofs are adapted from reference [22].

{\bf Proof of proposition 3.3.2}. The consistency of the equation
$(3.3.4)$ is proved in the following way,
\begin{displaymath}
<{{\nabla}}>_X f=<\pi _2 |_u  \nabla _{h _u (X)}  \pi^* _v
f>_u\, =
\end{displaymath}
\begin{displaymath}
=<\pi _2 |_u {h _u (X)}  (\pi^* _v f)>_u =<\pi _2 |_u \pi^* _u
(X  (f))>_u \, =<X f >_u\, =X f.
\end{displaymath}
The fourth equality holds because the definition of the horizontal
local basis $\{ \frac{{\delta}}{{\delta}x^0 }| _u ,...,
\frac{{\delta}}{{\delta}x^{n-1} }|_u \}$ in terms of
$\{\frac{\partial}{\partial x^i},\, i=0,...,n-1\}$ and
$\{\frac{\partial }{\partial y^i},\, i=0,...,n-1\}$.

We check the properties characterizing a linear covariant
derivative associated with the averaged connection $<{\nabla}>$:

\begin{enumerate}
\item $<{{\nabla}}>_X$ is a linear application acting on sections of $ {\bf TM}$:
\begin{displaymath}
<{{\nabla}}>_{X}(Y_1 +Y_2 )=<{{\nabla}}>_{X} Y_1
+<{{\nabla}}>_{X} Y_2
\end{displaymath}
\begin{displaymath}
<{{\nabla}}>_{X} \lambda Y= \lambda\tilde{{\nabla}}_{X} Y,
\end{displaymath}
\begin{equation}
\quad \forall \,\,  Y_1 , Y_2, Y\in {\bf \Gamma TM}, \lambda\in
\mathcal{{\bf R}}, \quad X\in {\bf T}_x {\bf M}.
\end{equation}
For the first equation, the proof consists in the following
calculation,
\begin{displaymath}
<{{\nabla}}>_{X }(Y_1 +Y_2 )=<\pi _2|_u  {\nabla}_{{h}_u
(X)} {\pi}^* _v (Y_1 +Y_2 )>=<\pi _2 |_u  {\nabla}_{{h}_u (X)}
{\pi}^* _v Y_1
>+
\end{displaymath}
\begin{displaymath}
+<\pi _2 |_u   {\nabla}_{{h}_u (X)}  {\pi}^* _v Y_2
>_v =<{{\nabla}}>_{X} Y_1 +<{{\nabla}}>_{X} Y_2 .
\end{displaymath}

For the second condition we have that
\begin{displaymath}
<{{\nabla}}>_{X } ({\lambda}Y)=<\pi _2 |_u {\nabla}_{{\iota}_u
(X)} {\pi}^*_v ({\lambda}Y)>_v ={\lambda}<\pi _2 |_u
{\nabla}_{{\iota}_u (X)} {\pi}^*_v Y>=\lambda<{{\nabla}}>_{X }
Y.
\end{displaymath}

\item $<{{\nabla}}>_{X } Y$ is a ${\bf \mathcal{F}}$-linear with
respect to $X$:
\begin{displaymath}
<{{\nabla}}>_{X_1 +X_2 } Y=<{{\nabla}}>_{X_1 }
Y+<{{\nabla}}>_{X_2 } Y,
\end{displaymath}
\begin{displaymath}
<{{\nabla}}>_{fX} (Y)=f(x)<{{\nabla}}>_{X } Y,
\end{displaymath}
\begin{equation}
\forall \, Y\in {\bf TM},v\in {\pi}^{-1}(z),\quad X,X_1,X_2 \in
{\bf T}_x {\bf M},\, f\in{\bf \mathcal{F}}({\bf M}).
\end{equation}
To prove the first equation is enough the following
calculation:
\begin{displaymath}
<{{\nabla}}>_{X_1 +X_2 }Y=<\pi _2 |_u ({\nabla}_{{h}_u
(X_1 +X_2 )}){\pi}^*_v Y>_v =
\end{displaymath}
\begin{displaymath}
=<\pi _2 |_u {\nabla}_{{h}_u (X_1 )}{\pi}^*_v Y >_v +<\pi _2
|_u {\nabla}_{{h _u}(X_2 )}{\pi}^*_v Y >=
\end{displaymath}
\begin{displaymath}
=(<{{\nabla}}>_{X_1 } Y)+(<{{\nabla}}>_{X_2 } Y).
\end{displaymath}

For the second condition the proof is similar.

\item The Leibnitz rule holds:
\begin{equation}
<{{\nabla}}>_{X } (fY)=df(X)Y+f<{{\nabla}}>_{X } Y,\quad
\forall \,\,  Y\in\Gamma {\bf TM},\, f\in {\bf \mathcal{F}}({\bf M}),\quad
X \in {\bf T}_x {\bf M},
\end{equation}
 where $df(X)$ is the action of the $1$-form
$df\in {\bf \Lambda}^1 {\bf M}$ on $X\in {\bf T}_x {\bf M}$ In
order to prove $(3.3.7)$ we use the following property:
\begin{displaymath}
{\pi}^*_v (fY)={\pi}^*_v f{\pi}^*_v Y,\quad \forall \,\, Y\in \Gamma {\bf
TM},\, f\in {\bf \mathcal{F}}({\bf M}).
\end{displaymath}
Then
\begin{displaymath}
<{\nabla}>_{X} (fY)=<\pi _2 |_u {\nabla}_{{h}_u
(X)}{\pi}^*_v (fY)>_u =<\pi _2 |_u {\nabla}_{{h}_u (X)}
{\pi}^*_v (f){\pi}^*_v Y>_u=
\end{displaymath}
\begin{displaymath}
=<\pi _2 |_u (\nabla _{{h}_u (X)}( {\pi}^* _v f)){\pi}^*_v
(Y)>_u+<\pi _2 |_u ({\pi}^* _u f){\nabla}_{{h}_u (X)}{\pi}^*_v
(Y)>_u=
\end{displaymath}
\begin{displaymath}
=<\pi _2 |_u ({{h}_u (X)}({\pi}^*_v f)) {\pi}^*_v (Y)>_u + f_x
<\pi _2 |_u {\nabla}_{{h}_u (X)} {\pi}^*_v (Y)>_u =
\end{displaymath}
\begin{displaymath}
=<(X_x f)\pi _2 |_u {\pi}^*_u (Y)>_u +f_x <\pi _2 |_v
{\nabla}_{{\iota}_u (X)}{\pi}^*_v (Y)>_u .
\end{displaymath}

For the first term we perform the following simplification,
\begin{displaymath}
<(X f)\pi _2 |_u  {\pi}^*_u (Y)>_u =(X f)<\pi _2 |_u {\pi}^*_u
(Y)>_u =
\end{displaymath}
\begin{displaymath}
=(X f)(< \pi _2 |_u {\pi}^*_u >_u )Y=(X f)(<I
>_u )Y .
\end{displaymath}

Returning to the above calculation, we obtain

\begin{displaymath}
\tilde{\nabla}_{X} (fY)=\tilde{\nabla} _{X}(f)Y+f
\tilde{\nabla}_{X}Y
=df(X)Y+f\tilde{{\nabla}}_{X }Y.
\end{displaymath}
\hfill$\Box$
\end{enumerate}
The generalization of $<\,^L\nabla>$ to higher order tensor bundles
is as usual.
\paragraph{}
{\bf Proof for Proposition 3.3.4 and corollary 3.3.5}
\begin{displaymath}
T_{<{\nabla}>}(X,Y)=<\pi _2 |_u \nabla _{h _u (X)} \pi^*
_w
>_u Y -<\pi _2 |_u \nabla _{h _u (Y)} \pi^* _w >_u X -[X,Y]=
\end{displaymath}
\begin{displaymath}
=<\pi _2 |_u  \nabla _{h _u (X)}  \pi^* _u >_u Y -<\pi _2 |_u
\nabla _{h _u (Y)} \pi^* _u >_u X
\end{displaymath}
\begin{displaymath}
 -<\pi _2 |_u  \pi^* _u [X,Y]>=
\end{displaymath}
\begin{displaymath}
=<\pi _2 |_u \big(\nabla _{h _u (X)}\pi^* Y -\nabla _{h _u
(Y)}\pi^*X -\pi^*[X,Y] \big) >_u \,=<T(X,Y)>.
\end{displaymath}
On the other hand:
\begin{displaymath}
T_{<{\nabla}>}(X,Y)=<\pi _2 |_u \nabla _{h_u (X)} \pi^*
_w
>_u Y -<\pi _2 |_u \nabla _{h _u (Y)} \pi^* _w >_u X -[X,Y]=
\end{displaymath}
\begin{displaymath}
=<\pi _2 |_u \nabla _{h _u (X)}  \pi^* _u >_u Y -<\pi _2 |_u
\nabla _{h _u (Y)} \pi^* _u >_u X
 -<\pi _2 |_u \pi^* _u [X,Y]>=
\end{displaymath}
\begin{displaymath}
=<\pi _2 |_u \big(\nabla _{h _u (X)}\pi^* Y -\nabla _{h _u
(Y)}\pi^*X -\pi^*[X,Y] \big) >_u \,=0.\hfil\Box
\end{displaymath}
\newpage
\section{Ordinary differential equations}

We state here a result on differential equations that we have used several times in the text.
It was used in [21] to show the local
existence and uniqueness of parameterized geodesics and similar results on the
existence and uniqueness of solutions of ordinary differential equations,
\begin{teorema}
Let $f_i(t,y,s)$ be a family of $n$ functions defined in $\mid
t\mid < \delta$ and $(y,s)\in {\bf D}$, where ${\bf D}$ is an open
set in ${\bf R}^{n+m}$. If $f_i(t,y,s)$ are continuous in $t$ and
differentiable of class $\mathcal{C}^1$ in $y$, then there exists
a unique family $\phi(t,y,s)$ of $n$ functions defined in $\mid
t\mid < \acute{\delta}$ and $(\eta , s)\in \mathcal{\acute{D}}$,
where $0<\acute{\delta}< \delta$ and $\acute{{\bf D}}$ is an open
subset of ${\bf D}$ such that
\begin{enumerate}
\item $\phi(t,\eta,s)$ is differentiable of class $\mathcal{C}^1$
in $t$ and $\eta$.

\item $\frac{\partial{\phi}_i}{\partial t}=f_i (t,
\phi(t,\eta,s),s).$

\item $\phi (0,\eta ,s)=\eta.$
\end{enumerate}
If $f (t,y,s)$ is differentiable of class $\mathcal{C}^p$,
$0\leq p\leq \infty$, in $y$ and $s$, then $\phi(t,\eta,s)$ is
differentiable of class $\mathcal{C}^{p+1}$ in $t$ and of class
$\mathcal{C}^q$ in $\eta$ and $s$.
\end{teorema}
The main use in the thesis was the following. Assume that the
geodesic equations are of the form
\begin{displaymath}
\frac{dy^i}{dt}=G^i(x,y,s);\quad \frac{dx^i}{dt}=y^i,\quad
i=1,...,n.
\end{displaymath}

We have applied the theorem to the case where s defines a smooth homotopy.
Therefore, we have to apply to the case where $s$ is
$1$-dimensional and $f=(y^i, G^i)$, with $i=1,...,2n$.

We have used this theorem to prove smoothness properties of the solutions of several differential equations.

\newpage
\section{Basic notions of asymptotic analysis}

In this {\it appendix} we follow the notation of [43, {\it chapter 1}].
\begin{definicion}
For two complex functions $f(z)$ and $g(z)$ we write $f(z)=\mathcal{O}(g(z))$ as $z\rightarrow z_0$
 iff there is a positive constants $K$ and $c$ such that for $0<|z-z_0|<c$ one has that $|f|\leq\,K |g|$.
\end{definicion}
\begin{definicion}
An  infinite sequence of functions $\{\Phi_n(z),\,
n=1,2...\,\}$ is an asymptotic sequence as $z\rightarrow z_0$ if
\begin{displaymath}
\lim_{z\rightarrow z_0} \frac{{\Phi}_{n+1}(z)}{\Phi_n(z)}=0,\,\quad \forall \,n \in {\bf N}.
\end{displaymath}
\end{definicion}
\begin{definicion}
Given the asymptotic sequence $\{\Phi_n(z),\, n=1,2...\,\}$ as
$z\rightarrow z_0$ the expression
\begin{displaymath}
f(z)=\sum^{N-1}_{n=1} a_n \Phi_n(z)+\mathcal{O}(\Phi_{N}(z))\, \textrm{as}\,
z\rightarrow z_0,
\end{displaymath}
is said to be an asymptotic expansion of the function $f(z)$ in terms of $\{\Phi_n(z),\,n=1,2... \}$.
\end{definicion}
Given an asymptotic sequence $\{\Phi_n(z),\,n=1,2... \}$, if for a given function $f$ the asymptotic expansion exists it is unique,
with coefficients given by
\begin{displaymath}
a_k=\lim_{z\rightarrow z_0} \Big\{\frac{f(z)-\sum^{k-1}_{n=1} a_n
\Phi_n(z)}{\Phi_k}\Big\}.
\end{displaymath}

Asymptotic expansions have the following elementary properties:
\begin{enumerate}
\item The first order term in the expansion is the leading term, which provides the major contribution to the series.

\item
The asymptotic expansion of a function depends on the choice of
the asymptotic sequence.

\item The asymptotic expansion as
$z\rightarrow z_0$ is a linear operation respect to the function which is being expanded: $f\mapsto (a_1,...,a_{N-1})$ is linear in $f$, $\forall N>1$.

\item If the derivative
of the function $f$ has an asymptotic expansion, the asymptotic
expansion of the derivative of a function is the derivative term by term of the
expansion of $f$.

\item The asymptotic expansion of the real
integral of a function is the real integral of the expansion, integrated term
by term.

\item The product of asymptotic expansions is in general
non-asymptotic. However, the product of asymptotic
expansions in power series around the same point is also an asymptotic expansion.
\end{enumerate}
If the asymptotic expansion is convergent, there is a convergence region such that the approximation of a function by an asymptotic series is becoming more accurate when we consider more terms in the expansion. If the asymptotic series is a power series and
divergent, the better accuracy attainable by an expansion consists on
taking the expansion
\begin{displaymath}
f(z)\sim \sum^{N}(z)_{n=1} a_n \Phi_n (z)
\end{displaymath}
such that $N(z)$ is the last term
that the magnitude of the terms $|a_n\Phi_n|$ is decreasing with $n$.
After this term $a_N \Phi_N$, the rest of the terms are
increasing. Usually, the error in the expansion in a power
series is of the order of the first term neglected.

\newpage
\section{Notion of almost-connection}

During our analysis of the Lorentz force equation,
 we have considered the associated non-linear connection on ${\bf TN}$ and the associated linear connections on $\pi^*{\bf TM}\longrightarrow {\bf \Sigma}$. However, strictly speaking the system of differential
equations $(4.1.1)$ does not define a non-linear connection on ${\bf T\Sigma}$.
The reason is the following. Let us fix a point $u\in {\bf \Sigma}$.
There exists a natural embedding
$e:{\bf \Sigma} \hookrightarrow {\bf TM}$. One can consider the sub-bundle $e({\bf \Sigma})\hookrightarrow {\bf TM}$
 and the {\it covariant derivative} $^LD$ acting on elements of $\Gamma{\bf T}e({\bf \Sigma})$.
This covariant derivative can be extended to derive sections
of the extended bundle $\bigsqcup _{u\in {\bf \Sigma}}{\bf T}_u {\bf N}$, which is a sub-bundle of ${\bf TM}$:
\begin{displaymath}
^L\hat{D}:\Gamma( \bigsqcup _{u\in {\bf \Sigma}}{\bf T}_u {\bf N})\times \Gamma(\bigsqcup _{u\in {\bf \Sigma}}{\bf T}_u {\bf N})
\longrightarrow \Gamma(\bigsqcup _{u\in {\bf \Sigma}}{\bf T}_u {\bf N}),
\end{displaymath}
\begin{displaymath}
^L\hat{D}_{\hat{X}} \hat{Y}=\, \hat{X}^i \frac{\partial \hat{Y}^j}{\partial x^i} \frac{\partial}{\partial x^j}+ \,^L\Gamma^i\,_{jk}\hat{Y}^k \hat{X}^j\,\frac{\partial}{\partial x^i},\quad \hat{X},\hat{Y}\in \Gamma( \bigsqcup _{u\in {\bf N}}{\bf T}_u {\bf TM}).
\end{displaymath}
From the definition of $^L D$, this operator cannot be extended in a smooth and
natural way to be a covariant derivative on ${\bf TM}$. This is because of the appearance
of factors $\sqrt{\eta(y,y)}$ in the connection coefficients: the function $\sqrt{\eta(y,y)}$ is not defined in the whole ${\bf TM}$. The non-extendibility of this function to {\bf TM} is the origin of the problem to extend the covariant derivative along arbitrary directions in {\bf TM}.

The above fact suggests the existence of a mathematical object which is a generalization of the ordinary notion of
covariant derivative in the sense that allows covariant derivatives along {\it outer directions
to the manifold}, although not obtained as a restriction of an {\it ambient covariant
derivative} operator.

{\bf Example}. The Lorentz connection was obtained from the Lorentz force equation assuming
that the torsion is zero. We obtained the following connection coefficients:
\begin{displaymath}
^L\Gamma^i\,_{jk} = \,^{\eta}\Gamma^i\,_{jk} + T^i\,\,_{jk}+L^i\,_{jk},
\end{displaymath}
\begin{displaymath}
L^i\,_{jk}=\frac{1}{2{\eta(y,y)}}({\bf F}^i\,_{j}y^m\eta_{mk}+ {\bf
F}^i\,_{k}y^m\eta_{mj}),
\end{displaymath}
\begin{displaymath}
T^i\,_{jk}={\bf F}^i\,_m
\frac{y^m}{\sqrt{\eta(y,y)}}\Big(\eta_{jk}-\frac{1}{\eta(y,y)}\eta_{js}
\eta_{kl}y^s y^l)\Big).
\end{displaymath}
This defines a rule to derive sections of $ {\bf T \Sigma}$ along directions of $ {\bf T \Sigma}$.
The same coefficients provide a rule to derive sections of
$\bigsqcup _{u\in {\bf \Sigma}}{\bf T}_u {\bf TM}$ along directions of ${\bf TM}$, but we can extend the definition of the covariant derivative acting on sections of {\bf TTM}.

A related issue is the following. On the unit hyperboloid recall that $T^i\,_{jk}y^j y^k =0$.
Therefore, if one considers instead an alternative connection $\tilde{\nabla}$ defined by the connection coefficients
\begin{displaymath}
^L\tilde{\Gamma}^i\,_{jk} = \,^{\eta}\Gamma^i\,_{jk}\, + L^i\,_{jk},
\end{displaymath}
the corresponding geodesic equation (parameterized by the proper time of the Lorentzian
metric $\eta$) is again the Lorentz force equation [49].
 Therefore we see that both $^L\nabla$ and $\tilde{\nabla}$ reproduce
 the Lorentz force and  both are torsion-free (the connection coefficients are symmetric
 in the lower indices). This is in contradiction with the fact that a connection of Berwald type (linear or non-linear) is
 determined by the set of all geodesics as parameterized curves on the base
  manifold {\bf M} and the torsion tensor (for linear connections the procedure can be seen in [35]; for non-linear connections, a procedure to define the connection is described for example in [34].

 A solution to this dilemma comes from the fact that the Lorentz force equation applies only to time-like
   trajectories for which the tangent velocity vector fields live on the unit hyperboloid ${\bf \Sigma}$; that is, we cannot extend the operator
\begin{displaymath}
^L\hat{D}:\Gamma \big( \bigsqcup _{u\in {\bf \Sigma}}{\bf T}_u {\bf TM}\big)\times \Gamma \big(\bigsqcup _{u\in {\bf \Sigma}}{\bf T}_u {\bf TM}\big)
\longrightarrow \Gamma\big(\bigsqcup _{u\in {\bf \Sigma}}{\bf T}_u {\bf TM}\big)
\end{displaymath}
to a genuine operator of the form
\begin{displaymath}
^LD:\Gamma({\bf TM})\times \Gamma({\bf TM})
\longrightarrow \Gamma({\bf TM}).
\end{displaymath}
In other words, we cannot extract enough information from the Lorentz force equation to determine a {\it projective connection} [44], because on the base manifold, we do not have information about all the possible geodesics. This is why we cannot determine the connection.

 One can be tempted to extend the Lorentz force equation to another equation valid for any kind of trajectory. This is partially accomplished by the averaged connection. However, these extensions could be non-natural or non unique and indeed hide a natural object such like {\it almost-connection}.

\paragraph{}

{\bf Preliminary Definition of Almost-Connection}

Let {\bf M} be a manifold of dimension $n$. An almost-connection
 is a spray vector field $\chi\in \Gamma{\bf TM}$ defined on a sub-bundle ${\bf TD}\hookrightarrow {\bf TTM}$, with ${\bf TD}$ a sub-bundle of arbitrary co-dimension. Associated with ${\bf D}$ is the corresponding almost projective covariant derivative on $\pi^*{\bf TM}$.

{\bf Examples.}
\begin{enumerate}
\item A projective connection (in the sense of Cartan [44]) is an almost connection such that ${\bf D}={\bf TM}$.

\item The Lorentz connection $^L\nabla$ provides an example where ${\bf \Sigma}={\bf D}\neq {\bf TM}$. One can define a {\it Koszul connection} acting on $\Gamma\big(\bigsqcup _{u\in {\bf \Sigma}}{\bf T}_u {\bf TM}\big)$.

\item The averaged covariant derivative $<\,^L\nabla>$ is an affine connection and therefore an almost connection in the above sense.
\end{enumerate}

One can consider the corresponding linear connections $^L\nabla$
 and $\tilde{\nabla}$ on $\pi^*{\bf TM}$. Generally, their averaged connections $<\,^L\nabla>$ and $<\,\tilde{\nabla}>$ (introduced in {\it section 4.7}) are not the same. However, we would like to have an averaged operation which is well defined for the objects in a given category. By definition this will be the category of {\it  almost-connections} and the corresponding morphisms. We require that the result of the averaging operation be the same for each representative belonging to the same almost connection,
\begin{definicion}
Let {\bf M} be a manifold of dimension $n$. A non-linear almost connection
 is the maximal set of semi-sprays $\chi$ defined on a sub-bundle ${\bf TD}\subset{\bf TTM}$ such that they have the same averaged linear covariant derivative $<\nabla>$ and the same torsion tensor
\begin{equation}
\hat{T}(\hat{X},\hat{Y}):=\,^L\hat{D}_{\hat{X}}\hat{Y}-
\,^L\hat{D}_{\hat{Y}}\hat{X}-\,[\hat{X},\hat{Y}],\quad \hat{X},\hat{Y}\in \Gamma{\bf TD}.
\end{equation}
\end{definicion}
Associated with ${\chi}$ over {\bf D} is the corresponding {\it almost-covariant derivative} on the bundle $\pi^*{\bf TM}\longrightarrow {\bf D}$. Any of the connections in the same almost-connection has the same averaged connection. The distance function $(4.6.3)$ can also being defined for almost-connections.

The general properties of {\it almost connections} are being explored in a separate work. Some of these properties are based on straightforward generalizations of the quantities associated with Koszul connections. For instance, the generalization of the curvature tensor is
\begin{equation}
\hat{R}(\hat{X},\hat{Y},\hat{Z}):=\,^L\hat{D}_{\hat{X}}\,^L\hat{D}_{\hat{Y}}\hat{Z}-\,^L\hat{D}_{\hat{Y}}\,^L\hat{D}_{\hat{X}}\hat{Z}-
\,^L\hat{D}_{[\hat{X},\hat{Y}]}\hat{Z},\quad \hat{X},\hat{Y}, \hat{Z}\in \Gamma {\bf TD}.
\end{equation}

It is not easy to handle a notion of parallel transport for almost-connections, since in general there will be initial conditions such that the auto-parallel curve goes out from the sub-bundle ${\bf D}$, even for arbitrary short-time parallel transports. Indeed, for auto-parallel curves whose initial velocity vector is not on ${\bf TD}$, the trajectory {\it goes out} from ${\bf D}$ after any finite time. This point is related with the notion of general connection [62]. However, the notion of almost-connection is even more general, since the solution of the projections on {\bf N} condition $\nabla_{h(\xi)}\pi^*(\xi)=0$ could not exists. For a generalized connection, this projection always exist.
\newpage
\section{Basic notions of Sobolev spaces}

Sobolev spaces are complete normed vector spaces (therefore Banach spaces) where the norms measure also the derivatives of the function. We have used Sobolev norms in {\it chapter 5} to introduce the bounds on some differential expressions appearing in the averaged Vlasov model. In this appendix we provide the basic notions of Sobolev norms and some additional notions of Sobolev spaces to understand the meaning of these expressions and its implications for further generalizations. We will follow references [41] and [57] because of their clarity in exposition. We assume that the theory of Lebesgue's integral holds.

Let ${\bf \Omega}$ be an open set of a manifold. There are some basic definitions:
\begin{definicion}
Two functions $f,g:{\bf \Omega}\longrightarrow {\bf R}\cup\{\pm \infty\}$ are equivalent iff they are equal almost everywhere on ${\bf \Omega}$, that is, the sub-set $A\subset {\bf \Omega}$ where $g\neq f$ is a null set.
\end{definicion}

\begin{definicion}
Let ${\bf \Omega}$ be open, $p\geq 1$ $(p\in {\bf R})$. $L^p({\bf \Omega})$ is the set of all Lebesgue measurable functions $f:{\bf \Omega} \longrightarrow {\bf R}\cup \{\pm \infty\}$ for which $|f|^p$ is integrable over ${\bf \Omega}$. For $f\in L^p({\bf \Omega})$ we set
\begin{displaymath}
\|f\|_{L^p({\bf \Omega})}:=\Big(\int_{{\bf \Omega}} |f|^p \,dx\big)^{\frac{1}{p}}.
\end{displaymath}
\end{definicion}
In order to define Sobolev norms, we introduce weak derivatives.
\begin{definicion}
Let $f\in L^1({\bf \Omega})$. A function $v\in {\bf L}^1({\bf \Omega})$ is called the weak derivative of $f$ in the direction $x^i$ if
\begin{displaymath}
\int_{{\bf \Omega}} v(x)\phi(x)\,dx=-\int_{{\bf \Omega}}f(x)\frac{\partial \phi(x)}{\partial x^i}\,dx.
\end{displaymath}
\end{definicion}
\begin{definicion}
Let $f\in L^1({\bf \Omega})$, ${\bf \beta}:=(\beta_1,...,\beta_d)$ with $\beta_i\geq 0$ $(i=1,...,d)$, $|{\bf \beta}|:=\sum^d_{i=1} \beta_i >0$. Then
\begin{displaymath}
D_{{\bf \beta}} \phi :=(\frac{\partial}{\partial x^1})^{\beta_1}\,\cdot\cdot\cdot(\frac{\partial}{\partial x^1})^{\beta_d}\phi,\quad f\in \mathcal{C}^{|{\bf \beta}|}({\bf \Omega}).
\end{displaymath}
\end{definicion}
\begin{definicion}
A function $v\in L^{1}({\bf \Omega})$ is called $\beta$-weak derivative of $f$ and written $v=D_{{\bf \beta}}f$ if
\begin{displaymath}
\int_{{\bf \Omega}} v(x)\phi(x)\,dx=\,(-1)^{|{\bf \beta}|}\int_{{\bf \Omega}}f(x)D_{{\bf \beta}}\phi(x)\,dx.
\end{displaymath}
\end{definicion}
We can now define Sobolev spaces and Sobolev norms:
\begin{definicion}
For $k\in {\bf N}$ a natural number, $1\leq p\leq \infty,$ we define the Sobolev space $\mathcal{W}^{k,p}(\Omega)$ by $\mathcal{W}^{k,p}(\Omega):=\{f\in L^p(\Omega)\,|\, D_{{\bf \beta}}f\,\, \textrm{exists and is in }L^p({\bf \Omega})               \}$. The Sobolev norms are defined by
\begin{equation}
\|f\|_{{k,p}}:=\big(\,\sum_{|{\bf \beta}|\leq k}\int_{{\bf \Omega}}|D_{{\bf \beta}}f(x)|\,dx\big)^{\frac{1}{p}},
\end{equation}
and by
\begin{equation}
\|f\|_{{k,p}}:=\,\sum_{|{\bf \beta}|\leq k} max\,sup_{x\in {\bf \Omega}}|D_{{\bf \beta}}f(x)|.
\end{equation}
\end{definicion}
\subsection{Basic properties of Sobolev spaces}

Some properties of Sobolev spaces are the following:
\begin{teorema}
Let $k\in {\bf N}$ be a natural number and $1\leq p< \infty$. Then the following is true
\begin{enumerate}
\item The normed space $(\mathcal{W}^{k,p}({\bf \Omega}), \|f\|_{{k,p}})$ is a Banach space.

\item Let $f\in \mathcal{W}^{k,p}({\bf \Omega})$ and $\theta\in \mathcal{C}^1({\bf R})$. Then Poincare's inequality holds:
\begin{equation}
|f|_{(0,p)}\leq C({\bf \Omega},p)|f|_{1,p}, \quad f\in (\mathcal{W}^{1,p}_c({\bf \Omega})),
\end{equation}
with $C({\bf \Omega},p)$ a constant and $\mathcal{W}^{1,p}_c({\bf \Omega})$ the completion
of the space of smooth functions on ${\bf \Omega}$ with compact support.

\item $\mathcal{W}^{p,p}_c({\bf \Omega})$ is a Hilbert space.
\end{enumerate}
\end{teorema}
There is a relevant result (Sobolev embedding theorem) which gives sufficient conditions to embed Sobolev spaces in $L^q$ spaces, in spaces of continuous functions or in spaces with some regularity conditions [41,57,58]. We do not need this theorem here, but it could be important for of further generalizations of the results of {\it chapter 5}.

\subsection{Sobolev spaces of functions defined on manifolds}

The discussion before of Sobolev spaces is restricted to open domains. One can extend the definition to manifolds. First one reduces to an open domain ${\bf U}\subset {\bf M}$. If the manifold is differentiable, there are partitions of the unity [32]. First, one can speak of $\mathcal{W}^{1,p}_c({\bf U})$. Using an atlas of the manifold and a associated partition of the unity, we can define the Sobolev norms $(\mathcal{W}^{k,p}({\bf M}), \|f\|_{{k,p}})$ from $(\mathcal{W}^{k,p}({\bf \Omega}), \|f\|_{{k,p}})$ in the usual way as the integrals are defined over manifolds from the local description using coordinates neighborhoods [32, {\it chapter 4}].

\subsection{Examples of Sobolev spaces}

There are some Sobolev norms that we have used in {\it chapter 5}. These are:
\begin{enumerate}
\item $(\mathcal{W}^{1,1}({\bf M}), \|f\|_{{1,1}})$. This is a Hilbert space, with a norm defined by the function
\begin{displaymath}
\|f\|_{\mathcal{W}^{1,1}}:=\big(\,\sum_{|{\bf \beta}|\leq 1}\int_{{\bf \Omega}}|D_{{\bf \beta}}f(x)|\,dx\big)=\int_{{\bf \Omega}}(|f(x)|\,+|\sum_i \partial_i f|)\,dx.
\end{displaymath}

\item $(\mathcal{W}^{0,2}({\bf M}), \|f\|_{{0,2}})$. The norm of a function is defined as
\begin{displaymath}
\|f\|_{{0,2}}:=\Big(\int_{{\bf \Omega}} \,|f(x)|^2\,dx\Big)^{\frac{1}{2}}.
\end{displaymath}
\end{enumerate}

\end{document}